\documentclass[11pt]{article}
\pdfoutput = 1

\usepackage[utf8]{inputenc}
\usepackage{color,graphicx}
\usepackage{epsfig}
\usepackage{amsmath}
\usepackage{verbatim}
\usepackage{amssymb}
\usepackage{mathabx}
\usepackage[mathcal]{eucal}
\usepackage{stmaryrd}
\usepackage{empheq}
\usepackage{esint}
\usepackage{physics}
\usepackage{amsfonts}
\usepackage{array}
\usepackage{setspace}
\usepackage{braket}
\usepackage{float}
\usepackage{url}
\usepackage{mathtools}
\usepackage{tensor}
\usepackage{bbold}
\usepackage{slashed}
\usepackage{tikz}
\usepackage{graphicx}
\usepackage{epstopdf}
\usepackage{subcaption}
\usepackage[labelfont=bf,font={sf}]{caption}
\usepackage{pgfplots}
\usepackage{mathrsfs}
\usepackage{fancybox}
\usepackage{eurosym}
\usepackage{tcolorbox}
\usepackage{tensor}
\allowdisplaybreaks
\usepackage[normalem]{ulem}
\usepackage{changepage}

\usepackage[margin = 2.2cm]{geometry}
\usepackage[ragged]{footmisc}
\usepackage[bookmarks=true,bookmarksnumbered=false,
hyperindex=true,bookmarksopen=true,hyperfigures=true,
colorlinks=true,linkcolor=webblue,citecolor=webgreen,urlcolor=webblue,breaklinks]{hyperref}
\definecolor{webgreen}{rgb}{0, 0.5, 0}
\definecolor{webblue}{rgb}{0, 0, 0.5}
\definecolor{webred}{rgb}{0.5, 0, 0}
\definecolor{darkgreen}{rgb}{0,0.5,0}

\newcommand{\dpi}{\mathcal{D}}

\newcommand{\psil}{\psi^{\mathsf{\scriptscriptstyle L}}}
\newcommand{\psir}{\psi^{\mathsf{\scriptscriptstyle R}}}

\def\ben{\begin{equation}}
\def\een{\end{equation}}

   \let\d=\delta 
   
     \let\r=v

\def\be{\begin{equation}}
\def\ee{\end{equation}}
\def\ba{\begin{array}}
\def\ea{\end{array}}

\def\dalemb#1#2{{\vbox{\hrule height .#2pt
       \hbox{\vrule width.#2pt height#1pt \kern#1pt
               \vrule width.#2pt}
       \hrule height.#2pt}}}

\newcommand{\bea}{\begin{eqnarray}}
\newcommand{\eea}{\end{eqnarray}}

\def\Im{{{\frak{Im}}}}
\def\Re{{{\frak{Re}}}}

\let\tilde=\widetilde

\def\rH{{\mathsf{H}}}

\renewcommand{\d}{\mathrm{d}}
\renewcommand{\i}{\mathrm{i}}

\numberwithin{equation}{section}

\begin{document}

\thispagestyle{empty}
\begin{center}
    ~\vspace{5mm}

     {\LARGE \bf 
   \begin{adjustwidth}{-0.5cm}{-1cm}
   An entropic puzzle in periodic dilaton gravity and DSSYK
   \end{adjustwidth}
   }
    
   \vspace{0.4in}

    \begin{adjustwidth}{-1cm}{-1cm}
     {\bf Andreas Blommaert$^{1,2}$, Adam Levine$^3$, Thomas G. Mertens$^4$, Jacopo Papalini$^4$, Klaas Parmentier$^5$}
    \end{adjustwidth}

    \vspace{0.4in}
    {$^1$Institute for Advanced Study, Princeton, NJ 08540, USA\\
    $^2$SISSA and INFN, Via Bonomea 265, 34127 Trieste, Italy\\
    $^3$Center for Theoretical Physics, Massachusetts Institute of Technology,\\ Cambridge, MA 02139, USA\\
    $^4$Department of Physics and Astronomy, Ghent University,\\
    Krijgslaan, 281-S9, 9000 Gent, Belgium\\
    $^5$Department of Physics, Columbia University, New York, NY 10027, USA}
    \vspace{0.1in}
    
    {\tt blommaert@ias.edu, thomas.mertens@ugent.be, 	jacopo.papalini@ugent.be, arlevine@mit.edu, k.parmentier@columbia.edu}
\end{center}

\vspace{0.4in}

\begin{abstract}
\noindent We study 2d dilaton gravity theories with a periodic potential, with special emphasis on sine dilaton gravity, which is holographically dual to double-scaled SYK. The periodicity of the potentials implies a symmetry under (discrete) shifts in the momentum conjugate to the length of geodesic slices. This results in divergences. The correct definition is to gauge this symmetry. This discretizes the geodesic lengths. Lengths below a certain threshold are null states. Because of these null states, the entropy deviates drastically from Bekenstein-Hawking and the Hilbert space becomes finite dimensional. The spacetimes have a periodic radial coordinate. These are toy models of 2d quantum cosmology with a normalizable wavefunction. We study two limiting dualities: one between flat space quantum gravity and the Heisenberg algebra, and one between topological gravity and the Gaussian matrix integral. We propose an exact density of states for certain classes of periodic dilaton gravity models.
\end{abstract}

\pagebreak
\setcounter{page}{1}
\setcounter{tocdepth}{2}
\tableofcontents

\section{Introduction}\label{sect:intro}
Consider models of 2d periodic dilaton gravity \cite{Grumiller:2002nm,henneaux1985quantum,Ikeda:1993fh,Louis-Martinez:1993bge,Witten:2020ert}, characterized by a periodic dilaton potential $V(\Phi)$
\begin{equation}
    I=\frac{1}{2}\int\d x \sqrt{g}(\Phi R + V(\Phi))\,+\text{ boundary}\,,\quad V(\Phi+2\pi a)=V(\Phi)\,.\label{1.1}
\end{equation}
The classical solutions are spacetimes with a black hole horizon and a cosmological horizon, analogous to Schwarzschild dS (see section \ref{sect:genericpotential}). The ADM energy of these spacetimes is bounded from below \emph{and} above \cite{Witten:2020ert}
\begin{equation}
    E(\Phi_2)-E(\Phi_1)=\int^{\Phi_2}_{\Phi_1}\d \Phi\, V(\Phi)\,.
\end{equation}
Therefore, these are candidate toy models for UV complete quantum gravity, and quantum cosmology. Our goal is to understand the quantization of these models. We will find that the periodic nature of the potential implies drastic changes as compared to the previously studied quantization of asymptotically AdS$_2$ models with potentials $V(\Phi)$ which approach $2\Phi$ for $\Phi\to\infty$ \cite{Maxfield:2020ale,Witten:2020wvy}. In particular, the periodicity forces us to gauge certain symmetries in the Hamiltonian formulation. The result is the \textbf{discretization of the length} of Cauchy slices. States with a discrete length below a certain cutoff $L<L_\text{min}$ become null, perhaps surprisingly. This has major consequences. The semiclassical entropy is much lower than the black hole area law \eqref{6.28} $S<S_\text{BH}=2\pi \Phi_h$:
\begin{equation}
    \begin{tikzpicture}[baseline={([yshift=-.5ex]current bounding box.center)}, scale=0.7]
 \pgftext{\includegraphics[scale=1]{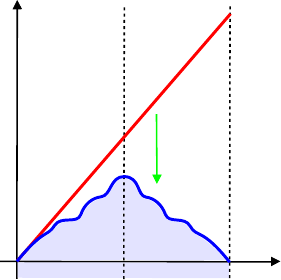}} at (0,0);
    \draw (2.8,-2.8) node {$\Phi_h$};
    \draw (1.5,-2.8) node {$\pi a$};
    \draw (-0.3,-2.8) node {$\pi a/2$};
    \draw (-2.1,-2.8) node {$0$};
    \draw (1.15,-1) node {\color{blue}$S$};
    \draw (1.2,0.05) node {\color{green}\text{gauge}};
    \draw (0.4,1.8) node {\color{red}$S_\text{BH}$};
  \end{tikzpicture}
\end{equation}
This result can be interpreted as some new type of \textbf{entropic paradox} in periodic dilaton gravity. Much like in the usual information paradox, at the quantum level the entropy reduces (at leading order) due to an overcounting of states in a naive semiclassical gravitational description \cite{Penington:2019kki}. What is missing (and why there is a puzzle) is some effective semiclassical rule that generalizes the Bekenstein-Hawking/QES formula \cite{Bekenstein:1973ur,Engelhardt:2014gca,Penington:2019npb,Almheiri:2019hni} to apply also in these models. Such a rule should explain ``intuitively'' why the entropy mismatches with black hole area. Finding such a generalized rule is outside our current scope, but we find it plausible that the deficit is explained by taking into account the effects on entropy of adding an ``observer'' in cosmological spacetimes \cite{Witten:2023xze}. See section 6.1 of \cite{Blommaert:2024ydx} for comments along these lines in this set-up.

In parallel to the statement that black holes should be described by a quantum system with $A/4G_\text{N}$ degrees of freedom, the ``central dogma for cosmological horizons'' has been proposed \cite{Susskind:2021yvs}. Our models clearly deviate from either statement even at the semi-classical level, raising a significant puzzle. This does not contradict earlier analysis, because our models describe black holes with extremely high energy. To resolve the puzzle and derive a generalization of the semi-classical entropy formula, one would need to understand a path integral (i.e.\,semiclassical) implementation of gauging the Hamiltonian symmetries. We speculate in the discussion \textbf{section \ref{sect:discnew}} that the zero mode of $\Phi$ may be compact in the path integral, but this is not an obvious symmetry of the theory on generic topology.

To derive these universal features, we will proceed on a practical level by first studying one example in great detail. This detailed understanding enables a rather straightforward generalization to generic dilaton potentials. The generalization is presented in \textbf{section \ref{sect:genericpotential}}. The example is \textbf{sine dilaton gravity}
\begin{equation}\label{1.4sdaction}
    I=\frac{1}{2}\int\d x \sqrt{g}(\Phi R + 2\sin (\Phi))\,+\text{ boundary}\,.
\end{equation}
This is a powerful example, because sine dilaton gravity is holographically dual to DSSYK \cite{Blommaert:2024ydx}. A part of deriving that duality was finding the location of the holographic screen, and the associated spacetime contour, which is not obvious for a periodic potential. We will export that lesson (and others) to general periodic dilaton gravity. 

We believe that these periodic dilaton gravities represent a new universality class of exactly solvable models that can be described as deformations of the ordinary Gaussian matrix integral (without double scaling). We present minor pieces of suggestive evidence, but do not prove this suggestion. 

Before summarizing the structure and results of this work in \textbf{section \ref{subsect:sumary}}, we present some minimal necessary background on sine dilaton gravity. More details can be found in \cite{Blommaert:2024ydx,Blommaert:2023wad,Blommaert:2023opb}.

\subsection{Recalling sine dilaton gravity}\label{ham}
Including a holographic counterterm we define sine dilaton gravity via the 2d gravitational path integral
\begin{equation}\label{sine_dilaton}
\int \mathcal{D} g \mathcal{D}\Phi\,\exp\bigg( \frac{1}{2}\int \mathrm{d} x \sqrt{g}\bigg(\Phi R+\frac{\sin(2\abs{\log q} \Phi)}{\abs{\log q}}\bigg)+\int \mathrm{d}\tau \sqrt{h} \bigg(\Phi K-\mathrm{i} \,\frac{e^{-\mathrm{i} \abs{\log q}\Phi}}{2\abs{\log q}}\bigg)\bigg).
\end{equation}
The parameter $\abs{\log q}$ stems from the DSSYK holographic description of this model \cite{Blommaert:2024ftn}. DSSYK is the quantum mechanics of $N$ Majorana fermions $\psi_i$ with the following Hamiltonian \cite{Cotler:2016fpe,Berkooz:2018qkz,Lin:2022rbf}\footnote{Following \cite{Blommaert:2023wad,Blommaert:2023opb,Blommaert:2024ftn} our conventions for $q$ differs from that used for instance in \cite{Lin:2022rbf} by $q^2=q_\text{there}$. Related works on a gravitational bulk interpretation of DSSYK includes \cite{Berkooz:2022mfk,Narovlansky:2023lfz,Verlinde:2024zrh,Berkooz:2024ofm,Lin:2023trc,Verlinde:2024znh,Almheiri:2024ayc,Aguilar-Gutierrez:2024oea}. Much recent progress on DSSYK has been reviewed in \cite{Berkooz:2024lgq}.}
\begin{equation}
    H_\text{SYK}=\i^{p/2}\sum_{i_1<\dots <i_p}J_{i_1\dots i_p}\psi_{i_1}\dots \psi_{i_p}\,,\quad \abs{\log q}=\frac{p^2}{N}\,,\quad 0<q<1\,.\label{1.6}
\end{equation}
This parameter $\abs{\log q}$ is also related to the central charge $c$ in the Liouville CFT rewriting of our sine dilaton gravity theory via \cite{Blommaert:2024ftn,Verlinde:2024zrh,Collier:2024kmo}
\begin{equation}
    \pi b^2=\i\abs{\log q}\,.\label{bb}
\end{equation}
For our purposes it is most intuitive to rescale $2\abs{\log q} \Phi \rightarrow \Phi$ and work with the action
\begin{equation}\label{Irescaled}
    I=\frac12\int \mathrm{d} x \sqrt{g}\,\big(\Phi R+2\sin(\Phi)\big)+\int \mathrm{d}\tau \sqrt{h}\,\big(\Phi K-\mathrm{i} \,e^{-\mathrm{i} \Phi/2}\big)\,.
\end{equation}
The parameter $\abs{\log q}$ then appears only in the quantization parameter $\hbar=2\abs{\log q}$.\footnote{Following usual conventions we still put $\hbar=1$ in the original gravity action. Reintroducing this ``bare'' $\hbar$ would lead to a canonical algebra involving the combination $2\abs{\log q}\hbar$ as indeed appeared in \cite{Blommaert:2023opb}.} Classical solutions are found using the general formulas for 2d dilaton gravity \cite{Witten:2020ert} in the gauge $\Phi=r$
\begin{equation}
    \d s^2=F(r)\d \tau^2+\frac{1}{F(r)}\d r^2\,,\quad F(r)=-2\cos(r)+2\cos(\theta)\,.\label{1.9}
\end{equation}
This spacetime has a black hole horizon at \(r=\theta\), so $\Phi_h=\theta$, and a cosmological horizon at \(r=2\pi-\theta\). According to DSSYK, the natural holographic metric contour to consider in this spacetime is complex
\begin{equation}
   \begin{tikzpicture}[baseline={([yshift=-.5ex]current bounding box.center)}, scale=0.7]
 \pgftext{\includegraphics[scale=1]{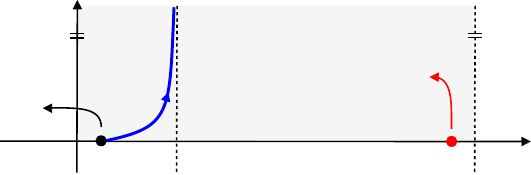}} at (0,0);
    \draw (1.7,-0.15) node {\color{red}horizon};
    \draw (1.7,0.55) node {\color{red}cosmo};
    \draw (-4.8,-0.25) node {horizon};
    \draw (5.6,-0.35) node {$r\sim r+2\pi$};
    \draw (3.7,-1.8) node {$2\pi$};
    \draw (-1.3,-1.8) node {$\pi/2$};
    \draw (-3.05,-1.8) node {$0$};
    \draw (-1.3,2.7) node {\color{blue}contour real $\rho$};
    \draw (-1.3,1.8) node {\color{blue}$\rho=\infty$};
  \end{tikzpicture}
\end{equation}
Even though complex metric contours are perfectly acceptable, and used widely in cosmology, they are not entirely standard for black hole physics. Therefore, let us pause to clearly motivate this point. For this and future purposes it is convenient to consider the Weyl rescaled metric
\begin{equation}
    \d s^2_\text{AdS}=\d s^2 e^{-\i \Phi}=F_\text{AdS}(\rho)\d \tau^2+\frac{1}{F_\text{AdS}(\rho)}\d \rho^2\,,\quad F_\text{AdS}(\rho)=\rho^2-\sin(\theta)^2\,,\quad e^{-\i \Phi}=\cos(\theta)-\i \rho\,.\label{1.11}
\end{equation}
This is the usual $R_\text{AdS}+2=0$ AdS$_2$ black hole, with an unusual (complex) dilaton profile. The metric contour which DSSYK is instructing us to consider is real $\rho$. To be more precise, we are told to choose the endpoint of the contour at $\rho\to \infty$. One way to appreciate this is via the group theoretic constraint that embeds the q-Schwarzian formulation of DSSYK in quantum mechanics on SU$_q$(1,1) \cite{Blommaert:2023wad,Blommaert:2023opb}, which implies indeed the usual asymptotically AdS boundary condition $\rho\to\infty$ \cite{Blommaert:2024ydx}
\begin{equation}
    \sqrt{h}\,e^{\i\Phi/2}=\i\,,\quad \Phi\to \frac{\pi}{2}+\i\infty\,.\label{2.3 bc}
\end{equation}
A more physical argument is that bilocal correlators in DSSYK \cite{Berkooz:2018jqr} are described in sine dilaton gravity by operators $e^{-\Delta \mathbf{L}}$ with $L$ the length of a slice measured in the AdS metric $\d s^2_\text{AdS}$
\begin{equation}
    L=\int \d s\,e^{-\i \Phi/2}\,.
\end{equation}
Correlators in DSSYK are real functions, because the length $L$ of geodesics is real in real AdS metrics. Therefore, the appropriate contour is real $\rho$.

An important observation is that the metric \eqref{1.9} is periodic in the radial direction. In \textbf{section \ref{sect:gauged}} and further we will be led to interpret the radial direction as compact by identifying $r\sim r+2\pi$.

We will be interested in \textbf{section \ref{sect:ungauge}} and \textbf{section \ref{sect:gauged}} in canonical quantization of this theory on global Cauchy slices, which for different boundary times $T$ on the AdS$_2$ Penrose diagram look as follows:
\begin{equation}
    \begin{tikzpicture}[baseline={([yshift=-.5ex]current bounding box.center)}, scale=0.7]
  \pgftext{\includegraphics[scale=1]{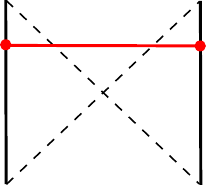}}
 at (0,0);
    \draw (2.25,0.7) node {$T/2$};
  \end{tikzpicture}
\end{equation}

The ADM energy $H$ and the holographically renormalized AdS length $L$ of these global Cauchy slices are parameterized in terms of the horizon location $\theta$ as
\begin{equation}\label{weyl}
H=-\cos(\theta)\,,\quad e^{-L}=\frac{\sin(\theta)^2}{\cosh \left(\sin(\theta) T/2\right)^2}.
\end{equation}
As discussed in \cite{Blommaert:2024ydx,Harlow:2018tqv} (and suppressing $\hbar=2\abs{\log q}$) the physical phase space is two dimensional 
\begin{equation}
    \omega=\d T\wedge \d H=\d L\wedge \d P
\end{equation}
Working out the conjugate momentum $P$ one can eliminate $T$ and find the quantum system
\begin{equation}\label{Hgrav}
\boxed{\mathbf{H}=-\cos(\mathbf{P})+\frac{1}{2}e^{\i\mathbf{P}}e^{-\mathbf{L}}\,,\quad \mathbf{P}\sim \mathbf{P}+2\pi}
\end{equation}
This is the transfer matrix of DSSYK \cite{Berkooz:2018jqr,Berkooz:2018qkz} with the identification $\mathbf{L}=2\abs{\log q}\textbf{n}$.

The real-time classical solutions of this system have two surprising properties. The (holographically renormalized) length is positive $L\geq 0$. Furthermore, the Hamiltonian \eqref{Hgrav} is a periodic function of $\mathbf{P}$. As we show in \textbf{section \ref{sect:genericpotential}}, this is a universal consequence of the periodicity of the dilaton potential.
\subsection{Summary and structure}\label{subsect:sumary}

In \textbf{section \ref{sect:ungauge}} we construct complete set of $L^2(L)$ eigenfunctions of $\mathbf{H}$ with the following spectrum
\begin{equation}
    \int_{-\infty}^{+\infty}\d L\left\langle \theta_1\rvert L\right\rangle\left\langle L\rvert \theta_2\right\rangle=\delta(\theta_1-\theta_2)\Big\slash \sin(\theta)\sinh\bigg(\frac{\pi \theta}{\abs{\log q}}\bigg)\,.\label{1.18}
\end{equation}
This leads to the semiclassical entropy $S_\text{BH}=2\pi \theta$. This is the usual Bekenstein-Hawking area formula $2\pi \Phi_h$ in dilaton gravity. However, the energy $E=-\cos(\theta)$ is periodic in $\theta$, which here takes values on the real axis. Because of this, the thermodynamic partition function is divergent as nothing suppresses large values of $\theta$. Therefore, this is not a reasonable quantum gravity theory. As we discuss in \textbf{section \ref{subsect:wavefung}} the ungauged theory corresponds to Liouville de Sitter gravity \cite{Verlinde:2024zrh, Verlinde:2024znh} or, equivalently, the analytic continuation of Liouville gravity \cite{Mertens:2020hbs, Fan:2021bwt, Blommaert:2023wad}. It is furthermore related (but not identical to) the complex Liouville string \cite{Collier:2024kmo}, see \textbf{section \ref{subsect:2.3}} for some brief comments.

In \textbf{section \ref{sect:gauged}} we prove that instead, the \emph{correct} definition of sine dilaton gravity (one that does not lead to divergences) involves interpreting the \textbf{periodicity in $\mathbf{P}$} of the theory as a \textbf{gauge redundancy}. We stress that naively, the gravitational path integral (and its classical solutions) reproduces the answers of the \emph{ungauged} theory with spectrum \eqref{1.18}. In order to understand what happens to the gravity theory upon gauging, we gauge this symmetry by the quantize first, constrain later approach. We introduce a projection operator such that $\mathbf{\Pi}\ket{\psi}$ are physical states\footnote{For the purpose of presentation we rescaled the length variable of the main text only in this subsection.}
\begin{equation}
    \mathbf{\Pi}=\sum_{m=-\infty}^{+\infty}e^{2\pi \i m \,\mathbf{L}}\Big\slash\sum_{m=-\infty}^{+\infty}1\,.
\end{equation}
Unsurprisingly, only states with discrete lengths survive the projection. Crucially, however, we also find that states with \textbf{negative renormalized length $\mathbf{L}$ are null}. In the ungauged theory, because $L$ is \emph{renormalized} length, states with \emph{negative} $L$ are very important, contributing massively to the entropy. In the gauged system we find the density of states
\begin{equation}
    \sum_{L=0}^{\infty}\left\langle \theta_1\rvert L\right\rangle\left\langle L\rvert \theta_2\right\rangle=\delta(\theta_1-\theta_2)\Big\slash (e^{\pm 2\i\theta};q^2)_\infty\,.\label{1.20}
\end{equation}
This leads to a quadratic entropy profile $S=2\pi \theta-2\theta^2$, so indeed the semiclassical entropy is much lower than predicted by the black hole area:
\begin{equation}
    \begin{tikzpicture}[baseline={([yshift=-.5ex]current bounding box.center)}, scale=0.7]
 \pgftext{\includegraphics[scale=1]{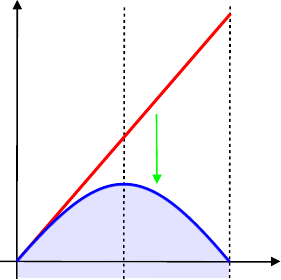}} at (0,0);
    \draw (2.3,-2.8) node {$\theta$};
    \draw (1.5,-2.8) node {$\pi$};
    \draw (-0.3,-2.8) node {$\pi/2$};
    \draw (-2.1,-2.8) node {$0$};
    \draw (1.15,-1) node {\color{blue}$S$};
    \draw (1.2,0.05) node {\color{green}\text{gauge}};
    \draw (0.4,1.8) node {\color{red}$S_\text{BH}$};
  \end{tikzpicture}
\end{equation}
This quadratic entropy profile is known in DSSYK \cite{Goel:2023svz}. The fact that it deviates from $S_\text{BH}$ has obscured several attempted gravity interpretations to design a theory of black holes \cite{Almheiri:2024xtw} or excitations in dS$_3$ \cite{Verlinde:2024znh} with the required spectrum. The answer is that sine dilaton gravity \emph{is} the bulk theory, but to compute the correct entropy one has to realize that the theory has a symmetry $P\to P+2\pi$ that is to be gauged. We stress that the fact that periodically identifying $P$ has such a big effect on the semiclassical entropy is unexpected (and was not realized for instance in \cite{Blommaert:2024ydx}). We also stress that gauging is \emph{not optional} as the ungauged theory is divergent, and therefore ill-defined.

The appearance of null states at $L<0$ shows in the behavior of the $L^2(L)$ wavefunctions at integers
\begin{equation}
    \begin{tikzpicture}[baseline={([yshift=-.5ex]current bounding box.center)}, scale=0.7]
    \pgftext{\includegraphics[scale=1]{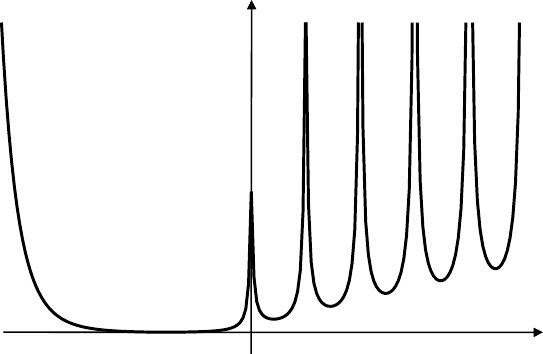}} at (0,0);
            \node at (-1.4,2.5) {$\left\langle L\vert \theta\right\rangle$};
            \node at (5.1,-2.65) {$L$};
            \node at (0.6,-3) {$1$};
            \node at (1.5,-3) {$2$};
            \node at (2.4,-3) {$3$};
            \node at (3.3,-3) {$4$};
    \end{tikzpicture}
\end{equation}
The poles at $L\geq 0$ become physical states, the divergence representing the volume of the gauge group $\mathbb{Z}$. The lack of divergences at $L<0$ leads to unphysical null states. In \textbf{section \ref{sect:genericpotential}} we show via explicit calculations that this analysis and its conclusions are valid for generic periodic potential $V(\Phi)$.

In the gauged theory, the radial direction $r$ is compact. There is a class of solutions to the equations of motion with metrics \eqref{1.9} and dilaton
\begin{equation}
    \Phi=r+2\pi m\,.
\end{equation}
These winding solutions are important and lead to a sum of saddles in the DSSYK spectral density $(e^{\pm 2\i\theta};q^2)_\infty$. In \textbf{section \ref{subleading}}, we show that these arise as classical saddles in the q-Schwarzian \cite{Blommaert:2023wad} (the path integral associated with the Hamiltonian system \eqref{Hgrav}).

We furthermore present two intriguing limits of this quantization of sine dilaton gravity. In \textbf{section \ref{sect:flathermite}} we consider a range of energies close to the top of the spectrum $\theta=\pi/2$. Sine dilaton gravity reduces to \textbf{flat space} dilaton quantum gravity. The ungauged theory has parabolic cylinder wavefunctions and the (usual) Hagedorn spectrum of the CGHS model \cite{Fiola:1994ir,Godet:2020xpk}. The gauged theory, instead, has Hermite wavefunctions and a Gaussian spectrum. We view this as a regularization of flat space quantum gravity due to a cosmological horizon ``at infinity''. This \textbf{Hermite model} was studied from the DSSYK side in \cite{Almheiri:2024xtw} (who proposed a different gravity interpretation, on which we comment).

In \textbf{section \ref{sect:topogravq0}}, we take the highly quantum limit $q\to 0$ (or $\abs{\log q}\to \infty$). Sine dilaton gravity reduces to \textbf{topological gravity} (which has zero action). At finite energies, the wavefunctions are Chebychev polynomials and the spectrum is the semicircle of a \textbf{Gaussian matrix integral}. At low energies, this reduces to the usual Airy model of topological gravity.\footnote{For some recent and less recent background material on the Airy model and topological gravity see \cite{kontsevich1992intersection,Dijkgraaf:2018vnm,Okuyama:2019xbv,Blommaert:2022lbh}.} This duality was studied in \cite{Gopakumar:2011ev, Gopakumar:2022djw}.

In the discussion \textbf{section \ref{sect:discnew}} we comment on a path integral interpretation of the gauging procedure as periodically identifying the dilaton zero mode. Such an identification is notoriously subtle \cite{Kapec:2020xaj,Verlinde:1986kw}. We briefly discuss the cosmological interpretation of our models, and mention that the gauging results in a normalizable wavefunction of the universe and a modified no-boundary proposal \cite{hartle1983wave}. This will be discussed in more detail elsewhere. In a companion paper \cite{wip}, we show that similar gaugings have the desired effect on amplitudes with EOW branes and on wormhole amplitudes in sine dilaton gravity. In the closed channel quantization we'll gauge $\Phi_h\to \Phi_h+2\pi m$. This discretizes the string spectrum and shows that our theory is the path integral formulation of ``q-deformed JT gravity'' \cite{Jafferis:2022wez}. We furthermore discuss the dS JT limit of periodic dilaton gravity in \textbf{section \ref{subsect:7.3dsjt}}. We summarize the three universal limits of periodic dilaton gravity as follows:
\begin{equation}
    \begin{tikzpicture}[baseline={([yshift=-.5ex]current bounding box.center)}, scale=0.7]
 \pgftext{\includegraphics[scale=1]{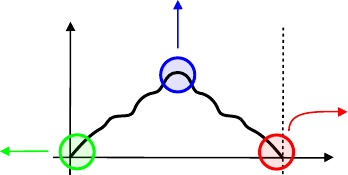}} at (0,0);
    \draw (3.1,-1.8) node {$\Phi_h$};
    \draw (1.8,-1.8) node {$\pi a$};
    \draw (-0,-1.8) node {$\pi a/2$};
    \draw (-1.75,-1.8) node {$0$};
    \draw (1.15,-0.1) node {$S$};
    \draw (0.1,2.7) node {\color{blue}flat space};
    \draw (0.1,2) node {\color{blue}section \ref{subsect:4.1}};
    \draw (4.4,-0.4) node {\color{red}dS space};
    \draw (4.4,-1.1) node {\color{red}section \ref{subsect:7.3dsjt}};
    \draw (-4.5,-1.1) node {\color{green}AdS space};
    \draw (-4.5,-1.8) node {\color{green}known};
  \end{tikzpicture}
\end{equation}

We will start with constructing a complete set of $L^2(L)$ eigenfunctions of $\mathbf{H}$.
\section{Ungauged quantization}\label{sect:ungauge}

There are ambiguities when quantizing a classical system. In this section, we will quantize the system as an analytic continuation of sinh dilaton gravity \cite{Mertens:2020hbs,Fan:2021bwt,Blommaert:2023wad}.
This aligns with the results of \cite{Blommaert:2024ydx}, where sine dilaton gravity was reformulated at the level of the Lagrangian as two Liouville CFTs with complex conjugate central charges, matching the analytic continuation of the non-critical string description of Liouville gravity \cite{Mertens:2020hbs,Fan:2021bwt,Blommaert:2023wad}.
Liouville CFT is characterized by a parameter $b$, which is related to our $q$ by \eqref{bb}. In this case, the quantization ambiguities can be fixed by demanding the system to have the usual Liouville $b \to 1/b$ symmetry. In this section we explore the quantum mechanical system defined in this way. This analytic continuation of Liouville gravity is Liouville de Sitter gravity \cite{Verlinde:2024zrh, Verlinde:2024znh}.

A key symmetry of classical sine dilaton gravity is $P\to P+2\pi$, which indeed leaves the Hamiltonian \eqref{Hgrav} invariant. In section \ref{sect:gauged} we gauge this symmetry. The distinction between how one deals with the modular $b\to 1/b$ symmetry and the shift $P\to P +2\pi$ symmetry is the central element in our analysis. One can view this as the physical translation of the mathematical construction of the q-Askey scheme in \cite{Lenells:2021zxo}, relating q-polynomials and a suitable continuous generalization thereof.\footnote{The system in section \ref{sect:gauged} is not symmetric under $b\to 1/b$.}

\subsection{Wavefunctions}\label{subsect:wavefung}

The global slice WdW wavefunctions of sine dilaton gravity are the solutions of the Schr\"odinger equation associated
with $\mathbf{H}$ \eqref{Hgrav}. Since this Hamiltonian is non-Hermitian, we need to separately find the left and right eigenfunctions, which will generally be different from each other \cite{Blommaert:2023opb,yao2018edge}. The equations look more symmetric in terms of a new length variable $\phi$ (which we use throughout this section)
\begin{equation}\label{def}
L=2 \phi-\abs{\log q}\,.
\end{equation}
The left respectively right eigenvectors of $\mathbf{H}$
\begin{equation}
    \left\langle \theta|\phi\right\rangle=\psil_\theta(\phi) \qquad \left\langle \phi|\theta\right\rangle=\psir_\theta(\phi)\,,
\end{equation}
are the solutions to the following Schr\"odinger difference equation $\mathbf{H}=E$
\begin{align}
2\cos(\theta)\psil_\theta(\phi)&=\psil_\theta(\phi+\abs{\log q})+(1-e^{\abs{\log q}}e^{-2\phi})\,\psil_\theta(\phi-\abs{\log q})\nonumber\\
2\cos(\theta)\psir_\theta(\phi)&=\psir_\theta(\phi-\abs{\log q})+(1-e^{-\abs{\log q}} e^{-2\phi})\,\psir_\theta(\phi+\abs{\log q})\,.\label{7.2}
\end{align}
We parametrize $E=-\cos(\theta)$; this energy range follows from the $L\to\infty$ asymptotic free motion region of the Hamiltonian \eqref{Hgrav}. Difference equations do not have unique solutions. However if we choose to enforce the $b\to 1/b$ symmetry of Liouville theory, one does obtain a unique solution \cite{Fan:2021bwt}. Here $b\to 1/b$ acts as
\begin{equation}\label{2.14qdual}
     \log q\rightarrow \log q_\text{dual}=-\frac{\pi^2}{\log q} \,.
\end{equation}
To impose this symmetry one demands that the wavefunctions simultaneously solve the dual equations to \eqref{7.2}
\begin{align}
2\cos(\theta/b^2)\psil_\theta(\phi)&=\psil_\theta(\phi-\pi^2/\abs{\log q})+(1-e^{-\pi^2/\abs{\log q}}e^{-2\phi})\,\psil_\theta(\phi+\pi^2/\abs{\log q})\nonumber\\
2\cos(\theta/b^2)\psir_\theta(\phi)&=\psir_\theta(\phi+\pi^2/\abs{\log q})+(1-e^{\pi^2/\abs{\log q}} e^{-2\phi})\,\psir_\theta(\phi-\pi^2/\abs{\log q})\,.\label{2.5}
\end{align}
Because the two recursion relations \eqref{7.2} and \eqref{2.5} have incommensurate periods, this leads to a unique solution \cite{Fan:2021bwt}. We now determine these solutions.

The recursion relations are easily solved in the Fourier domain
\begin{equation}
    \psil_\theta(\phi)=\frac{1}{2\pi}\int_{-\infty}^{+\infty}\d p\,e^{-\i p\phi}\,\psil_\theta(p)\,,\quad 2(\cos(\theta)-\cos(p\abs{\log q}))\,\psi^{\mathsf{\scriptscriptstyle L}}_\theta(p)=-e^{-\abs{\log q}}e^{\i p \abs{\log q}}\,\psil_\theta(p+2\i)\,.\label{7.4}
\end{equation}
The solution to \eqref{7.4} can be built as a product of an exponential and double sine functions $S_{b}$ \cite{Kharchev:2001rs,ip2015tensor}
\begin{equation}\label{fourier}
\psil_\theta(p)=S_b\bigg(-\i \frac{b p}{2}\pm\i \frac{b}{2}\frac{\theta}{\abs{\log q}}\bigg)\exp\bigg(-\abs{\log q}\frac{p^2}{4}+\frac{\theta^2}{4\abs{\log q}}-\frac{\pi p}{2}\bigg)\,,\quad S_b(x+b)=2\sin(\pi b x) S_b(x)\,.
\end{equation}
Therefore the left wavefunction is
\begin{equation}\label{left}
    \boxed{\psil_\theta(\phi)=\frac{1}{2\pi}\int^{+\infty+\i \epsilon}_{-\infty+\i \epsilon}\d p\, e^{-\i p \phi}\,S_b\bigg(-\i \frac{b p}{2}\pm\i \frac{b}{2}\frac{\theta}{\abs{\log q}}\bigg)\exp\bigg(-\abs{\log q}\frac{p^2}{4}+\frac{\theta^2}{4\abs{\log q}}-\frac{\pi p}{2}\bigg)}
\end{equation}
The contour avoids the poles of the double sine functions which are located at\footnote{In deriving the RHS of \eqref{7.4} we used a contour shift that assumed no poles between $p \in \mathbb{R}$ and $p\in 2\i+ \mathbb{R}$, which is true.}
\begin{equation}
    p=\frac{\pm \theta-2\pi m}{\abs{\log q}}-2\i n\,, \qquad m,n =0,1,2,\hdots\label{polesleftwf}
\end{equation}
One checks that \eqref{left} is the unique solution to \eqref{7.2} that also satisfies the dual recursion relation \eqref{2.5} by in the Fourier domain exploiting the fact that $S_b(x)$ diagonalizes also the dual shifts by $1/b$ \cite{Kharchev:2001rs,ip2015tensor}. In a similar way one obtains the right eigenfunction\footnote{The contour $\gamma$ is slightly above the real axis such that all poles are on one side, but goes down in the complex plane to $c-i\infty$ to the right of the rightmost pole (for arbitrary large enough constant $c$). This makes the integral over $\gamma$ convergent.}
\begin{equation}\label{right}
    \psir_\theta(\phi)=\frac{1}{2\pi}\int_{\gamma}\d p\, e^{-\i p \phi}\,S_b\bigg(-\i \frac{b p}{2}\pm\i \frac{b}{2}\frac{\theta}{\abs{\log q}}\bigg)\exp\bigg(\abs{\log q}\frac{p^2}{4}-\frac{\theta^2}{4\abs{\log q}}-\frac{\pi p}{2}\bigg)\,.
\end{equation}

These results \eqref{left} and \eqref{right} can alternatively be obtained by analytically continuing the Liouville gravity eigenfunctions \cite{Mertens:2020hbs,Fan:2021bwt}. The wavefunctions \( \psi_\alpha(\phi) \) of Liouville gravity are solutions to
\begin{equation}
\label{eq:deliouv}
    2 \cosh(\alpha) \psi_\alpha(\phi) = \psi_\alpha(\phi + \i \pi b^2) + (1 - e^{\i \pi b^2} e^{-2\phi}) \psi_\alpha(\phi - \i \pi b^2),
\end{equation}
Making the following identifications
\begin{equation}
    \psir_\theta(\phi) \to \psi_\alpha(\phi), \quad \psil_\theta(\phi) \to \psi_\alpha(\phi)^*,
\end{equation}
and analytically continuing $b$ as in \eqref{bb} along with $\theta=\i\alpha$ indeed takes \eqref{eq:deliouv} to the recursion relations \eqref{7.2}.

\subsection{Gravitational partition function}
We would now like to know the entropy of this quantum theory. For this we compute the inner product between left-and right eigenvectors. It is convenient \cite{Mertens:2020hbs} to compute the overlap \( \left\langle \theta_1 | \theta_2 \right\rangle \) by considering the limit \( \Delta \rightarrow 0 \) of the matrix element of an operator \( e^{-2\Delta \phi} \)
\begin{equation}\label{spectral}
\left\langle \theta_1 | \theta_2 \right\rangle = \lim_{\Delta \rightarrow 0} \bra{\theta_1}e^{-2\Delta \phi}\ket{\theta_2} = \lim_{\Delta \rightarrow 0} \int_{-\infty}^{+\infty} \mathrm{d}\phi \, e^{-2 \Delta \phi} \, \psil_{\theta_1}(\phi) \psir_{\theta_2}(\phi)\,.
\end{equation}
This calculation is straightforward and is presented is appendix \ref{app:ortho}. The result is
\begin{align}
\left\langle \theta_1 \middle| \theta_2 \right\rangle = \delta(\theta_1-\theta_2)\Big\slash \sin(\theta)\sinh\bigg(\frac{\pi \theta}{\abs{\log q}}\bigg)\,,\label{2.14}
\end{align}
with corresponding completeness relation
\begin{equation}
    \mathbf{1} = \int_{-\infty}^{+\infty} \d\theta \, \rho(\theta) \ket{\theta}\bra{\theta}\,,\quad \rho(\theta)=\sin(\theta)\sinh\bigg(\frac{\pi\theta}{\abs{\log q}}\bigg)\,.\label{density}
\end{equation}
We stress that although the energy variable is periodic $E=-\cos(\theta)$, the completeness relation for the wavefunctions \eqref{left} involves $\theta$ on the real axis. Indeed, the wavefunctions \eqref{left} have no symmetry under $\theta\to \theta+2\pi$. This spectrum \eqref{density} can equivalently be interpreted as the direct analytic continuation of the Liouville gravity density of states (the Virasoro S-matrix) by setting 
$\theta=\i 2\pi b s$ in
\begin{equation}
\d s \, \rho(s) = \d s\,\sinh (2\pi b s) \sinh (2\pi s/b)\,.
\end{equation}

The partition function of this system is to be computed as a transition amplitude in the $L=-\infty$ state \cite{Blommaert:2023wad,Blommaert:2024ydx,Blommaert:2018oro,Lin:2022zxd,Lin:2022rbf,Belaey:2024dde}. Indeed, the initial state in gravity reflects the ``absence of structure'' and implements a smooth gluing of the left-and right boundary. This smooth boundary may be defined by a vanishing AdS geodesic length $L_\text{geo}$. Because of holographic renormalization, $L=L_\text{geo}-\infty$. Hence, the smooth state is indeed $\ket{L=-\infty}$, or in the notation \eqref{def} $\ket{\phi=-\infty}$:
\begin{equation}
   \begin{tikzpicture}[baseline={([yshift=-.5ex]current bounding box.center)}, scale=0.7]
 \pgftext{\includegraphics[scale=1]{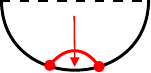}} at (0,0);
    \draw (-0.3,-0.9) node {\color{red}$L=-\infty$};
  \end{tikzpicture} \label{3.25bbb}
\end{equation}
Therefore the partition function of this quantum mechanics is
\begin{equation}\label{2.41}
Z_{\mathrm{ungauged}}(\beta)=\bra{\phi=-\infty} e^{-\beta \mathbf{H}/\hbar}\ket{\phi=-\infty}=\int_{-\infty}^{+\infty} \mathrm{d}\theta \rho (\theta)\, \psil_\theta(-\infty)\,\psir_\theta(-\infty)\, e^{\beta \cos (\theta)/2\abs{\log q}}\,.
\end{equation}
The spectrum is the product of what we called $\rho(\theta)$ in \eqref{density}, and the behavior of the wavefunctions in the region $\phi\to -\infty$. 
The entropy is determined uniquely by selecting the unique tracial state \cite{Penington:2023dql}, in casu the smooth state $\ket{\phi=-\infty}$ \cite{Kolchmeyer:2023gwa}. We have already normalized our wavefunctions \eqref{left} and \eqref{right} such that
\begin{equation}
    \psil_\theta(\phi)\,\psir_\theta(\phi)\to 1\,,\quad \phi\to -\infty\,,\label{7.42norm}
\end{equation}
as demonstrated in appendix \ref{app:wf}. Therefore the gravitational partition function \eqref{2.41} becomes
\begin{equation}
\label{eq:fakepf}
\boxed{
Z_\text{ungauged}(\beta) = \int_{-\infty}^{+\infty} d\theta \sin(\theta) \, \sinh\bigg(\frac{\pi \theta}{|\log q|}\bigg) e^{\beta \cos\theta/2\abs{\log q}}}
\end{equation}
We note that this density of states can alternatively be derived by considering the asymptotics of $\psil_\theta(\phi)$ for $\phi \to +\infty$, where the wavefunctions are in the WKB regime. We demonstrate in appendix \ref{app:wf} that this indeed leads to the same equation for the spectrum \eqref{density}.

This expression for the partition function \eqref{eq:fakepf}, which we obtained directly from canonical quantization of the Hamiltonian \eqref{Hgrav}, matches with the semiclassical calculation of the disk path integral of sine dilaton gravity, the derivation of which is found in section 2.1 of \cite{Blommaert:2024ydx}. We found here that the semiclassical approximation is \emph{exact} (taking into account the contribution from the ``inner'' horizon at $-\theta$ \cite{Kruthoff:2024gxc}). For related comments see section \ref{subleading} and \ref{sec:gasofdef}. The semi-classical thermodynamics of \eqref{eq:fakepf} is indeed the standard Bekenstein-Hawking thermodynamics of the black hole solution \eqref{1.9} \cite{Blommaert:2024ydx}:
\begin{equation}\label{fake}
S_\text{BH}=\frac{\pi \theta}{\abs{\log q}}\,,\quad \beta_\text{BH}=\frac{2\pi}{\sin (\theta)}\,.
\end{equation}

However, the expression \eqref{eq:fakepf} has fundamental issues. Because of the periodicity of $E$, the partition function is manifestly divergent! Indeed, the density of states for fixed energy $E=-\cos(\theta)$ diverges
\begin{equation}
\rho(E) = \sum_{m=-\infty}^{+\infty} \, \sinh\bigg(\frac{\pi (\arccos(E)+2\pi m)}{|\log q|}\bigg)\,.\label{2.22}
\end{equation}
Therefore, even though the current description reproduces gravitational semiclassics, it is not a healthy system. The progress in this section as compared to \cite{Blommaert:2024ydx} is that we see that the semiclassical thermodynamical results \eqref{fake} indeed arise from an exact quantization of \eqref{Hgrav}. This sharpens the tension between DSSYK and the ``naive'' sine dilaton quantum gravity. If we are to find a reasonable definition of sine dilaton quantum gravity, we have to drastically reduce the number of physical states. 

In the next section \ref{sect:gauged} we will propose a different quantization scheme of sine dilaton gravity, which does match with DSSYK, and hence elevates the above tension. This leaves us with a puzzle: we need a new semiclassical rule which replaces the Bekenstein-Hawking/QES formula (or the no-boundary state), and explains why the spectrum of quantum sine dilaton gravity \eqref{3.34} deviates a lot from $A/4 G_\text{N}$.

It is useful at this point to compare our above quantization of sine dilaton gravity with that obtained in \cite{Collier:2024kmo}. Technically, their result seems to be related to ours by an analytic continuation of $\theta = ix,\,\, x\in \mathbb{R}$, which leads to a convergent partition function. We present some formulas and more speculation on this in the conclusion in section \ref{subsect:2.3}.

\section{Gauged quantization}\label{sect:gauged}
In this section we present a more correct, fully quantum mechanical definition of the theory which does not suffer from divergences. Recall that the Hamiltonian \eqref{Hgrav} is a periodic function of momentum $P$
\begin{equation}
\mathbf{H}=-\cos(\mathbf{P})+\frac{1}{2}e^{\i\mathbf{P}}e^{-\mathbf{L}}\,.\label{3.1}
\end{equation}
We propose that the correct definition of sine dilaton gravity involves interpreting the periodicity in $P$ of the theory as a gauge redundancy. We stress that naively, the gravitational path integral reproduces the answers of the ungauged theory of section \ref{sect:ungauge}.

It is known from the DSSYK description \cite{Berkooz:2018jqr, Berkooz:2018qkz} that the Schr\"odinger equation \eqref{3.1} is also solved by the q-Hermite polynomials at positive integers. In \textbf{section \ref{discrete}} we show that the q-Hermite polynomials are encoded in the continuous wavefunctions $\left\langle L\rvert \theta \right\rangle$ at integer values of $L$. In \textbf{section \ref{gauge}} we show that this Hermite component is projected upon by gauging the $P\to P+2\pi$ of the system. Surprisingly, as shown in \textbf{section \ref{gauge}}, states with integer but \emph{negative} renormalized length $L$ become null upon gauging $P\to P+2\pi$. This has dramatic consequences, as we show in \textbf{section \ref{subsect:infopuzzle}}: because of these null states, the entropy of the quantum theory deviates at leading order from the black hole area law $S=2\pi\theta$ and results in a finite partition function. There are so many null states that the theory is thermodynamically finite even at infinite temperature $\beta=0$. In this sense, this model is UV complete. In \textbf{section \ref{subleading}} we discuss an application of the identification: new saddles in gravity and the q-Schwarzian. The periodic identification of $P$ is the direct consequence of the periodicity of the potential, as we explain in section \ref{sect:genericpotential}. We speculate in the path integral interpretation of gauging $P$ shifts in the discussion section \ref{sect:discnew}.

\subsection{q-Hermite polynomials from ungauged wavefunctions}\label{discrete}
\begin{figure}[t]
\centering
\begin{tikzpicture}
              \node[inner sep = 0pt] at (0,0) {\includegraphics[width=0.5\linewidth]{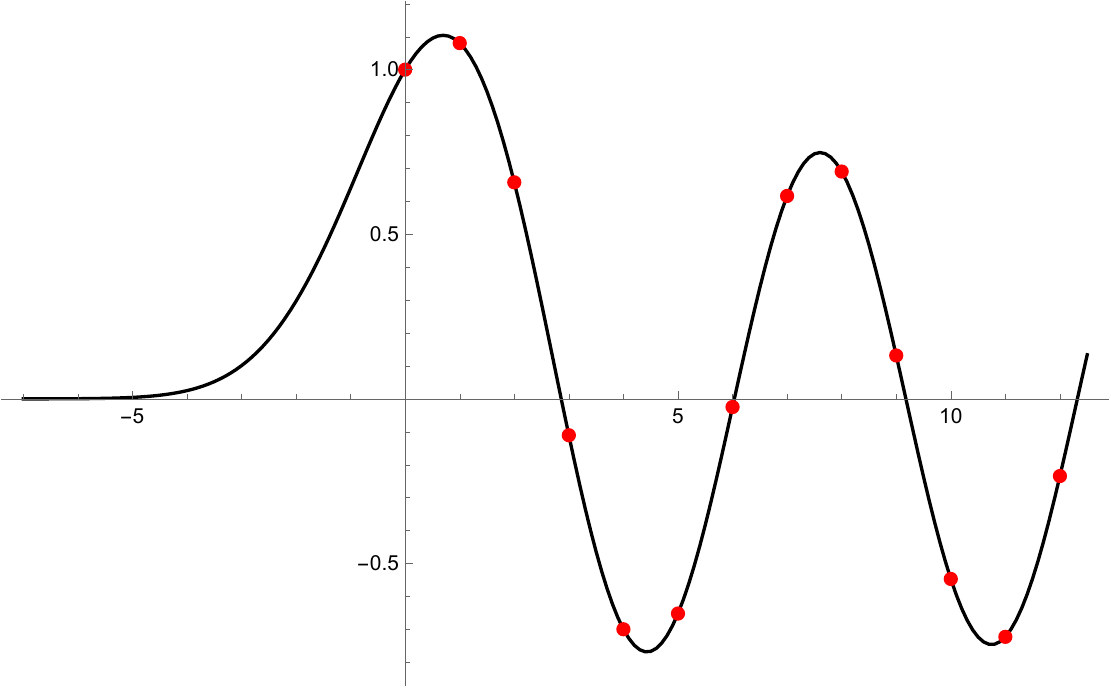}};
              \node at (-1.8,2.8) {\text{$\psil_\theta(n)$}};
              \node at (4.8,-0.7) {$n$};
\end{tikzpicture}
\caption{The left wavefunction \eqref{left1} interpolates between the q-Hermite polynomials \eqref{left2}, which are indicated by red dots at positive integer values of $n$.
}
\label{fig:psil}
\end{figure}

\begin{figure}[t]
\centering
\begin{tikzpicture}
              \node[inner sep = 0pt] at (0,0) {\includegraphics[width=0.55\linewidth]{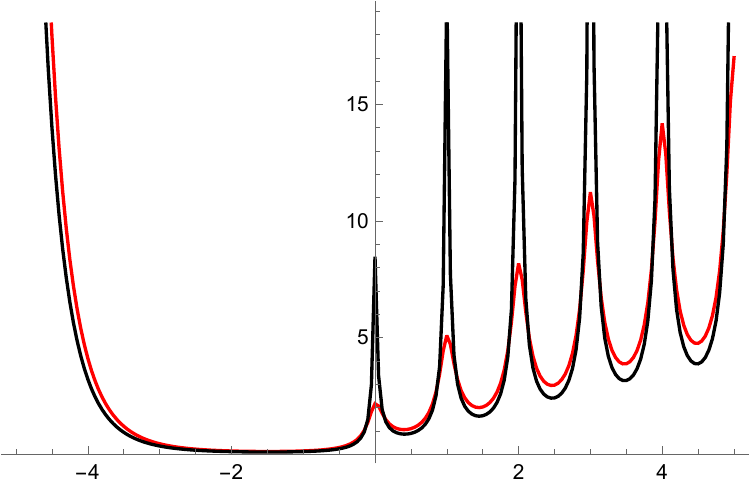}};
              \node at (-1.4,3) {\text{$|\psir_\theta(n+\i\epsilon)|$}};
              \node at (5.2,-3.1) {\text{$n$}};
\end{tikzpicture}
\caption{The right wave function $\psir_\theta(n+\i \epsilon)$ with $q=0.4$, $\theta=0.1$, and $\epsilon=0.1$ and $0.02$ in red and black respectively. Crucially, the wavefunction diverges only at positive integers as explained around \eqref{inf}. }
\label{fig:psirpoles}
\end{figure}

Introducing a rescaled length variable $L=2\abs{\log q} n$ as (recall \eqref{def})
\begin{equation}\label{7.33}
    \phi - \frac{|\log q|}{2} = n |\log q|\,,
\end{equation}
and expressing the double sine function $S_b(z)$ in terms of q-Pochhammer symbols via Faddeev's quantum dilogarithm $\phi_b(z)$ \cite{ip2015tensor, kashaev2001quantum}, the left eigenfunction \eqref{left} simplifies to\footnote{The relevant identities are\begin{equation}\label{master}
    S_b(x) = \frac{e^{\i \frac{\pi}{2} x (b + 1/b - x)}}{\phi_b\left(\i x - \i \frac{b + 1/b}{2}\right)}\,, \qquad \phi_b(z) = \frac{\left(e^{2\pi b \left(z + \i \frac{b + 1/b}{2}\right)}; q^2\right)_{\infty}}{\left(e^{\frac{2\pi}{b} \left(z - \i \frac{b + 1/b}{2}\right)}; q_\text{dual}^{-2}\right)_{\infty}}\,.
\end{equation}}
\begin{equation}\label{left1}
\psil_\theta(n) = \frac{1}{2 \pi}\int_{-\infty+\i \epsilon}^{+\infty+\i \epsilon} \d p \, e^{-\i p n} \, \frac{(q_\text{dual}^{-2} e^{\pi(p\pm\theta)/\abs{\log q}}; q_\text{dual}^{-2})_\infty}{(e^{\i (p \pm  \theta)}; q^2)_\infty} \exp\left(-\frac{p^2}{2 |\log q|}\right)\,.
\end{equation}
Here $q_\text{dual}$ was defined in \eqref{2.14qdual}. For positive integer $n$, we show in appendix \ref{derivation} that this eigenfunction simplifies tremendously to a q-Hermite polynomial
\begin{equation}\label{left2}
\psil_\theta(n) = \frac{(q^{-2}_\text{dual}; q^{-2}_\text{dual})_\infty}{(q^2; q^2)_\infty} \exp\left(-\frac{\theta^2}{2 |\log q|}\right) \rH_n\left(\cos(\theta)| q^2\right)\,,\quad n\in \mathbb{N}\,.
\end{equation}
We also verified this relation numerically, as shown in Figure \ref{fig:psil}, where one can observe how the continuous wavefunction in \eqref{left1} smoothly interpolates between the q-Hermite polynomials at integer points.
\indent The right eigenvectors \eqref{right} are analogously rewritten as
\begin{equation}\label{right1}
\psir_\theta(n) = \frac{1}{2 \pi }\exp\left(-\frac{\theta^2}{2 |\log q|}\right) \int_{\gamma} \d p \, e^{-\i p n} \, \frac{(q_\text{dual}^{-2} e^{\pi(p\pm\theta)/\abs{\log q}}; q_\text{dual}^{-2})_\infty}{(e^{\i (p \pm  \theta)}; q^2)_\infty}\, .
\end{equation}
We are again interested in this wavefunction for $n$ close to integers. It is useful to stretch the contour $\gamma$ in \eqref{right1} along the real axis as much as possible, decomposing it in the limit into the real line $\tilde{\gamma} = \mathbb{R}+\i \epsilon$, and a downwards vertical segment at infinity. 
The contribution from $\tilde{\gamma}$ will yield an infinite contribution, whereas the vertical segment stays finite and can therefore be neglected. Indeed, for integer $n$ one can sum over $p\to p+2\pi$ images in the integrand to discover a \emph{divergence}\footnote{In the last step we used the known Fourier transform or generating function of the q-Hermite \eqref{int_rep} and the asymptotics
\begin{equation}
        (a q^{2-2m};q^2)_\infty \to 1\,,\quad m\to -\infty\,.
\end{equation}
}
\begin{align}\label{inf}
    \psir_\theta(n)&=\exp\bigg(-\frac{\theta^2}{2\abs{\log q}}\bigg)\int_{\tilde{\gamma}}\frac{\d p}{2\pi}\,e^{-\i p n}\,\frac{(q_\text{dual}^{-2} e^{\pi(p\pm\theta)/\abs{\log q}};q_\text{dual}^{-2})_\infty}{(e^{\i(p\pm \theta)};q^2)_\infty}\nonumber\\
    &=\exp\bigg(-\frac{\theta^2}{2\abs{\log q}}\bigg)\int_{0}^{2\pi}\frac{\d p}{2\pi}\,\frac{e^{-\i pn}}{(e^{\i(p\pm \theta)};q^2)_\infty}\sum_{m=-\infty}^{+\infty}(q_\text{dual}^{2m-2} e^{\pi(p\pm\theta)/\abs{\log q}};q_\text{dual}^{-2})_\infty\nonumber\\
    &=\exp\bigg(-\frac{\theta^2}{2\abs{\log q}}\bigg)\frac{\textsf{H}_n\left(\cos(\theta)| q^2\right)}{(q^2;q^2)_n}\bigg(\sum_{m=-\infty}^{+\infty}1\bigg)\,+\text{ finite}\,,\quad n\in \mathbb{N}\,.
\end{align}
This is a simple derivation, repeated more carefully in appendix \ref{app:reg}, of Theorem 7.9 of \cite{Lenells:2021zxo}. Note that at negative integers the divergence cancels exactly with a zero in $1/(q^2;q^2)_n$ . This behavior of the right wavefunction, with poles \emph{only} at positive integers was numerically checked in Figure \ref{fig:psirpoles}. 
In summary, at integer lengths the ungauged wavefunctions encode the q-Hermite polynomials
\begin{equation}
\psil_{\theta_1}(n) \psir_{\theta_2}(n)\Big/\sum_{m=-\infty}^{+\infty} 1 = \frac{(q^{-2}_\text{dual}; q^{-2}_\text{dual})_\infty}{(q^2; q^2)_\infty(q^2; q^2)_n} \exp\left(-\frac{\theta_1^2 + \theta_2^2}{2 \abs{\log q}}\right)  \rH_n\left(\cos(\theta_1)| q^2\right) \, \rH_n\left(\cos(\theta_2) | q^2\right)\,. \label{7.63prod}
\end{equation}
Crucially, this expression vanishes for negative integers. In subsection \ref{gauge} we show that the expression on the right is projected upon by gauging $P\to P+2\pi$ in the Hamiltonian \eqref{3.1}. The $\infty$ is the volume of the gauge group, and the fact that this vanishes at $n<0$ will signal that states with $n<0$ are null.

Before proceeding we make three comments concerning this observation.
\begin{enumerate}
    \item The divergences at non-negative integers can be appreciated directly from the Hamiltonian \eqref{3.1}. The term multiplying $e^{\i P}$ vanishes at $L=0$. Hence if $\psir_\theta (n) \neq 0$ for $n<0$, the recursion relation requires $\psir_\theta(0) = \infty$, which in turn requires $\psir_\theta(n\geq 0) = \infty$. An exception occurs if the terms involving $\psir_\theta(n=-1)$ and $\psir_\theta(n=-2)$ conspire to cancel in this recursion relation. This special case does not happen for $\psir_\theta(n)$, but does occur for $\psil_\theta(n)$, see Figure \ref{fig:psil}.
    Conversely, a finite answer at positive integers (as in the right of \eqref{7.63prod}) implies vanishing wavefunctions at negative $n$. This is important to understand what happens for generic periodic dilaton potentials in section \ref{sect:genericpotential}.
    \item The expressions at integer $n$ are not symmetric under $b\rightarrow 1/b$. This makes sense, as the projection operator constructed in section \ref{gauge} does not commute with the dual ``Hamiltonian''. Indeed, the transformation $b\to 1/b$ maps the integer $n$ lattice \eqref{7.33} to another dual lattice.\footnote{One way of saying this is that, in the Liouville formulation, the line operators which we insert with the projector \eqref{pro} have $b$-dependent weights and therefore explicitly break the $b\to 1/b$ symmetry of Liouville conformal field theory.} This matches the mathematical perspective of \cite{Lenells:2021zxo}.
    \item We obtained our gauged wavefunctions \eqref{7.63prod} by selecting the $\mathbf{\Pi}=1$ component of the continuous wavefunctions of section \ref{sect:ungauge}. The latter were eigenfunctions of the $b\to 1/b$ dual $\mathbf{H}_\text{dual}$ of $\mathbf{H}$, with constrained eigenvalues $\mathbf{H}=\mathbf{H}_\text{dual}=E$. The most general solutions to the Schr\"odinger equation would have different eigenvalues for $\mathbf{H}_\text{dual}$ and $\mathbf{H}$. So one could wonder if the fact that our gauged wavefunctions have no support on negative integers is only a feature of the sector $\mathbf{H}=\mathbf{H}_\text{dual}=E$ or is a feature of the theory as a whole. We prove in appendix \ref{app:B5nonegativelengths} that it is a feature of the full theory. Our wavefunctions \eqref{inf} and \eqref{left2} are the only solutions to $\mathbf{H}=E$ at integer lengths, no reasonable solutions have support on $n<0$. Projecting from any other sector of the theory other than $\mathbf{H}=\mathbf{H}_\text{dual}=E$ would lead to the same conclusion. The gauged quantization is thus unique.
\end{enumerate}
\subsection{Negative renormalized length states become null upon gauging}\label{gauge}
With the technical result \eqref{7.63prod} in hand, we construct a new theory by gauging $P\to P+2\pi$. Unsurprisingly, this discretizes $L$. A much more important effect is the unusual fact that states with a negative renormalized AdS length $L$ become null states, as we now demonstrate.

To gauge a symmetry, one constructs a projection operator $\mathbf{\Pi}$ as sum over the symmetry generators, and inserts it along all non-trivial cycles of the manifold. In our case, the projection operator is formally
\begin{equation}\label{pro}
    \mathbf{\Pi}=\sum_{m=-\infty}^{+\infty}e^{2\pi \i m \mathbf{n}}\Big\slash\sum_{m=-\infty}^{+\infty}1\,.
\end{equation}
The $\infty$ normalization is fixed by the projector property $\mathbf{\Pi}^2=\mathbf{\Pi}$,
and $e^{2\pi i m \mathbf{n}}$ generates $P\to P+2\pi m$. The physical Hilbert space of the gauged theory and its operators are constructed from the invariant sector of the ungauged theory as usual:
\begin{equation}
\ket{\psi_\text{invariant}}=\mathbf{\Pi}\ket{\psi}\,,\quad \mathbf{O}_\text{invariant}=\mathbf{\Pi}\,\mathbf{O}\,\mathbf{\Pi}\,.
\end{equation}
The physical Hilbert space follows after projecting out null states $\ket{\psi_\text{null}}$, which satisfy $\bra{\psi}\mathbf{\Pi}\ket{\psi_\text{null}}=0$ for all $\ket{\psi}$. We now demonstrate that this procedure takes the theory of section \ref{sect:ungauge} to the chord quantum mechanics of DSSYK \cite{Berkooz:2018qkz,Lin:2022rbf}. The key technical identity is \eqref{7.63prod}. 

Recall \eqref{density}, namely that in the ungauged theory 
\begin{equation}\label{overlap1}
    \left\langle \theta_1 \middle| \theta_2 \right\rangle = \int_{-\infty}^{+\infty} \d n \, \psil_{\theta_1}(n) \, \psir_{\theta_2}(n) =\delta(\theta_1-\theta_2)\Big\slash \sin(\theta_1)\sinh\bigg(\frac{\pi \theta_1}{\abs{\log q}}\bigg)\,.
\end{equation}
This results in the completeness relation
\begin{equation}
    \mathbf{1} = \int_{-\infty}^{+\infty} \d\theta \, \sin(\theta) \sinh\left(\frac{\pi \theta}{|\log q|}\right) \ket{\theta}\bra{\theta}\,.\label{3.14}
\end{equation}
In the gauged theory we are supposed to consider inner products of invariant states $\ket{\theta_\text{invariant}}=\mathbf{\Pi}\ket{\theta}$
with wavefunctions that sample the continuous wavefunctions \eqref{left1}
\begin{equation}
\bra{\theta}\mathbf{\Pi}\ket{n} = \sum_{j=-\infty}^{+\infty} \delta(n-j)\,\psil_\theta(n)\Big\slash \sum_{m=-\infty}^{+\infty} 1\,.\label{3.15}
\end{equation}
The infinite normalization factor in $\mathbf{\Pi}$ combines nicely with the divergences in the ungauged wavefunctions at positive integers \eqref{7.63prod} to give a finite inner product on invariant states
\begin{align}
    \left\langle\theta_{1\,\text{invariant}}\vert \theta_{2\,\text{invariant}}\right\rangle &=\bra{\theta_1}\mathbf{\Pi}\ket{\theta_2}=\int_{-\infty}^{+\infty}\d n\, \sum_{j=-\infty}^{+\infty}\delta(n-j)\psil_{\theta_1}(n)\psir_{\theta_2}(n) \bigg/\sum_{m=-\infty}^{+\infty}1\nonumber\\&= \frac{(q^{-2}_\text{dual};q^{-2}_\text{dual})_\infty}{(q^2;q^2)_\infty}\exp\bigg(-\frac{\theta_1^2+\theta_2^2}{2\abs{\log q}}\bigg)\sum_{j=0}^{+\infty}\frac{1}{(q^2;q^2)_j}\rH_j(\cos(\theta_1)|q^2)\,\rH_j(\cos(\theta_2)|q^2)\nonumber\\&=\frac{(q^{-2}_\text{dual};q^{-2}_\text{dual})_\infty}{(q^2;q^2)_\infty}\exp\bigg(-\frac{\theta_1^2+\theta_2^2}{2\abs{\log q}}\bigg)\sum_{\pm}\sum_{m=-\infty}^{+\infty}\delta(\theta_1 \pm \theta_2+2\pi m)\Big\slash (e^{\pm 2\i\theta}; q^2)_\infty\,.\label{3.16}
\end{align}
The crucial step is the second equality, which relies on our earlier result in \eqref{7.63prod}. As shown in \eqref{7.63prod}, the product \( \psil_{\theta_1}(n) \psir_{\theta_2}(n) \) diverges for \( n \geq 0 \), but remains finite for \( n < 0 \). The divergence for \( n \geq 0 \) is the same infinity that appears in the normalization of $\mathbf{\Pi}$, giving a finite contribution \eqref{7.63prod}. Crucially, states with $n<0$ do not contribute to this inner product. In the last line we used q-Hermite orthogonality \eqref{standard}. We can absorb the prefactors in \eqref{3.16} into the normalization of \( \ket{\theta_\text{invariant}} \), resulting in
\begin{equation}
    \left\langle \theta_{1\,\text{invariant}} \middle| \theta_{2\,\text{invariant}} \right\rangle = \sum_{\pm}\sum_{m=-\infty}^{+\infty}\delta(\theta_1 \pm \theta_2+2\pi m)\Big\slash (e^{\pm 2\i\theta}; q^2)_\infty\,.\label{3.17}
\end{equation}
We find that upon projection, states with $\theta$ differing by $\pi m$ are no longer orthogonal. Their difference is null. Projecting out these null states leads to the following physical identity operator
\begin{equation}\label{3.33}
    \boxed{\mathbf{1}_\text{phys} = \int_0^{\pi} \d\theta \, (e^{\pm 2\i\theta}; q^2)_\infty \ket{\theta_\text{invariant}} \bra{\theta_\text{invariant}}}\,.
\end{equation}
This is the familiar completeness relation in the chord Hilbert space of DSSYK \cite{Berkooz:2018qkz,Lin:2022rbf}. The physical consequences of the physical density of states $(e^{\pm 2\i\theta}; q^2)_\infty$ are discussed in section \ref{subsect:infopuzzle}. First, we clarify in more detail why negative $n$ states become null.

Accounting for the change of normalization between \eqref{3.16} and \eqref{3.17}, we observe that the invariant right wavefunctions become simply the familiar DSSYK wavefunctions \cite{Lin:2022rbf}
\begin{equation}
    \left\langle n_\text{invariant}\rvert \theta_\text{invariant}\right\rangle=\frac{\rH_n(\cos(\theta)|q^2)}{(q^2;q^2)_n}\,,\quad n\in \mathbb{Z}\,.
\end{equation}
This vanishes for negative integers. Using the projected left wavefunctions \eqref{3.15}, inserting the physical identity \eqref{3.33}, and using standard q-Hermite completeness one finds Kronecker delta normalized states
\begin{equation}
    \left\langle n_{1\,\text{invariant}}\vert n_{2\,\text{invariant}}\right\rangle =\int_{0}^{\pi} \d \theta\, (e^{\pm 2\i\theta}; q^2)_\infty \rH_{n_1}(\cos(\theta)|q^2)\frac{\rH_{n_2}(\cos(\theta)|q^2)}{(q^2;q^2)_{n_2}}=\delta_{n_1n_2}\,,\quad n_1\in \mathbb{N}\,.\label{3.20}
\end{equation}
This inner product vanishes for negative integers, and for non-integer values of the renormalized length. The conclusion is that after projecting out null states, the physical identity is
\begin{equation}
\mathbf{1}_\text{phys}=\sum_{n=0}^{+\infty}\ket{n_\text{invariant}}\bra{n_\text{invariant}}\,.
\end{equation}
As usual, gauging $P\to P+2\pi$ discretizes the conjugate renormalized AdS length. The surprise is that negative renormalized length states become null. This follows from the behavior of the wavefunctions in Figure \ref{fig:psirpoles} and as discussed below \eqref{7.63prod} can be directly inferred from the specific form of the Hamiltonian $H$ \eqref{3.1} (leading to a truncating recursion relation). Unlike the discretization (which is semiclassically not visible), the fact that $n<0$ states do not survive the gauging has a major effect on the semiclassical (leading order in $\abs{\log q}$) entropy of this theory, as we discuss next in section \ref{subsect:infopuzzle}. Before proceeding we make three comments.
\begin{enumerate}
\item The projection operator corresponds in the gravitational path integral to inserting (non-minimally coupled) matter particles with imaginary weights\footnote{See \cite{Blommaert:2023wad,Blommaert:2024ydx} for more background on these matter line operators.}
    \begin{equation}
        \mathbf{\Pi}=\sum_{m=-\infty}^{+\infty}e^{-\Delta_m\mathbf{L}}\,,\quad \Delta_m=\frac{\i \pi m}{\abs{\log q}}\,.\label{3.24}
    \end{equation}
    In the Liouville formulation \cite{Blommaert:2024ftn,Verlinde:2024zrh,Collier:2024kmo} these line operators correspond to degenerate boundary chiral vertex operators with weights (notation follows (2.23) in \cite{Mertens:2020hbs} and (2.33) in \cite{Blommaert:2023wad})
    \begin{equation}
        \beta_\text{M}=\frac{m}{b}\,.
    \end{equation}
    The breaking of $b\to 1/b$ symmetry of the theory discussed below \eqref{7.63prod} occurs from the Liouville perspective explicitly because one inserts operators with $b$-dependent $\beta_M$. It would be interesting to understand better the role of the remaining degenerate line operators after the projection, which evaluate to $(-1)^\mathbf{n}$ and $q^{-m \mathbf{n}}$. This would span a q-polynomial algebra of operators ``internal'' to the gravity theory.\footnote{Non-degenerate operators are usually somewhat less natural for instance from the matrix integral perspective \cite{Mertens:2020hbs,Mertens:2020pfe,Blommaert:2020seb}.} It seems plausible that this projection takes one from a two-matrix integral \cite{Collier:2024kmo} to our finite cut single matrix integral \cite{wip,Jafferis:2022wez}.

    Degenerate operators can also be identified in DSSYK in the two-point function \eqref{m_element} as the poles of $S_b(2b\Delta)$, leading to a Dirac-delta matrix elements with $\Delta$-dependent (complex) energy shifts. This encodes the simple OPE of Virasoro degenerate operators. The exact two-point function of DSSYK has this structure for
    \begin{equation}
        \Delta=-\frac{m}{2}\,,\quad m\geq 0\,.
    \end{equation}
    These are precisely the degenerate Liouville operators that evaluate to $q^{-m \mathbf{n}}$ which we mentioned earlier. 
    \item Alternatively \eqref{3.20} is obtained by inserting the ungauged completeness relation \eqref{3.14}
    \begin{equation}\label{3.47}
        \left\langle n_{1\,\text{invariant}}\vert n_{2\,\text{invariant}}\right\rangle =\int_{-\infty}^{+\infty} \d \theta \sin(\theta) \sinh\left(\frac{\pi \theta}{|\log q|}\right) \psil_{\theta}(n_1) \psir_{\theta}(n_2)\Big/\sum_{m=-\infty}^{+\infty}1\,.
    \end{equation}
    For integer $n$ the q-Hermite parts of the wavefunctions \eqref{7.63prod} are invariant under $\theta\to 2\pi-\theta$. The density of states can then be computed on the interval $0<\theta<\pi$ by summing over images, and including the normalization factors in \eqref{left2} and \eqref{left1}. The technical identity underlying this is
    \begin{align}
        \boxed{\sum_{m=-\infty}^{+\infty} 2 \sin (\theta) \sinh \frac{\pi(\theta+2\pi m)}{\abs{\log q}}e^{-\frac{\left(\theta+2\pi m\right)^2}{\abs{\log q}}}=e^{\frac{\pi^2}{4 \abs{\log q}}} \sin (\theta) \sum_{k=-\infty}^{+\infty}(-1)^k e^{-\frac{\left(\theta-\pi k-\pi/2\right)^2}{\abs{\log q}}}}
        \label{3.22}
    \end{align}
    The expression on the RHS can now be directly read as the DSSYK spectral density $(e^{\pm 2\i \theta};q^2)_\infty$ \cite{Verlinde:2024znh}.

    \item The above reasoning can be reversed. One could attempt, as in \cite{Blommaert:2024ydx}, to impose that wavefunctions vanish for $L<0$. The Schr\"odinger recursion equation following from $H$ \eqref{3.1} then implies that the wavefunctions must vanish for $L>0$ \emph{except} at positive integer points (which are decoupled). Indeed, the recursion relation leaves the value at $L=0$ arbitrary in this case, such that the wavefunctions at positive integers are multiples of the q-Hermite polynomials.\footnote{Solutions to differential (rather than difference) equations can also be strictly zero at $L<0$ and non-zero at $L>0$ if the differential equation has a singular point at $L=0$, but discretization only occurs for difference equations.}$^,$\footnote{In the case of Liouville gravity \eqref{eq:deliouv} where $b$ is real, it is impossible for the coefficient of the last term in \eqref{eq:deliouv} to become zero. Hence discretization is impossible for Liouville gravity.} Thus, at the quantum level, demanding positive lengths or demanding $P\sim P+2\pi$ are equivalent constraints:
    \begin{equation}
        P\sim P+2\pi\quad \Leftrightarrow \quad L\geq 0\,.\label{3.23}
    \end{equation}
\end{enumerate}
The interpretation of the gauged theory as obtained from the ungauged theory of section \ref{sect:ungauge} by actively inserting symmetry lines \eqref{3.24} in the original ungauged path integral, gives a natural route to extend gauging to higher genus surfaces.\footnote{We thank Shota Komatsu for explaining this.} We'll comment briefly on higher genus in the discussion section \ref{sect:discnew}.

\subsection{An entropic paradox}\label{subsect:infopuzzle}
As discussed around \eqref{2.41} the gravitational disk partition function in the ungauged theory is computed as a transition amplitude in the $L=-\infty$ state. This is a universal fact in 2d dilaton gravity \cite{Blommaert:2023wad,Blommaert:2024ydx,Blommaert:2018oro,Lin:2022zxd,Lin:2022rbf}. Indeed, the initial state in gravity naturally reflects the ``absence of structure'' and implements a smooth gluing together of the left and right boundary. For any 2d dilaton gravity we can consider a Weyl rescaled holographically renormalized AdS length $L$. The smooth boundary may be defined by a vanishing AdS geodesic length $L_\text{geo}$. Because of holographic renormalization $L=L_\text{geo}-\infty$. Hence, the smooth state is $\ket{L=-\infty}$
\begin{equation}
   \begin{tikzpicture}[baseline={([yshift=-.5ex]current bounding box.center)}, scale=0.7]
 \pgftext{\includegraphics[scale=1]{DSSYK19.pdf}} at (0,0);
    \draw (-0.3,-0.9) node {\color{red}$L=-\infty$};
  \end{tikzpicture} \label{3.25}
\end{equation}
We emphasize that this reproduces the black hole area law for gravitational entropy. For ungauged sine dilaton gravity we have, according to equation \eqref{eq:fakepf}
\begin{equation}
    Z_\text{ungauged}(\beta)=\bra{L=-\infty}e^{-\beta \mathbf{H}}\ket{L=-\infty}=\int_{-\infty}^{+\infty} d\theta \sin(\theta)\sinh\left(\frac{\pi \theta}{\abs{\log q}}\right) \exp\bigg(\beta \frac{\cos(\theta)}{2\abs{\log q}}\bigg)\,.\label{3.26}
\end{equation}
This indeed matches with the universal semiclassical black hole entropy in 2d dilaton gravity\footnote{See \cite{Blommaert:2023vbz,Dong:2022ilf} for recent discussion. Recall furthermore $\theta=\Phi_h$ and $\hbar=2\abs{\log q}$. The sum of $\theta\to \pm \theta + 2\pi m$ adds a formal $\infty$ to the entropy which was suppressed here.}
\begin{equation}
    S_\text{BH}=2\pi \Phi_h\,.
\end{equation}
This expressing makes clear how significant the restriction to $L\geq 0$ is. For instance, the smooth state in the ungauged theory becomes null in the gauged theory
\begin{equation}
    \bra{L=-\infty}\mathbf{\Pi}\,e^{-\beta \mathbf{H}}\mathbf{\Pi}\ket{L=-\infty}=0\,.
\end{equation}
Therefore a different smooth state in the gauged theory is required. What condition selects the unique smooth state to compute a partition function in the gauged quantum mechanical system?

The answer is that the correct state is $\ket{L=0}$. The smooth state is the unique \emph{tracial state} of the quantum mechanical system \cite{Penington:2023dql}. This is the state $\ket{\text{smooth}}$ with the property $\bra{\text{smooth}}\mathbf{A B}\ket{\text{smooth}}=\bra{\text{smooth}}\mathbf{B A}\ket{\text{smooth}}$, such that it defines a trace on the operator algebra, or in other words satisfies the $\beta=0$ KMS condition. In dilaton gravity without gauging shifts in $P$ this state is $\ket{L=-\infty}$ \cite{Penington:2023dql,Kolchmeyer:2023gwa}. In that case, a convenient way to derive the cyclicity of the state is precisely the gravitational picture \eqref{3.25}. Without such a tracial state, the entropy (density of states) of a continuous quantum mechanical system has no meaning.\footnote{By changing the normalization of states, one can at will change the notion of spectral density in a completeness relation.} The cyclicity of the state $\ket{L=-\infty}$ proves that the disk in gravity computes an entropy \cite{Penington:2023dql}. The state is chosen by the QM system.

In the gauged quantum mechanics, the tracial state is instead $\ket{L=0}$ \cite{Xu:2024hoc}. To prove this requires understanding the Hilbert space of Cauchy slices with matter excitations. Because sine dilaton gravity is quantum equivalent to DSSYK \cite{Blommaert:2024ydx} (after gauging), we may use the result of \cite{Xu:2024hoc}. A first principles gravity understanding of the Hilbert space of matter along the lines of \cite{Kolchmeyer:2023gwa} is beyond our current scope, but would be valuable for several reasons.\footnote{For instance, 
 to understand the appearance of the Hagedorn divergences in \cite{Jafferis:2022wez}.} As initial progress in this direction, we study EOW branes in \cite{wip}. The cyclicity of $\ket{L=0}$ is rather obvious in the chord description of this Hilbert space \cite{Lin:2022rbf,Berkooz:2018jqr}. Indeed, in the chord diagrams the initial state is not a ``special point'' along the boundary trajectory, therefore the initial state is smooth (or the identity state). 

From the gravity point of view, one piece of intuition why $\ket{L=0}$ is the correct smooth state, is to describe the smooth state classically as some EOW brane with infinite tension (or infinite mass) \cite{Gao:2021uro,Penington:2019kki}. This indeed gives the correct answer in both the ungauged and the gauged scenario, projecting on the lowest length state
\begin{align}
    \ket{\mu=\infty}&=\int_{-\infty}^{+\infty} \d L\, e^{-\infty L}\ket{L}=\ket{L=-\infty}\,,\quad\mathbf{\Pi}\ket{\mu=\infty}=\ket{L=0_\text{invariant}}\label{3.30}
\end{align}
We remark that classically, in sine dilaton gravity, the configuration $L=0$ is already special. It is the lowest value attained in the Lorentzian classical solution \eqref{weyl}. This happens because $\sin(\theta)=2\pi/\beta_\text{fake}$ is bounded from above (unlike in non-periodic dilaton gravity where the effective inverse temperature can become arbitrarily low). In this sense, perhaps it is reassuring that a correct quantization restricts to values of $L$ that are actually reached in the real phase space.

In summary, the correct trace is
\begin{equation}
    Z_\text{gauged}(\beta)= \bra{L=0_\text{invariant}}e^{-\beta \mathbf{H}} \ket{L=0_\text{invariant}}=\int_0^\pi\d\theta\,(e^{\pm 2\i\theta};q^2)_\infty \exp\bigg(\beta \frac{\cos(\theta)}{2\abs{\log q}}\bigg)\,.
\end{equation}
This can be rewritten using \eqref{3.22} as
\begin{equation}
\label{eq:truepf}
Z_{\mathrm{gauged}}(\beta) = \int_{-\infty}^{+\infty} d\theta \sin(\theta) \, \sinh\left(\frac{\pi \theta}{|\log q|}\right) \exp\bigg(-\frac{\theta^2}{\abs{\log q}}\bigg)\exp\bigg(\beta \frac{\cos(\theta)}{2\abs{\log q}} \bigg)\,.
\end{equation}
Hence, comparing to \eqref{3.26}, gauging introduces a Gaussian damping factor, leading to the finite semiclassical entropy
\begin{equation}
S=2\pi\Phi_h-2\Phi_h^2 \,\,< \,\, S_\text{BH}=2\pi \Phi_h\,.
    \label{3.34}
\end{equation}
This dramatic departure from the black hole area law in sine dilaton gravity presents a new entropy puzzle. In the Hamiltonian description, the unphysical null states $L<0$ are responsible for this decrease in entropy. Getting rid of this redundancy in the description of the system explains that in the previous system of section \ref{sect:ungauge} we were overcounting a lot of states. 

The resulting $\Tr(\mathbf{1})$ computes the dimension of the Hilbert space of the original SYK model and is finite (unlike in the ungauged theory)\footnote{We have suppressed overall constants scaling with $2^N$ throughout.}
\begin{equation}
    Z_\text{gauged}(0)=\dim \mathcal{H}_\text{SYK}\,.\label{3.34bb}
\end{equation}
 
As one might expect, semiclassically the discretization is invisible. But more is true: using \eqref{left2} and \eqref{inf} one finds the gauged partition function is encoded in a transition matrix element of the \emph{ungauged} theory (in the same sense that wavefunctions were in section \ref{discrete})
\begin{equation}
    Z_\text{gauged}(\beta)=\bra{L=0}e^{-\beta \mathbf{H}} \ket{L=0}\Big/\sum_{m=-\infty}^{+\infty}1\,.\label{disk_gauged}
\end{equation}
The infinite prefactor is to be interpreted as the volume of the gauge group $V_\text{gauge}$. Therefore, the exact disk partition function for the gauged theory may be expressed as a q-Schwarzian \cite{Blommaert:2023wad,Blommaert:2023opb} path integral
\begin{equation}\label{4.5}
Z_{\mathrm{gauged}}(\beta) =\int_{L(0)=L(\beta)=0} \frac{\dpi L \dpi P}{V_{\mathrm{gauge}}} \exp\bigg\{\frac{1}{2\abs{\log q}}\int_0^\beta \d u \bigg(\i\, P \frac{\d}{\d u}L+\cos(P)-\frac{1}{2}e^{\i P}e^{-L}  \bigg) \bigg\}\,.
\end{equation}

Before proceeding, we briefly discuss the JT gravity limit. Upon rescaling the dilaton $2\abs{\log q}\Phi\to \Phi$ between \eqref{sine_dilaton} and \eqref{Irescaled}, we have implicitly also rescaled the metric \eqref{1.9}, to be compared with equation (5.12) in \cite{Blommaert:2023opb}. This affects the notion of the holographically renormalized length $L$
\begin{equation}
   L_\text{JT}=L+2\log(2\abs{\log q})\to L-\infty\,,\quad \abs{\log q}\to 0\,.
\end{equation}
The simplest way to see this is to directly rescale $\theta\to 2\abs{\log q}\Phi_{h\,\text{JT}}$ and $2\abs{\log q} T\to T_\text{JT}$ in \eqref{weyl}. Thus the quantum constraint $L\geq 0$ is geometrically not very constraining for the JT gravity disk amplitude, as we also see from the fact that the entropy is actually linear in $\Phi_{h\,\text{JT}}$ in this regime. For wavefunction orthogonality see (2.17) and (2.18) in \cite{Blommaert:2023wad}.

More generally, we point out that the constraint $L\geq 0$ in terms of the original metric in \eqref{sine_dilaton} reads
\begin{equation}
    L_\text{original}\geq 2\log(2\abs{\log q})\to -\infty\,,\quad \hbar=2\abs{\log q}\hbar_\text{bare}\,.
\end{equation}
This effect naively does not look very constraining, and may naively be ignored semiclassically. But in sine dilaton gravity we consider very heavy (or very high energy) black holes with 
\begin{equation}
    \Phi_{h\,\text{original}}\sim 1/2\abs{\log q}\,.
\end{equation}
For such very high energy black holes, as we emphasized, the effect \emph{does} become semiclassically visible. It severely constrains the Euclidean disk \eqref{3.25}, changing the entropy \eqref{3.34} at leading order in $1/\abs{\log q}$. At such high energies, we are seeing that the UV completeness of the model changes the bulk picture significantly. More generally, the bulk picture might sensitively depend on the precise UV completion.

\subsection{Summing over gravitational saddles}\label{subleading}
A semiclassical approximation of the q-Schwarzian path integral \eqref{4.5} was studied in \cite{Blommaert:2024ydx}, and led to the leading $k=0$ term in the density of states \eqref{3.22} of the gauged theory. The purpose of this section is to refine and extend that analysis. Surprisingly, we will find that the exact partition function of sine dilaton gravity \eqref{eq:truepf} follows from a one-loop q-Schwarzian path integral calculation.

The classical solutions are \cite{Blommaert:2024ydx}
\begin{equation}\label{sol}
    e^{- L} = \frac{\sin (\theta)^2}{ \sin ( \sin (\theta) \tau /2 + \theta )^2}\,,\quad e^{- \i P} =  \frac{\sin(\theta)}{\tan \left(\sin(\theta)\tau/2+\theta\right)}+\cos(\theta), \qquad 0<\tau<\beta= \frac{2\pi-4\theta}{\sin\theta}\,.
\end{equation}
The entropy was computed using a contour integral of the symplectic term in the action
\begin{equation}\label{non_ana}
S=-\frac{\left(\theta-\pi/2\right)^2}{\abs{\log q}}\,.
\end{equation} 
During this calculation it was assumed that \cite{Blommaert:2024ydx}
\begin{equation}
    0<\theta < \pi\,.
\end{equation}
We can generate additional solutions to the EOM by analytically continuing $\theta \to  \pm \theta + 2\pi m$.\footnote{In addition new solutions are generated by time independent shifts $P \rightarrow P + 2\pi l$, with $l$-independent on-shell actions. This infinite sum over $l$ cancels the volume of the gauge group in \eqref{4.5}. Indeed, by definition, upon gauging we consider shifts $P\to P+2\pi l$ as redundant.}
These solutions have identical ADM energies $E=-\cos(\theta)$, but should be considered on different time intervals
\begin{equation}
\label{eq:win}
0<\tau<\frac{2\pi\mp 4\theta -8\pi m}{\pm\sin\theta}.
\end{equation}
Unlike the $m=0$ solutions \eqref{sol}, these solutions do not satisfy $L\geq 0$. The on-shell entropy \eqref{non_ana} of these winding saddles analytically continues from \eqref{non_ana} to
\begin{equation}
    S=-\frac{\left(\theta+\pi n-\pi/2\right)^2}{\abs{\log q}}\,.\label{3.42}
\end{equation}
We wrote $n=2m$ or $n=2m+1$, combining both sectors $\pm \theta + 2\pi m$ into one expression. We suggest a computation of this entropy as coming from winding contours associated with the solutions \eqref{sol} in the q-Schwarzian quantum mechanics in appendix \ref{app:contours}.

To get some intuition into the physical origin of these new solutions, we can consider them separately for fixed $m$ in the JT (or Schwarzian) limit where $\theta \to 0$, $\tau \to \infty$ with $\theta\tau$ fixed. For the Schwarzian time reparametrization $F(\tau) = -\cot \frac{\pi}{\beta} \tau$,\footnote{This is the PSL$(2,\mathbb{R})$ $F \to -1/F$ mapping of the usual $F(\tau) = \tan \frac{\pi}{\beta}\tau$ thermal reparametrization.} and identifying $e^{-L} \sim F'$ in the Schwarzian limit, the classical solution \eqref{sol} with periodicity \eqref{eq:win} matches that of the $(2\abs{m}+1)$ wound Schwarzian solution, where $F(\tau)$ winds this many times around the thermal boundary circle. The $\pm$ sign simply changes the orientation of the Schwarzian time reparametrization.\footnote{These can all be identified with the exceptional elliptic Virasoro coadjoint orbits \cite{Mertens:2019tcm}.} The resulting on-shell entropy \eqref{3.42} picks up the linear in $\theta$ term, matching with the $n$-winding Schwarzian density of states $\sim e^{2\pi n \sqrt{E}} - e^{-2\pi n \sqrt{E}}$ indeed.

We now take into account the one-loop determinant computed by the quadratic fluctuations around each saddle in the q-Schwarzian theory \eqref{4.5}. This one-loop factor is computed in \cite{Papalini} using a generalization of the Gelfand-Yaglom theorem. Taking that result on $\theta \to \pm \theta +2\pi m$, we write:
\begin{equation}\label{dete}
        \det \bigg(\left.\frac{\delta^2 S}{\delta \Phi_i \delta \Phi_j}\right|_{\Phi=\Phi_{\mathrm{cl}}}\bigg)^{-1/2}=\sin(\pm\theta)\,\frac{e^{\pm\frac{1}{2}(\pi/2\mp\theta -2\pi m) \cot \left(\theta\right)}}{\sqrt{1\pm\left(\pi/2\mp\theta-2\pi m\right)\cot(\theta)}}\,.
\end{equation}
Everything except the first factor $\sin(\pm\theta)$ is a one-loop factor from transforming from fixed $\theta$ to fixed $\beta$ in \cite{Papalini}, since we work at fixed $\theta$ we should ignore that piece. Thus we obtain as one-loop approximation to the spectrum
\begin{equation}\label{Z}
\boxed{\rho_\text{one-loop}(\theta)=\sin(\theta)\sum_{n=-\infty}^{+\infty} (-1)^n e^{-\frac{1}{\abs{\log q}}\left(\theta+\pi n-\frac{\pi}{2}\right)^2}=(e^{\pm 2\i \theta};q^2)_\infty}
\end{equation}
Remarkably, this is the exact spectral density \eqref{3.22} of gauged sine dilaton gravity, and of DSSYK. 

While the canonical partition function for DSSYK is not one-loop exact, this analysis suggests the microcanonical partition function actually is! Our approach is similar to Gutzwiller's trace formula \cite{Gutzwiller:1971fy}, a semiclassical construction that expresses the quantum mechanical density of states in terms of a sum of periodic orbits and the functional determinant around each orbit. The trace formula can be derived starting from the exact WKB method \cite{Sueishi:2020rug}. In this context, the $(-1)^n$ factor is the Maslov index, which counts the number of negative eigenvalues found in the expansion around a periodic orbit. In general, it would be very interesting to better understand if the exactness of this semiclassical analysis can be explained in terms of a localization argument for sine dilaton gravity or the q-Schwarzian. This microcanonical one-loop exactness is not just limited to this model. It was shown in \cite{Kruthoff:2024gxc,Griguolo:2023aem} that this holds for JT gravity as well. Liouville gravity also has this property. Indeed, the Liouville gravity disk partition function is
\begin{equation}
Z(\beta) \sim \int_{0}^{\infty} \d (\cosh(2\pi b s)) \sinh(2\pi s/b) ~e^{-\beta \cosh(2\pi b s)} \sim \frac{1}{\beta}K_{1/b^2}(\beta).
\end{equation}
Whereas $Z(\beta)$ is clearly not one-loop exact, this is true for the microcanonical partition function $\sinh \frac{2\pi s}{b}$ as a sum of two saddle contributions where the second saddle comes from the time-reversed winding-one solution, just like for JT gravity. One can wonder whether this is then true more generally in dilaton gravity; we will present some results on this in the conclusion \ref{sec:gasofdef}.

To understand which classical spacetimes correspond with the new q-Schwarzian saddles with periodicity \eqref{eq:win}, it is more convenient to consider instead the state $\ket{n=-\infty}$, where the solutions are simply the usual (no-defect \cite{Blommaert:2024ydx}) AdS$_2$ solutions
\begin{equation}
    e^{- L} = \frac{\sin (\theta)^2}{ \sin ( \sin (\theta) \tau /2)^2}, \qquad e^{- \i P} =  \frac{\sin(\theta)}{\tan \left(\sin(\theta)\tau/2\right)}+\cos(\theta), \qquad \tau=0\to\beta = \frac{2\pi}{\sin\theta}.\label{3.49}
\end{equation}
This corresponds with computing non-minimally coupled matter correlators $e^{-\Delta L}$ in the spacetime
\begin{equation}
    \d s^2=F(r)\d \tau^2+\frac{1}{F(r)}\d r^2\,,\quad F(r)=-2\cos(r)+2\cos(\theta)\,.\label{3.50}
\end{equation}
However, the q-Schwarzian solution with $n=2 m$ involves a globally shifted value of the dilaton\footnote{Odd integers $n$ correspond with spacetime contours extending to the interior horizon at $r=-\theta$ \cite{Kruthoff:2024gxc}.}
\begin{equation}
    \Phi=r+2\pi m \label{3.51}
\end{equation}
The spacetime contour for the radial coordinate $r$ in \eqref{3.50} is identical to the one in \eqref{1.11}
\begin{equation}
    e^{-\i r}=\cos(\theta)-\i \rho\,,\quad \sin(\theta)<\rho<+\infty
   \begin{tikzpicture}[baseline={([yshift=-.5ex]current bounding box.center)}, scale=0.7]
 \pgftext{\includegraphics[scale=1]{gauge6.pdf}} at (0,0);
    \draw (1.7,-0.15) node {\color{red}horizon};
    \draw (1.7,0.55) node {\color{red}cosmo};
    \draw (-4.8,-0.25) node {horizon};
    \draw (5.6,-0.35) node {$r\sim r+2\pi$};
    \draw (3.7,-1.8) node {$2\pi$};
    \draw (-1.3,-1.8) node {$\pi/2$};
    \draw (-3.05,-1.8) node {$0$};
    \draw (-1.3,2.7) node {\color{blue}contour real $\rho$};
    \draw (-1.3,1.8) node {\color{blue}$\rho=\infty$};
  \end{tikzpicture}
\end{equation}
Indeed, also the boundary conditions \eqref{2.3 bc} allow for constants shifts of $\Phi$
\begin{equation}
    \sqrt{h}e^{\i \Phi/2}=\pm \i\,,\quad \Phi\to \frac{\pi}{2}+2\pi m+\i\infty\,. 
\end{equation}
Using the same Weyl-transformation as in \eqref{1.11}, one brings \eqref{3.50} in AdS$_2$ form. The structure of the solutions \eqref{3.49} indeed uniquely determines that a real AdS$_2$ metric is probed in the bulk. The on-shell gravitational action in section 4.2 of \cite{Blommaert:2024ydx} is modified because of the change in the dilaton $\Phi_h=\theta+2\pi m$, and reproduces \eqref{3.42}. 
\section{Limit 1. Heisenberg algebra and flat space JT gravity}\label{sect:flathermite}
In this section we consider a $\beta\to 0$ or $\Phi\to \pi/2$ limit of sine dilaton gravity, where the model reduces to (regulated) flat space quantum gravity, as we demonstrate in section \textbf{section \ref{subsect:4.1}}. The ungauged model has a Hagedorn divergence, whereas the gauged system has a Gaussian energy spectrum. We show in \textbf{section \ref{subsect:4.2}} and in \textbf{section \ref{subsect:4.3}} how this follows from canonical quantization. The classical system reduces to the ordinary harmonic oscillator.\footnote{From the DSSYK perspective this was discussed in \cite{Almheiri:2024xtw}.} Ungauged wavefunctions are parabolic cylinder functions. The gauged wavefunctions (obtained by projection) are the ordinary Hermite functions.

In addition to the application to flat space and the technical simplification, this model may be useful because it captures the essence of the entropic paradox in section \ref{subsect:infopuzzle} in a perhaps sharper way. Which semi-classical degrees of freedom are counted in the entropy formula \eqref{3.34}?

\subsection{Flat space JT gravity from gauged sine dilaton gravity}\label{subsect:4.1}
Recall the action of sine dilaton gravity \eqref{1.4sdaction} (we will suppress boundary terms throughout this section)
\begin{equation}
    I=\frac{1}{2}\int\d x \sqrt{g}(\Phi R + 2\sin (\Phi))
\end{equation}
Expanding the dilaton field as $\Phi\to \frac{\pi}{2} + \sqrt{|\log q|}\Phi $ with $\abs{\log q}\to 0$ and rescaling the metric conveniently this becomes
\begin{equation}\label{Sflat}
\boxed{I=\frac{1}{2}\int \d x\sqrt{g}(\Phi R+2-\abs{\log q}\Phi^2)\,\, \to \,\, \frac{1}{2}\int \d x\sqrt{g}(\Phi R+2)}
\end{equation}
where we ignored the subleading term in the $\abs{\log q} \to 0$ limit for now. We have also changed our notion of $\hbar$ to $\hbar=2\sqrt{\log q}$. 
Integrating out $\Phi$ localizes on flat metrics, thus in this limit sine dilaton gravity is simply flat JT quantum gravity
\begin{equation}
    R=0\,.
\end{equation}
The classical solutions are directly obtained as the limit of the sine dilaton solutions \eqref{1.9}. With $\Phi=r$ one obtains\footnote{The holographic boundary is at $\Phi = i\infty$ in these new variables according to equation \eqref{2.3 bc}.}
\begin{equation}
    \d s^2=F_\text{flat}(r)\d \tau^2+\frac{1}{F_\text{flat}(r)}\d r^2\,,\quad F_\text{flat}(r)=2r-2E\,.\label{4.4flat}
\end{equation}
Here the ADM energy $E=\Phi_h$. Using $r=E+\rho^2/2$, this becomes the usual Rindler metric $\rho^2\d\tau^2+\d\rho^2$ with temperature
\begin{equation}
    \beta=2\pi\,.
\end{equation}
The fact that the temperature of the horizon is independent of the ADM energy $E$ signals a Hagedorn divergence. This can also be appreciated by taking this limit in the exact partition function of ungauged sine dilaton gravity \eqref{eq:fakepf}, which becomes
\begin{equation}
    Z(\beta)=\int_{-\infty}^{+\infty}\d E \exp\bigg(\frac{2\pi E}{2\sqrt{\log q}}\bigg) \exp\bigg( -\frac{\beta E}{2\sqrt{\log q}}\bigg)\,.
    \label{4.6}
\end{equation}
This Hagedorn spectrum is the usual flat space result, derived first in the CGHS model \cite{Fiola:1994ir,Godet:2020xpk}, which is related to flat space JT gravity by a Weyl transformation. See also \cite{Stanford:2020qhm}. 

Such a degeneracy is usually unwanted. This is where the first subleading correction in \eqref{Sflat} comes into play. Taking the leading $\abs{\log q}$ corrections into account one finds an almost flat metric $R=2\abs{\log q} \Phi$
\begin{equation}
    F(r)=2r-\frac{\abs{\log q}}{3}r^3-2E\label{4.7}\,,
\end{equation}
leading to the regularized thermodynamic relations
\begin{equation}
    S=\frac{2\pi}{\sqrt{\log q}} \text{ arctanh}(\sqrt{\abs{\log q}}E)\,,\quad \frac{2\pi}{\beta}=1-\abs{\log q}\frac{E^2}{2}\,.\label{4.8}
\end{equation}
This correction removes the Hagedorn divergence, but does not render the partition function convergent, because the energy $E$ remains unbounded from below. This stems from the divergence in the original ungauged partition function \eqref{eq:fakepf}. To obtain a finite sensible quantum theory of flat space quantum gravity, we show in section \ref{subsect:4.2} that the Hamiltonian $H$ still has a $P\to P+2\pi$ symmetry in this limit. Gauging that symmetry results in a reasonable theory.

\begin{figure}[t]
 \centering
\includegraphics[width=0.5\textwidth]{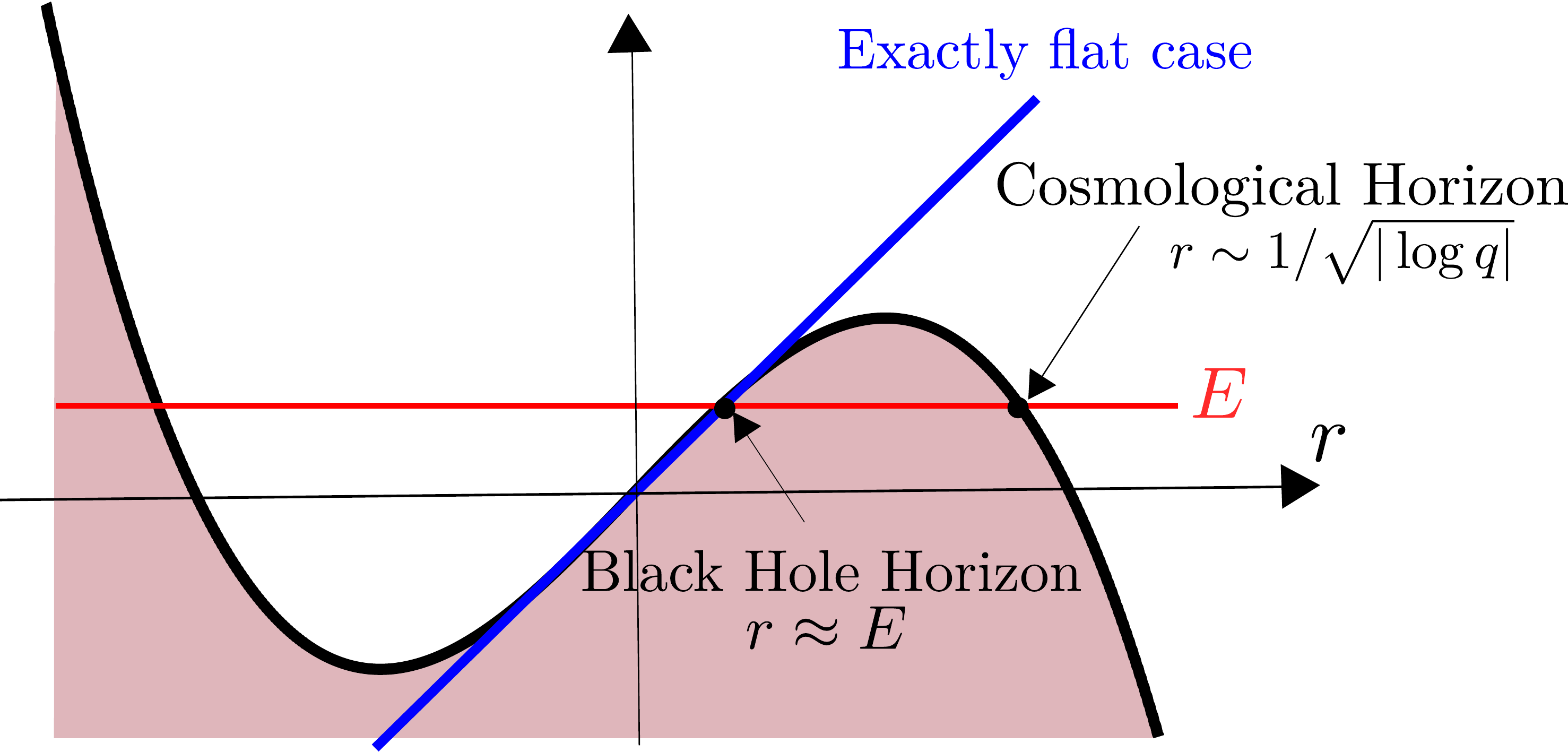}
\caption{Cartoon of the horizon function \eqref{4.7}. Flat space gets regulated by a cosmological horizon, which is infinitely far away in the strict $\abs{\log q}\to 0$ limit of exactly flat space \eqref{4.4flat} (blue). The light red region is the normal signature region outside of the black hole horizon.
}
\label{fig:cubic}
\end{figure}

Before doing that, we comment on the physical way in which the subleading correction in the metric \eqref{4.7} regulates flat space in order to make sense of it in quantum gravity. The point is that the corrected metric \eqref{4.7} still has a cosmological horizon at $r\sim 1/\sqrt{\abs{\log q}}\to\infty$.\footnote{We should of course not trust the metric \eqref{4.7} is this regime as a good approximation to our original sine dilaton gravity model, since higher order corrections become relevant there. This only affects the order one prefactor of the cosmological horizon location which is at $r=\pi/\sqrt{\log q}-E$.} This cosmological horizon is very far away. Regularizing flat space with a cosmological horizon very far away is perhaps not unreasonable, given the current state of our universe. We sketch the structure of the horizons in Figure \ref{fig:cubic}.
\subsection{Flat space JT gravity and the harmonic oscillator}\label{subsect:4.2}
In order to discuss the gauging procedure that cures the divergences in the partition function \eqref{4.6}, we take the relevant limit of the phase space of sine dilaton gravity \eqref{weyl} spanned by $L$ and its conjugate $P$, with the latter given by
\begin{equation} \label{mom}
e^{-\mathrm{i} P}=-\mathrm{i}\sin(\theta)\tanh(\sin(\theta)T/2)+\cos(\theta)\,.
\end{equation}
We consider the following scaling of parameters
\begin{equation}
    \theta=\pi/2-\sqrt{\abs{\log q}}E\,,\quad L=2\abs{\log q}n\,,\quad T=2\sqrt{\log q}k
\end{equation}
In this limit one finds the phase space (in this equation all $\abs{\log q}$ dependence has been made explicit)
\begin{equation}
    H=\frac{E}{2\sqrt{\abs{\log q}}}\,,\quad n=\frac{E^2}{2}+\frac{k^2}{2}\,,\quad \omega=\d E\wedge \d k\,.
\end{equation}
This is the harmonic oscillator where $E$ plays the role of the ``position'' operator, and $n$ is the energy of the harmonic oscillator. We can rewrite this Hamiltonian in terms of $n$ and its conjugate $P$ as\footnote{We suppressed a prefactor $1/2\sqrt{\abs{\log q}}$ again.}
\begin{equation}
    \mathbf{H}=\cos(\mathbf{P})\sqrt{2\mathbf{n}}\,.\label{4.12}
\end{equation}
This is the system discussed by the authors of \cite{Almheiri:2024xtw} as the high temperature limit of DSSYK. Given the duality between sine dilaton gravity and DSSYK \cite{Blommaert:2024ydx}, we had to find this system. The point is that this is recovered as the canonical description of (regularized) flat space JT gravity \eqref{Sflat}. Moreover, the ``naive'' quantization of this gravity theory corresponding to the almost-Hagedorn thermodynamics \eqref{4.8}, would correspond with quantization where one one does not gauge $P\to P+2\pi$, and considers $n$ on the entire real axis. For the harmonic oscillator we usually do not consider this, but it is possible to quantize the gravity theory like this nonetheless (see section \ref{subsect:4.3}), following section \ref{sect:ungauge}. 

To stress the gravitational nature of \eqref{4.12}, we note that we can immediately find the Hamiltonian \eqref{4.12} from canonical quantization of the (regularized) flat space theory without prior reference to sine dilaton gravity. Using the equation for the momentum in general dilaton gravity which we derive later in \eqref{eqn:momentumintegral}
\begin{align}
&P= \int^H \frac{dH}{\sqrt{V(W^{-1}(H))^2-4e^{-L}}}=  \int^{W^{-1}(H)} \frac{d \Phi\, V(\Phi)}{\sqrt{V^2(\Phi) - 4e^{-L}}},
\end{align}
applied to the regularized flat space potential in \eqref{Sflat} with data
\begin{equation}
    V(\Phi)=2-\abs{\log q}\Phi^2\,,\quad e^{-L}=1-2\abs{\log q}n\,,\quad W^{-1}(H)=H+\text{ subleading}\,,
\end{equation}
and working to leading order in $\abs{\log q}\to 0$ one indeed recovers the correct relation \eqref{4.12}
\begin{align}
    & P =\int^{H}_{-\sqrt{2n}} \frac{\d\Phi}{\sqrt{2n - \Phi^2}} = \text{arccos}(H/\sqrt{2n}).
\end{align}
Notice that we are focusing on extremely short lengths in sine dilaton gravity for finite $n$ and $\abs{\log q}\to 0$.

The most important point in terms of a gravity interpretation is that the Hamiltonian \eqref{4.12}, much like the original sine dilaton gravity Hamiltonian \eqref{Hgrav}, has a $P\to P+2\pi$ symmetry which should be gauged, leading to the usual HO quantization. This causes a huge decrease in entropy. Indeed, the HO quantization leads to a gauged partition function with the usual Gaussian density \cite{Almheiri:2024xtw}
\begin{equation}\label{flatq1disk}
\boxed{Z(\beta)_{\text{gauged}} =\int_{-\infty}^{+\infty} \d x\,\exp\bigg(-\frac{E^2}{2}-\beta \frac{E}{2\sqrt{\abs{\log q}}}\bigg)}
\end{equation}
Thus the entropy of the theory is
\begin{equation}
    S=-E^2<S_\text{BH}=\frac{\pi E}{\sqrt{\log q}}\,.
\end{equation}
We emphasize that the state with maximal entropy does not correspond to the black hole with maximal area $E\to\infty$. As anticipated, the gauged system \eqref{flatq1disk} is indeed convergent. We can view this limiting case as an extreme scenario where the original $A/4G$ term is completely gone, only the new ``observer'' contribution remains.\footnote{The correction factor to the entropy \eqref{3.34} in this limit indeed has a contribution that may be interpreted as canceling perfectly the $A/4G_\text{N}$ term (ignoring constant contributions to the entropy)
\begin{equation}
-\frac{\theta^2}{\log q}=\frac{\pi E}{\abs{\log q}^{1/2}}- E^2=-\frac{A}{4 G_\text{N}}-E^2\,.
\end{equation}} 
We do not confidently know the physical interpretation of what $-E^2/2$ is counting, but we proposed in \cite{Blommaert:2024ydx} a concrete explanation in terms of an observer entropy contribution in this context. In parallel to the statement that black holes should be described by a quantum system with $A/4G_\text{N}$ degrees of freedom, the ``central dogma for cosmological horizons'' has been proposed \cite{Susskind:2021yvs}. Our models clearly deviate from \emph{both} statements even at the semi-classical level, raising a significant puzzle.

Finally, we note that \cite{Almheiri:2024xtw} proposed a dilaton potential of the form
\begin{equation}
V(\Phi) = U_0\left(1-\frac{96\pi^2(\Phi-\Phi_h)}{\beta^4 U_0}\right)^{-1}
\end{equation}
for some constant $U_0$. This depends explicitly on the inverse temperature $\beta$ resulting in a non-local field theory. Instead we have an ordinary dilaton gravity model \eqref{Sflat}, with a gauge redundancy, resulting in the same Gaussian thermodynamics \eqref{flatq1disk}.

\subsection{Gauged and ungauged wavefunctions}\label{subsect:4.3}
In this section we discuss the wavefunctions of the ungauged flat space quantum gravity, and how these reduce to the usual Hermite functions of the HO upon gauging (along the lines of section \ref{sect:gauged}). By doing a canonical transformation, the Schr\"odinger equation \eqref{4.12} becomes
\begin{equation}
    \mathbf{H}=\frac{1}{2}e^{-\i \mathbf{P}}+\frac{1}{2}e^{\i \mathbf{P}}\,2\mathbf{n}=E\,. 
\end{equation}
This can also be obtained from the sine dilaton Hamiltonian \eqref{Hgrav} upon shifting $P$ by a constant and taking the limit $\abs{\log q}\to 0$. The left-eigenfunctions are parabolic cylinder function. These are encoded in the left-eigenfunctions \eqref{left1} of sine dilaton gravity in the relevant limit $\abs{\log q}\to 0$\footnote{The associated disk has Hagedorn spectrum \eqref{4.6} and thus not the regulated version \eqref{4.8}.}
\begin{equation}\label{eq:psilqto1}
    \frac{(q^2;q^2)_\infty}{|\log q|^{\frac{n}{2}}} \,\exp\bigg(\frac{\theta^2}{2\abs{\log q}}\bigg)\,\psil_\theta(n)\to 2^\frac{n}{2}e^{E^2/2}\mathsf{D}_n(\sqrt{2}E)\,.
\end{equation}
At integer values these parabolic cylinder functions become Hermite polynomials (see \href{https://mathworld.wolfram.com/ParabolicCylinderFunction.html}{wolfram})\footnote{We thank Ohad Mamroud for discussions related with this.}
\begin{equation}\label{Dnvsherm}
   2^\frac{n}{2} e^{E^2/2} \mathsf{D}_n(\sqrt{2}E) = \rH_n(E)\,, \quad n \in \mathbb{N}\,.
\end{equation}
This is consistent with the values of the left wavefunctions in sine dilaton gravity at integers \eqref{left2}, and the known classical $\abs{\log q}\to 0$ limit of the q-Hermite functions
\begin{equation}
  |\log q|^{-\frac{n}{2}} \rH_n(\cos\theta|q^2)\to \rH_n(E)\, .
\end{equation}
Hermite functions are orthonormal with a Gaussian measure, resulting in the partition function \eqref{flatq1disk}. These relations are shown numerically in Figure \ref{fig:hermlim} and \eqref{eq:psilqto1} is derived analytically in appendix \ref{app:flatwave}.

\begin{figure}[t]
    \centering
    \begin{subfigure}{0.45\textwidth}
    \centering
 \includegraphics[width=\textwidth]{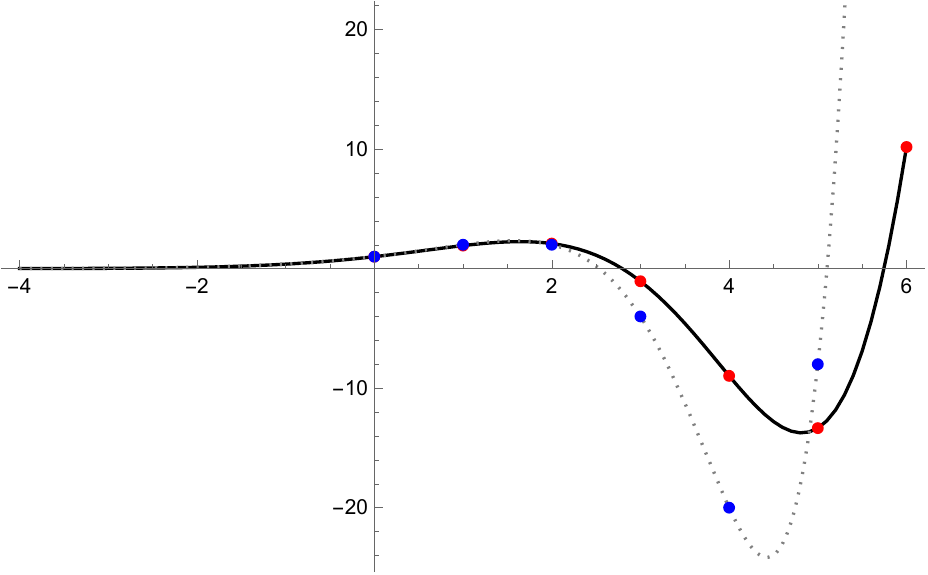}
    \end{subfigure}
    \begin{subfigure}{0.43\textwidth} 
    \centering 
\includegraphics[width=\textwidth]{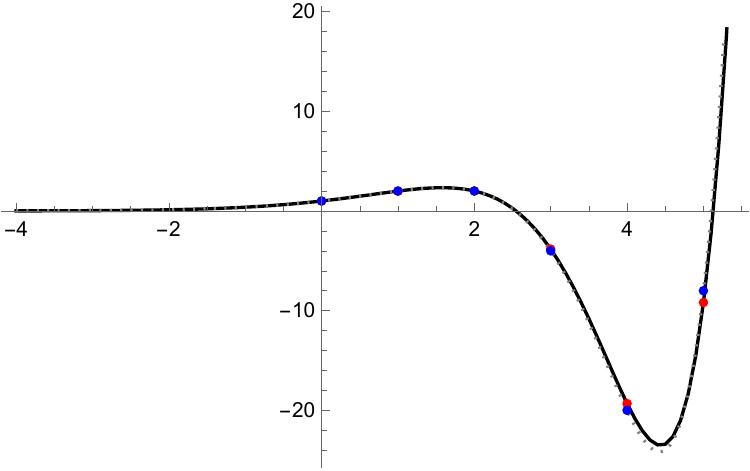}
    \end{subfigure}
    \caption{As $q\to 1$ the rescaled $\psil_\theta(n)$ \eqref{eq:psilqto1}  (black) becomes a parabolic cylinder function (dotted gray). At integer $n$ these functions are $q$-Hermite (red dots) respectively ordinary Hermite polynomials (blue dots). To illustrate the limit we considered $q=0.8$ (left) and $q=0.99$ (right).}
    \label{fig:hermlim}
\end{figure}
\section{Limit 2. Gaussian matrices and topological gravity }\label{sect:topogravq0}
In this section we briefly discuss a second limit of our sine dilaton gravity quantization that has made an appearance in recent literature \cite{Gopakumar:2011ev, Gopakumar:2022djw}. We consider the very quantum limit $\abs{\log q}\to \infty$, which can be thought of as $\hbar\to\infty$. In this limit the weight in the gravitational path integral \eqref{Irescaled} becomes trivial
\begin{equation}
    \int\mathcal{D}g\mathcal{D}\Phi\exp\bigg(S_0 \chi + \frac{1}{2|\log q|}\int \d x \sqrt{g}(\Phi R+2\sin(\Phi))+\text{ boundary}\bigg) \quad \to \quad \int\mathcal{D}g\,e^{S_0 \chi}\,.\label{5.1}
\end{equation}
Off shell configurations are not suppressed. We can make sense of amplitudes in this theory by taking the corresponding limit in the gauged sine dilaton gravity amplitudes of section \ref{sect:gauged}. Since the dilaton is a periodic variable its contributions are finite, and can be absorbed in $S_0=N\log 2$. This topological field theory is ordinary Einstein-Hilbert gravity. Somewhat surprisingly, this has different amplitudes than the model with the same action studied in \cite{Marolf:2020xie}. Instead, the theory that we obtain is the usual description of topological gravity. By this, we mean the $(2,1)$ minimal model (also called ``Airy'' model) underlying intersection theory on the moduli space of curves. See for instance \cite{kontsevich1992intersection,witten1990two,Okuyama:2019xbv,Blommaert:2022lbh} for some background. There is indeed a description
of topological gravity as pure Einstein-Hilbert gravity \cite{Dijkgraaf:1991qh} (section 7.2). Let us see how this appears out of our formalism.

The gauged spectral density \eqref{3.33} of sine dilaton gravity reduces to the famous Wigner semicircle distribution of the Gaussian matrix model \cite{mehta2004random}
\begin{equation}\label{chebmeasure}
    \d \theta (e^{\pm2\i \theta};q^2)_\infty \to \d \theta\,\frac{2}{\pi} \sin^2\theta = \d E\,\frac{2}{\pi} \sqrt{1-E^2}\,,\quad E=\cos(\theta)\,.
\end{equation}
At the level of the gravitational wavefunctions we find the Chebyshev polynomials of the second kind
\begin{equation}
    \rH_n(\cos\theta|q^2) \to  \sum_{k=0}^n e^{\i (n-2k)\theta} = \frac{\sin((n+1)\theta)}{\sin\theta} = \mathsf{U}_n(E)\,.
\end{equation}
These are indeed orthogonal with respect to the semicircle density. We can also check this at the level of the recursion relations \eqref{7.2}, which indeed become those of the Chebychev polynomials:
\begin{equation}
    \mathbf{H}\to\cos(\mathbf{P})\,,\quad\psi_E(n+1)+\psi_E(n-1) = 2E\, \psi_E(n)\,.
\end{equation}
We keep $n$ finite when sending $q\to 0$, which means that the lengths of the surfaces become extremely large $L\to \infty$ with $L/2\abs{\log q}=n$. Similarly we scale $\beta\to \infty$ with $\beta/2\abs{\log q}$ finite. In this large lengths limit the Riemann surfaces counted in \eqref{5.1} degenerate into thin long strips called ribbon graphs \cite{Blommaert:2022ucs,Stanford:2022fdt}. For instance a three holed sphere with boundary lengths $L_1,L_2,L_3$ degenerates into one of the following ribbon-like Riemann surfaces:
\begin{equation}
        \begin{tikzpicture}[baseline={([yshift=-.5ex]current bounding box.center)}, scale=0.7]
 \pgftext{\includegraphics[scale=1]{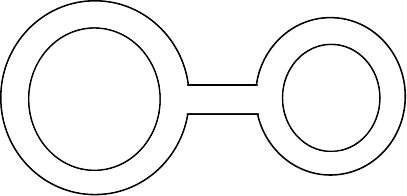}} at (0,0);
 \draw (0.4, -0.8) node {$L_1$};
     \draw (2.2,0) node {$L_2$};
     \draw (-1.8, 0) node {$L_3$};
  \end{tikzpicture}\quad\text{or}\quad
        \begin{tikzpicture}[baseline={([yshift=-.5ex]current bounding box.center)}, scale=0.7]
 \pgftext{\includegraphics[scale=1]{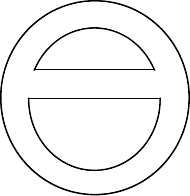}} at (0,0);
     \draw (0, 0.8) node {$L_1$};
     \draw (0, -0.6) node {$L_2$};
     \draw (2.2, 0) node {$L_3$};
  \end{tikzpicture}
    \end{equation}

For finite $n$, these ribbons are quantized into chord diagrams. Morally, the chords are the Gaussian Wick contractions between boundaries in the Gaussian matrix integral.\footnote{More precisely there is an integral transform between these numbers \cite{kontsevich1992intersection}.} The wormhole in this limit matches also with the eigenvalue correlation \cite{Blommaert:2021gha} in a Gaussian matrix integral, as we show in \cite{wip}. We fail to see any reason to believe that this correspondence would not extend to arbitrary genus, because this is essentially known for topological gravity \cite{Dijkgraaf:1991qh,Gopakumar:2011ev, Gopakumar:2022djw}.

The strict Airy model is recovered in the limit $n\to \infty$ with $n \theta$ finite. The length of the ribbons can then be treated as a continuum, and the associated path integrals with trivial action (thus computing volumes of moduli space) compute intersection numbers \cite{kontsevich1992intersection}. This is indeed the known low-energy limit of the Gaussian matrix integral \cite{Saad:2019lba,Maldacena:2004sn,Blommaert:2019wfy}. The semicircle density becomes $\sqrt{E}$ and the gravitational wavefunctions become $\psi_E(L)=\sin(L\sqrt{E})/\sqrt{E}$.

The fact that this theory appears as a limit of sine dilaton gravity should help in understanding how to define our theory on more complicated topologies. Together with the wormhole amplitude computed in \cite{wip}, this limit is strong evidence that non-perturbatively (gauged) sine dilaton gravity is the matrix integral that \cite{Jafferis:2022wez} called q-deformed JT gravity. We comment more on other topologies in the discussion section \ref{sect:discnew}.

\section{Periodic dilaton gravity}\label{sect:genericpotential}
In this section, we consider generic periodic dilaton gravity \eqref{1.1}. We'll argue in \textbf{section \ref{subsect:6.1momsym}} that the Hamiltonian, when expressed in terms of the effective Weyl rescaled AdS geodesic length $L$ of global slices, universally exhibits periodicity in the conjugate momentum $P$. Just as in sine dilaton gravity, in order to avoid divergences, this symmetry is to be gauged. This discretizes the geometry. In \textbf{section \ref{subsect:6.2entropy}} we show that the Schr\"odinger recursion relation ends at the minimal lengths $L_\text{min}$ of the Lorentzian phase space. So, the pole structure in the wavefunctions resembles that of Figure \ref{fig:psirpoles}, notably with no poles at $L<L_\text{min}$. Therefore states with $L<L_\text{min}$ are null states. Assuming the tracial state remains the infinite tension brane \eqref{3.30}, one finds an entropy profile $S<S_\text{BH}$ reminiscent of a finite cut matrix integral.
\subsection{Momentum shift symmetry}\label{subsect:6.1momsym}
Recall the action of periodic dilaton gravity \eqref{1.1}\footnote{For a generic dilaton gravity model, the boundary term that reproduces classical thermodynamics is \cite{Blommaert:2024ydx}
\begin{equation}
\int \d\tau \sqrt{h}\,\left(\Phi K-\sqrt{W(\Phi)}\right), \qquad W(\Phi_\text{bdy}) = \int^{\Phi_{\text{bdy}}}\d \Phi\,V(\Phi)\,.
\end{equation}}
\begin{equation}
    I=\frac{1}{2}\int\d x \sqrt{g}(\Phi R + V(\Phi))\,+\text{ boundary}\,,\quad V(\Phi+2\pi a)=V(\Phi)\,.
\end{equation}
In order to quantize this theory on global slices, and inspired by the sine dilaton gravity construction \eqref{1.11}, we first find a Weyl transformation that takes us from the physical metric to the auxiliary AdS$_2$ metric. This means we search for $\omega$ such that\footnote{The spacetime contour is $r_h<r<\infty$ with the holographic screen at $r\to +\infty$ as for sine dilaton gravity.}
\begin{align}
\left(W(\Phi) - W(\Phi_h)\right) \d\tau^2 + \frac{\d\Phi^2}{W(\Phi) - W(\Phi_h)} = e^{2\omega} \bigg((r^2 - r_h^2)\,\d\tau^2 + \frac{\d r^2}{r^2 - r_h^2}\bigg)\,.\label{6.3bb}
\end{align}
This equation implies that
\begin{align}
\Phi'(r) = e^{2\omega} = \frac{W(\Phi) - W(\Phi_h)}{r^2 -r_h^2}\,.
\end{align}
Demanding that $\Phi(r_h) = \Phi_h$, we observe that this equation implies
\begin{align}\label{eqn:Vrrelation}
V(\Phi_h) = 2r_h.
\end{align}

\begin{figure}
    \centering
    \begin{subfigure}{0.45\textwidth}
    \centering
 \includegraphics[width=\textwidth]{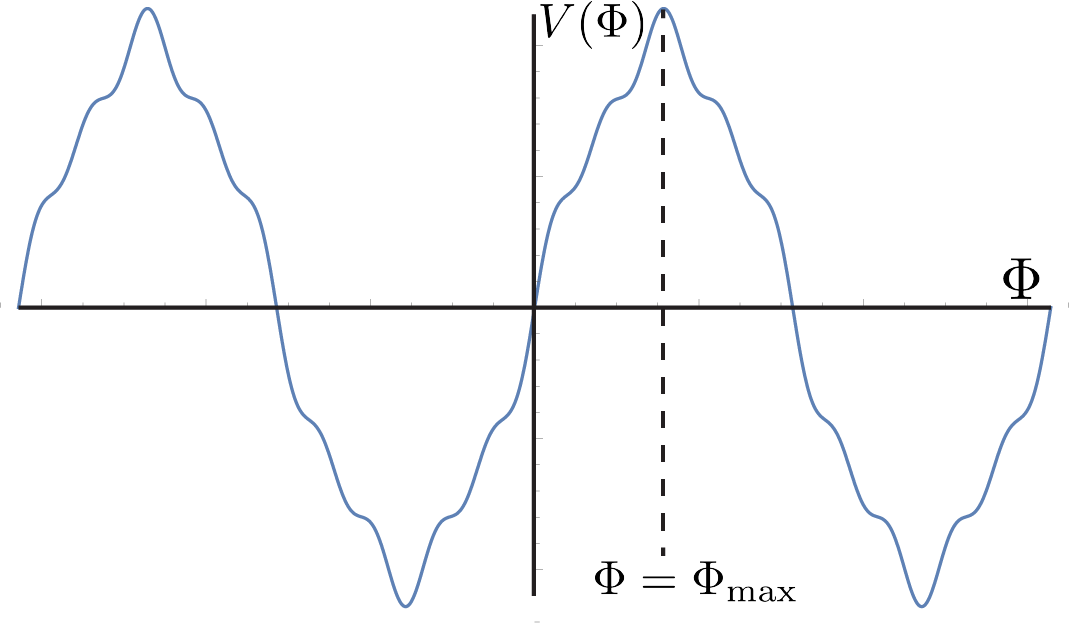}
    \end{subfigure}
    \begin{subfigure}{0.43\textwidth} 
    \centering 
\includegraphics[width=\textwidth]{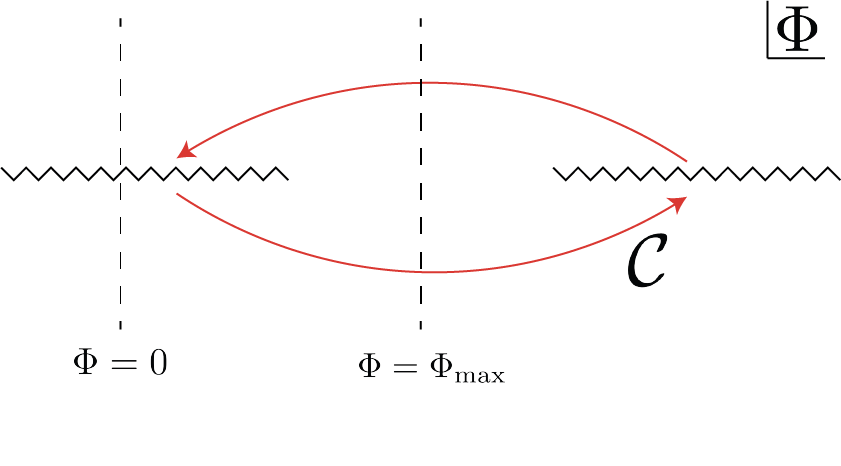}
    \end{subfigure}
    \caption{On the left, we plot an example of a potential $V(\Phi)$ that obeys the assumptions we use. This potential takes the form $V(\Phi) = \sin(\Phi) + c_9 \sin(9 \Phi)$ for some small enough $c_9$. On the right, we demonstrate the qualitative branch cut structure for an example potential in the integral \eqref{eqn:momentumintegral}. The two horizontal branch cuts connect the branch points at $V(\Phi) = \pm e^{-L/2}$. The space on the real $\Phi$ line between the branch cuts is where the integrand of \eqref{eqn:momentumintegral} takes real values. Since the integrand is uniformized on the double cover of the complex plane, the contour $\mathcal{C}$ (in red) is a closed cycle which starts on the principal branch, goes to the second sheet and then comes back.}
    \label{fig:periodicpotential}
\end{figure}

Using this conformal mapping, we can describe the phase space of solutions in terms of the length of geodesics in this auxiliary spacetime. The regularized length $L$ of the geodesic in the AdS$_2$ spacetime with horizon at $r_h$ that reaches the holographic AdS$_2$ boundaries at Lorentzian time $\tau=\i T$ is
\begin{align}
e^{-L}=\frac{V(\Phi_h)^2}{\cosh(V(\Phi_h) T/2)^2}\,.\label{6.6}
\end{align}
This is to be compared with \eqref{weyl}. The symplectic form for this phase space is simply \cite{Harlow:2018tqv}
\begin{align}
\omega = \delta T \wedge \delta H\,,\quad H=W(\Phi_h)=\int^{\Phi_h}_{\Phi_\text{max}}\d \Phi\,V(\Phi)\,,\quad V(\Phi)\leq V(\Phi_\text{max})=V_\text{max}\,.\label{6.7}
\end{align}
Here $\Phi_\text{max}$ is the place where the potential reaches its maximum. As shown in Figure \ref{fig:periodicpotential}, in the models of our interest $\Phi_\text{max}=a\pi/2$. Using \eqref{6.6} we can solve for $T=T(L,H)$. Plugging this into $\omega$, we find the momentum $P$ conjugate to $L$
\begin{align}\label{eqn:momentumintegral}
P= \int^H \frac{\d E}{\sqrt{V(W^{-1}(E))^2-e^{-L}}}=  \int^{W^{-1}(H)}_{\Phi_0(L)} \frac{d \Phi\, V(\Phi)}{\sqrt{V^2(\Phi) - e^{-L}}}\,,\quad V(\Phi_0(L))=e^{-L/2}\,.
\end{align}
For $L\to\infty$ this relation simplifies to
\begin{align}
H = W(P).
\end{align}
This means that we might be tempted to identify $P$ with the horizon value of the dilaton $\Phi_h$. Perhaps more importantly, we observe that at $L\to\infty$ periodicity of $V(\Phi)$ implies periodicity of the Hamiltonian under $P\to P+2\pi a$. As we discuss in section \ref{subsect:6.2entropy}, we must treat this as a redundancy for the quantum theory to make sense
\begin{equation}
    \boxed{\mathbf{P}\sim \mathbf{P}+2\pi a}
\end{equation}
In the remainder of this subsection we explain why this generalizes to finite $L$.

In principle one should compute the integral \eqref{eqn:momentumintegral} explicitly, and solve the resulting equation $P=P(H,L)$ for $H=H(P,L)$. For generic $V(\Phi)$ this is difficult. However, one can prove the periodicity of $H(P,L)$ under shifts of $P$ from the integral representation \eqref{eqn:momentumintegral} itself. For the argument, we restrict to potentials which are symmetric around a point $\Phi_\text{max}$ and anti-symmetric around $\Phi=0$.\footnote{Constant shifts in $\Phi$ only change $S_0$ so this choice is not restrictive.} This makes $V(\Phi)$ periodic with $2\pi a=4\Phi_\text{max}$. We also assume there are no other local extrema. These assumptions are only sufficient, not necessary. An example of such a potential is shown on the left in Fig. \ref{fig:periodicpotential}.\footnote{Such potentials have the form
\begin{align}
    V(\Phi) = \sum_{n = 1\, \text{mod}\, 4} c_n \sin(n \Phi/a).\label{6.11}
\end{align}
The assumption that $V(\Phi)$ have no other extrema besides at $\pm m \Phi_{max}$ also places a non-trivial constraint on the $c_n$'s.} We leave
a full exploration of what types of potentials lead to a real shift symmetry in the Hamiltonian for future study.\footnote{The results of section \ref{sect:flathermite} suggest that a symmetric maximum would be sufficient too.}

As illustrated in Figure \ref{fig:periodicpotential}, the integrand in \eqref{eqn:momentumintegral} has branchpoints at $\Phi_0(L)$ and $\pi a-\Phi_0(L)$. The integrand is single-valued on a double cover of the complex $\Phi$ plane. Thus, we can go around cycles on this Riemann surface, starting and ending at the same value of $W(\Phi) = H$. We can then always add such a cycle, starting and ending at $W^{-1}(H)$, to the integral in \eqref{eqn:momentumintegral} without changing the value of $H$. The integral around this cycle then corresponds to a shift in the momentum which preserves $H$. With our assumptions, the only interesting cycle $\mathcal{C}$ is illustrated in Figure \ref{fig:periodicpotential}.\footnote{For our anti-symmetric potentials the integral around the branchcut vanishes.} The Hamiltonian $H(P,L)$ is then invariant under the following real momentum shifts
\begin{align}
    H(P,L) = H\left(P+2\pi a f(L),L\right)\,,
\end{align}
with
\begin{equation}
    2\pi a f(L) = \oint_{\mathcal{C}} \frac{d\Phi\,V(\Phi)}{\sqrt{V(\Phi)^2 - e^{-L}}} = 2\int_{\Phi_0(L)}^{\pi a-\Phi_0(L)} \frac{d\Phi\,V(\Phi)}{\sqrt{V(\Phi)^2 - e^{-L}}}\,,\quad f(\infty)=1\,.
\end{equation}
Performing the canonical transformation
\begin{equation}
    P=\tilde{P} f(L)\,,\quad \tilde{L} = \int^L_{L_\text{min}}\d L' f(L')\,,\quad V(\Phi_\text{max})=e^{-L_\text{min}/2}\,,\label{6.16}
\end{equation}
The identification becomes $\tilde{P}\sim \tilde{P}+2\pi a$, which upon quantization discretizes $\tilde{L}=\hbar n/a$ with $n$ integer.\footnote{Recall that in sine dilaton gravity we absorbed $\hbar=2\abs{\log q}$.} We stress that this is a universal consequence of having a potential with $\Phi\to \Phi+2\pi a$ symmetry. We now argue that, just like in sine dilaton gravity, in the quantum theory states with $n<0$ become null, with major effects on the semiclassical entropy profile.
\subsection{An entropic paradox}\label{subsect:6.2entropy}
Because of the symmetry $\tilde{P}\sim \tilde{P}+2\pi a$, the Schr\"odinger equation $\mathbf{H}=E$ becomes a difference equation that relates wavefunctions $\psi_E(\tilde{L})$ at $\tilde{L}+\hbar m/a$. Upon gauging the symmetry, this is a recursion relation for $\psi_E(n)$. Not gauging is not an option, even semiclassically the ungauged partition function diverges, analogously to \eqref{3.26}\footnote{This equation is correct to leading order in the semiclassical limit $\hbar\to 0$. However it seems plausible that this is actually the exact equation, see the discussion section \ref{sec:gasofdef}.}
\begin{equation}
\label{eq:Zunexact}
    Z_\text{ungauged}(\beta)=\int_{-\infty}^{+\infty}\d \Phi_h V(\Phi_h)\sinh(2\pi \Phi_h/\hbar)\,e^{-\beta W(\Phi_h)/\hbar}\,.
\end{equation}
This diverges again due to the periodicity of the energy $E=W(\Phi_h)$. Consider the minimal value $L_\text{min}$ that is reached in the classical phase space \eqref{6.6}
\begin{equation}
    e^{-L_\text{min}/2}=V_\text{max}\,.
\end{equation}
At this point in phase space
\begin{equation}
    H(P,L_\text{min})=0\,,
\end{equation}
for all momenta. This is quite obvious from equation \eqref{6.6} and \eqref{6.7} since the minimal length is achieved at $T=0$ and $\Phi_h=\Phi_\text{max}$ such that indeed according to \eqref{6.7} the Hamiltonian vanishes. This can also be understood from the $E$ integral representation \eqref{eqn:momentumintegral}, which develops a simple pole at $E=0$ at $L_\text{min}$ such that for finite $P$ indeed $H=0$.\footnote{The sine dilaton Hamiltonian in this formulation is related to equation \eqref{Hgrav} by a canonical shift in $P$ that renders the Hamiltonian real
\begin{equation}
    H=-\cos(P)\sqrt{1-e^{-L}}\,.\label{6.18}
\end{equation}
This indeed vanishes at $L=0$.
}

One can now repeat the argument made in section \ref{sect:gauged} below \eqref{7.63prod} and around \eqref{3.23}. According to \eqref{6.16}, $L=L_\text{min}$ corresponds to $n=0$. For $\psi_E(0)\neq 0$, the Schr\"odinger equation for $\psi_E(0)$ implies that $\psi_E(1)$ is divergent.\footnote{Actually since now the Hamiltonian may have a whole Laurent series in $e^{\i P}$, the requirement is that at least one of the terms appearing in the series is divergent. But then all wavefunction components at larger $n$ are necessarily also divergent.} More in general, one could in principle find consistent solutions to the recursion relation by imposing that $\psi_E(n\leq 0)=0$. This corresponds with the fact that in the gauged theory we are truely quantizing the Lorentzian phase space, which has a restriction $L\geq L_\text{min}$ as discussed in the introduction section \ref{sect:intro} below \eqref{Hgrav}. States with $L< L_\text{min}$ are null, and the infinite tension state \eqref{3.30} becomes the minimal length state that is not null
\begin{equation}
    \mathbf{\Pi}\ket{\mu=\infty}=\ket{L=L_\text{min}}\,.
\end{equation}
By analogy we conjecture that the infinite tension state $\ket{n=0}$ is again the unique tracial state of the quantum mechanics to which one can associate an entropy. It would be interesting to make this precise. In what is left of this section we compute this entropy in the leading semiclassical approximation.

Consider the trace
\begin{equation}
    \bra{\mathbf{L}=L_\text{min}}e^{-\beta \mathbf{H}}\ket{\mathbf{L}=L_\text{min}}=\int_{0}^{\pi a}\d W(\Phi_h)\, e^{S(\Phi_h)/\hbar-\beta W(\Phi_h)/\hbar}\,.\label{6.20}
\end{equation}
The task is to compute $S(\Phi_h)$ to leading order in $1/\hbar$. This calculation will follow a logic similar to the one used in \cite{Blommaert:2024ydx}. For this we consider the Euclidean version of the renormalized geodesic lengths \eqref{6.6}, with a boundary condition indicative of the fact that we compute a transition matrix element between states $\ket{\mathbf{L}=L_\text{min}}$:
\begin{equation}
    L(0)=L(\beta)=L_\text{min}\,.
\end{equation}
The boundary condition $L(0)=L_\text{min}$ fixes the solution to
\begin{equation}
    e^{-L}=\frac{V(\Phi_h)^2}{\sin^2(V(\Phi_h)\tau/2-(\pi-\text{arcsin}(V(\Phi_h)/V_\text{max}))}\,.\label{6.22bb}
\end{equation}
Indeed, the constant $\pi-\text{arcsin}(V(\Phi_h)/V_\text{max})$ implies $e^{-L(0)}=V_\text{max}^2$. The boundary condition $L(\beta)=0$ fixes the inverse temperature $\beta$ of the solution to be
\begin{equation}
    \beta V(\Phi_h)=2\pi-4\text{arcsin}(V(\Phi_h)/V_\text{max})=\frac{\d S}{\d \Phi_h}\,.\label{6.23bb}
\end{equation}
This can be integrated to obtain the desired semiclassical entropy profile
\begin{equation}
    \boxed{S=2\pi \Phi_h-4\int_0^{\Phi_h}\d \Phi\,\,\text{arcsin}(V(\Phi)/V_\text{max})\,\,<\,\,S_\text{BH}=2\pi \Phi_h}\label{6.28}
\end{equation}
This equation passes some sanity checks. The entropy vanishes at the extrema of the spectrum
\begin{equation}
    S(0)=S(\pi a)=0
\end{equation}
and is symmetric around the maximum $S(\pi a /2)=S_\text{max}$. This shares all the important physical features of the sine dilaton gravity entropy profile \eqref{3.34}.

We remark that the classical solution \eqref{6.22bb} corresponds (along the lines of section 4.2 in \cite{Blommaert:2024ydx}) with the effective AdS renormalized geodesic length $L$ computed in a geometry with two conical {\color{blue}half-defects}
\begin{equation}
    \begin{tikzpicture}[baseline={([yshift=-.5ex]current bounding box.center)}, scale=0.7]
 \pgftext{\includegraphics[scale=1]{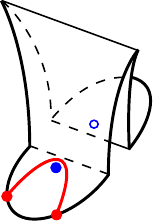}} at (0,0);
    \draw (0.3,-2.3) node {\color{red}geodesic};
  \end{tikzpicture}\quad e^{-L}=\frac{V(\Phi_h)^2}{\sin^2(V(\Phi_h)\tau/2-(\pi-\text{arcsin}(V(\Phi_h)/V_\text{max}))}\,.\label{4.46 analyticcont}
\end{equation}
The two half defects with opening angles $2\pi-2 \text{arcsin}(V(\Phi_h)/V_\text{max})$, which one can think of as preparing the state $\ket{\mathbf{L}=L_\text{min}}$, classically merge into one defect with opening angle
\begin{equation}
    \gamma=2\pi-4\text{arcsin}(V(\Phi_h)/V_\text{max})\,.
\end{equation}
Following equation (4.28) of \cite{Blommaert:2024ydx}, this leads to the same thermodynamics \eqref{6.23bb} indeed.

In summary, we have shown that the physically relevant features of the quantization of sine dilaton gravity which we have emphasized throughout this work (gauging, discretization, null states, an entropic puzzle) generalize to generic periodic dilaton gravity models. Near the entropic maximum, the models are universally described by the flat space JT gravity model of section \ref{sect:flathermite}. The end of the spectra become AdS respectively dS (see the discussion section \ref{sect:discnew}) JT gravity. Thus, we have defined a new universality class of dilaton gravity models. We believe these are all toy models for 2d quantum cosmology with a normalizable wavefunction of the universe, see the discussion section \ref{sect:7.5}.

It would be interesting to understand if these models are somehow all exactly solvable. One could imagine working perturbatively around sine dilaton gravity using $c_n=\epsilon_n$ in \eqref{6.11} following the spirit of the defect expansion of \cite{Maxfield:2020ale,Witten:2020wvy}. Because these theories all have finite spectral support one could imagine describing them all as perturbations of the Gaussian matrix integral (see for instance \cite{Blommaert:2021gha}), perhaps by expanding in critical potentials \cite{gross1990nonperturbative}. One could find the potentials that reproduces the entropy \eqref{6.28} and hope that connects with the gravity analysis. From the path integral point of view an expansion in $\epsilon_n\sin(n \Phi)$ around sine dilaton gravity might not be too horrible either. These are still defects, although uncommon ones. If we learn how to compute exact defect amplitudes in sine dilaton gravity, this seems possible. Some progress on the latter is reported in \cite{wip}.

The spacetimes \eqref{6.3bb} have black hole horizons at $\Phi_h+2\pi a m$ and cosmological (or inner \cite{Kruthoff:2024gxc}) horizons at $-\Phi_h+2\pi m$. Indeed $\Phi\to \Phi+2\pi a m$ generates new solutions of the periodic dilaton gravity equations of motion. Based on our experience with summing over solutions in sine dilaton gravity \eqref{3.51} and the associated surprising one-loop exactness of the density of states \eqref{Z}, we will conjecture the following exact result for the disk partition function of periodic dilaton gravity
\begin{equation}
    \text{conjecture:}\qquad Z_\text{exact}(\beta)=\int_0^{\pi a}\d \Phi_h\,V(\Phi_h)\sum_{n=-\infty}^{+\infty}(-1)^n e^{S(\Phi_h+\pi a n)/\hbar}\,e^{-\beta W(\Phi_h)/\hbar}\,,
\end{equation}
with $S(\Phi_h+\pi a n)$ the suitable analytic continuation of \eqref{6.28}. For the cosmological horizons, we used the fact that $S(\Phi_h+\pi a)=S(-\Phi_h)$. Globally, the entropy decreases in the same way with $n$ as for sine dilaton gravity. For instance $S(2\pi a m)=4\pi^2 a (m-2m^2)$. So the sum converges and $n=0$ is dominant. Whereas we have no proof, it is reassuring that, at least for classes of dilaton potentials, such equations are indeed exact (see also section \ref{sec:gasofdef}).
\section{Concluding remarks}\label{sect:discnew}
We have demonstrated how canonically quantizing sine dilaton gravity, and more generally any dilaton gravity with a periodic potential reveals a momentum shift symmetry of the Hamiltonian. The existence of a shift symmetry is tied with the fact that the classical spacetimes have a periodic radial coordinate and both a black hole and a cosmological horizon. Gauging this symmetry leads to a large set of null states and discretizes the bulk geometry. The symmetry encodes the fact that the energy is bounded both from above and below. This restricts small-scale observations, thus quantizing the bulk geometry. We expect these to be generic features of UV complete systems.

We end this paper with several concluding remarks.

\subsection{Gauging the path integral?}
Consider again the action of sine dilaton gravity
\begin{equation}
    \frac{1}{2 \abs{\log q}}\left\{\frac12\int \mathrm{d} x \sqrt{g}\,\big(\Phi R+2\sin(\Phi)\big)+\int \mathrm{d}\tau \sqrt{h}\,\big(\Phi K-\mathrm{i} \,e^{-\mathrm{i} \Phi/2}\big)\right\}.
\end{equation}
Naively if one integrates over $\Phi$ this theory seems ill-defined. Indeed we can shift globally $\Phi\to \Phi+2\pi m$ leaving the potential unchanged, however this generates a divergent sum
\begin{equation}
    \sum_{m=-\infty}^{+\infty}\exp\bigg(\frac{\pi m}{2\abs{\log q}}\int\d x\sqrt{g}R+\text{boundary}\bigg)
\end{equation}
This is the same divergence that we see in the density of states \eqref{2.22}. In that case the divergence arises because of shifts $\Phi_h\to \Phi_h+2\pi m$. On the level of the path integral it is not immediately obvious how to cure the divergence, except if the Euler character vanishes. However, we found that in the canonical formulation of the theory there is a way to cure the divergence by gauging a symmetry $P\to P+2\pi$. So at the level of the quantum mechanics we have understood how to properly deal with periodic dilaton gravity. 

In a companion paper \cite{wip} we show that similar gaugings have the desired effect on wormholes and on EOW brane amplitudes. In the closed channel quantization we in fact gauge exactly $\Phi_h\to \Phi_h+2\pi m$. This suggests that we are gauging the zero mode of $\Phi$ in the path integral. Such identification in the path integral are unfortunately notoriously subtle \cite{Kapec:2020xaj,Verlinde:1986kw}.\footnote{We thank Herman Verlinde and Nathan Seiberg for discussions on this point.} It would be very interesting to understand in detail what is the precise path integral formulation that implements our Hamiltonian gaugings (and those of \cite{wip}), and the associated insertions of networks of line projection operators \eqref{3.24}. One natural way to expect to make progress on this is to understand the amplitudes of the gauged theory on higher genus surfaces, the resulting structure may then hint at what the precise path integral definition is. A consistent category based on a discrete spectrum of Virasoro states would be very useful.

It is not impossible that the correct definition is just to gauge the zero mode of the dilaton. Indeed, the selection rules on closed channel primaries \cite{wip} prohibit the naive no-boundary state itself, instead ``decomposing'' the disk into cylinder topologies \eqref{7.12} with $\chi=0$, which are fine.\footnote{We thank Edward Witten for pointing this out.} Higher topologies may also not be so dramatic. Perhaps one could trick the gauged path integral into counting light-cone diagrams, in the same sense that the Lorentzian AdS$_2$ JT path integral can count them despite naively not allowing topology change (because of localization to smooth spacetimes when integrating out $\Phi$), see \cite{Marolf:2022ybi,Blommaert:2023vbz,Usatyuk:2022afj}. We leave this for future investigations.

\subsection{Relation to other work}\label{subsect:2.3}
Sine dilaton gravity is equal to two copies of Liouville CFT with complex conjugate central charges \cite{Blommaert:2024ydx}. Our quantization presented in Section \ref{sect:ungauge} is based on that Liouville formulation (since we imposed the $b\to 1/b$ symmetry) and can be viewed (as we explained) as an analytic continuation from the case with real central charges. 

There is a different proposal \cite{Collier:2024kmo} for quantizing sine dilaton gravity via the Liouville formulation, dubbed the ``complex Liouville string''. It would be interesting to understand in more detail how these two approaches are related. The density of states in their quantization scheme is equation (11) in \cite{Collier:2024kmo}
\begin{equation}
\rho_\text{CLS}(E) =\sin(-\i b^2 \text{arccosh} (E/2))\,,\quad 2<E<+\infty\,.
\end{equation}
To better compare with our story, we set $b \to 1/b$ and introduce the parameterization of the spectrum $2<E<\infty$
\begin{equation}
    E=2\cosh(x)\,.
\end{equation}
The above density of states then leads to the canonical partition function
\begin{equation}
    Z_\text{CLS}(\beta)=\int_{-\infty}^{+\infty}\d x\,\sinh(x) \sin\bigg(\frac{\pi x}{\abs{\log q}} \bigg)\,e^{-\beta \cosh(x)/2\abs{\log q}}\,.\label{2.25}
\end{equation}
This is essentially our equation \eqref{eq:fakepf} upon $\theta\to \i x$, but with real $x$. Along the lines of section 2.1 in \cite{Blommaert:2024ydx}, one can understand that this is obtained from the sine dilaton gravity semiclassical path integral by counting contributions from spacetimes \eqref{1.9} with complex values of the dilaton at the horizon
\begin{equation}
    \Phi_h=\pm \i x\,.
\end{equation}
This leads indeed to the sine density in \eqref{2.25}, which reflects the imaginary entropy of these solutions $S_\text{BH}=\pm \i \pi x/\abs{\log q}$. 

For the moment, we can interpret \eqref{2.25} as a second way to cure the divergence in the naive theory \eqref{eq:fakepf}. Indeed, along the real $x$ contour, the integral \eqref{2.25} is convergent. 
It would be interesting to see if there is a reasonable set of gravitational wavefunctions (eigenfunctions of $\mathbf{H}$ in \eqref{Hgrav}) that directly lead to the spectrum \eqref{2.25}.\footnote{Perhaps this involves choosing a different metric contour (holographic screen) such that $L$ and $P$ have complex solutions.} In our setup, it would be interesting to understand if there is an analogue to the Liouville equation (2.63) in \cite{Collier:2024wsboundaries} for the theory of section \ref{sect:gauged}. It hence seems plausible that \cite{Collier:2024kmo} and our gauged theory of section \ref{sect:gauged} correspond to quantizing different sectors of the classical phase space of sine dilaton gravity, analogous to different sectors of the classical phase space of dS JT gravity \cite{Held:2024rmg, AloHarJeff24}.

\subsection{Dilaton gravity and gas of defects}
\label{sec:gasofdef}
In this section we provide some evidence that equation \eqref{eq:Zunexact} for the partition function of (ungauged) dilaton gravity 
\begin{equation}
    Z_\text{ungauged}(\beta)=\int_{-\infty}^{+\infty}\d \Phi_h V(\Phi_h)\sinh(2\pi \Phi_h/\hbar)\,e^{-\beta W(\Phi_h)/\hbar}\,,\label{7.7}
\end{equation}
which is motivated semiclassically, may be the exact answer. We leave a more detailed investigation to future work. We start from the equations that follow from the gas-of-defects expansion \cite{Maxfield:2020ale,Witten:2020wvy}, formulated explicitly in formula (2.15) of \cite{Kruthoff:2024gxc}\footnote{We assume for the sake of the present argument that $E_0=0$. The integration contour $\mathcal{C}$ is typically taken on the right of all non-analyticities of the integrand.}
\begin{align}\label{gas}
\rho(E)=\frac{1}{2 \pi G_\text{N}}\int_{\mathcal{C}} \frac{\mathrm{d}\Phi}{2 \pi \i}  \ e^{2\pi \Phi/G_\text{N}} \mathrm{arctanh}(\sqrt{E/W(\Phi)})=\frac{1}{8 \pi^2}\int_{\mathcal{C}} \frac{\mathrm{d}\Phi}{2 \pi \i}  \ e^{2\pi \Phi/G_\text{N}} \frac{W'(\Phi)}{W(\Phi)-E} \frac{\sqrt{E}}{\sqrt{W(\Phi)}}\,.
\end{align}
In the second step we integrated by parts. Introducing the energy variable $M=W(\Phi)$, this becomes
\begin{equation}\label{W(x)}
\rho(E)=\frac{1}{8 \pi^2}\int_{\mathcal{C}} \frac{\mathrm{d} M}{2 \pi \i}  \ e^{2\pi W^{-1}(M)/G_\text{N}} \frac{1}{M-E} \frac{\sqrt{E}}{\sqrt{M}}\,.
\end{equation}
The residue at $M=E$ contributes the usual semiclassical approximation
\begin{equation}
\rho(E)\supset\frac{1}{8 \pi^2} e^{2\pi W^{-1}(E)/G_\text{N}}\,. 
\end{equation}
However, it is important to recall that the change of variables $\Phi=W^{-1}(M)$ is in general non invertible. So in \eqref{W(x)} one should sum over the contributions coming from all regions where the function is invertible. In the cases of JT or Liouville gravity, for instance, $\Phi=\pm \sqrt{M}$ respectively $\Phi=\pm \mathrm{arccosh} (M)$. Adding the two branches and including the minus sign from the measure, leads to the exact spectral densities for JT \cite{Stanford:2017thb} respectively Liouville gravity \cite{Mertens:2020hbs}. In sine dilaton gravity we have $\Phi=\pm \mathrm{arccos}(M)+2\pi n$, which leads indeed to the ungauged spectral density \eqref{2.22}. For general dilaton gravity this leads to
\begin{equation}
\rho(E) = \sum_{W(E_i)=W(E)} (-1)^{\text{sign}(W'(W^{-1}(E_i)))}\frac{1}{8 \pi^2} e^{2\pi W^{-1}(E_i)/G_\text{N}}\,. \label{7.11}
\end{equation}
We sum over all images $E_\text{i}$ such that $W(E)=W(E_i)$. This expression could be negative or divergent for physical values of $E$. If $W(\Phi)$ is a continuous function with $W(\Phi\to +\infty) \to +\infty$, $\rho(E)$ is automatically positive everywhere. Divergences can occur if there are an infinite number of images $E_i$. This happens for all periodic pre-potentials $W(\Phi)$, which therefore have to be gauged (as discussed in section \ref{subsect:6.2entropy}).

Equation \eqref{7.11} matches with \eqref{7.7}. The reason that we do not declare victory is that it is not quite clear what is the correct contour $\mathcal{C}$ in \eqref{gas} \cite{Kruthoff:2024gxc}. There may be contributions from the $\sqrt{M}$ branchcut in the integrand of \ref{W(x)}, picking up the discontinuity for $M<0$. For certain classes of dilaton potentials (including JT, sinh and (ungauged) sine dilaton gravity) this contribution vanishes. But, in general it does not, and this could lead to corrections to \eqref{7.11}.

\subsection{dS JT gravity from gauged sine dilaton gravity}\label{subsect:7.3dsjt}
It is well understood that sine dilaton gravity \eqref{1.4sdaction} reduces to JT gravity upon scaling $\Phi=2\abs{\log q} \Phi_\text{AdS}$ with $\abs{\log q}\to 0$ whilst keeping $\Phi_\text{AdS}$ finite. In section \ref{sect:flathermite} we expanded around the maximum of the sine dilaton gravity spectrum \eqref{3.34}. Scaling $\Phi=\pi/2+\sqrt{\abs{\log q}} \Phi_\text{flat}$ we obtained a quantum theory of flat space JT gravity \eqref{Sflat}. There is a third limit where one recovers dS JT gravity, by zooming in on the upper edge of the spectrum $\Phi_h=\pi$. Indeed, scaling
\begin{equation}
    \Phi=\pi-2\abs{\log q}\Phi_\text{dS}\,,\quad \abs{\log q}\to 0\,,
\end{equation}
the sine dilaton gravity action reduces to
\begin{equation}
    \boxed{I\to -\frac{1}{2}\int \d x\sqrt{g}\Phi_\text{dS}(R-2)}
\end{equation}
The sine dilaton gravity metric becomes\footnote{This is indeed the limit of the real $\rho$ contour in \eqref{1.11}.}
\begin{equation}
    \d s^2=F_\text{dS}(r)\d \tau^2+\frac{1}{F_\text{dS}(r)}\d r^2\,,\quad F_\text{dS}(r)=r_h^2-r^2\,,\quad r_h<r<+\infty\,.
\end{equation}
Along this contour the metric is $-$AdS$_2$, indeed the conformal transformation \eqref{1.11} becomes simply
\begin{equation}
    \d s^2_\text{AdS}=-\d s^2\,.
\end{equation}
This complex geometry has been considered before as a way to quantize dS JT gravity \cite{Maldacena:2019cbz,Cotler:2019nbi}. Upon defining $r=r_h\cosh(\rho)$ and shifting $\rho\to \i \pi/2+T$ this metric becomes ordinary expanding dS space \cite{Maldacena:2019cbz}. In higher dimensions, there are analogs of such a complex geometry, which amount to considering de Sitter space using the Maldacena contour \cite{Castro:2012gc}. We see here that our method of canonical quantization naturally forces us to consider de Sitter space on such a contour. 

A (to us) surprising observation is that the gauged partition function of sine dilaton gravity \eqref{eq:truepf} reduces in this regime to\footnote{Technically this is divergent, however as we will explain it seems more natural now to consider complex values of $\beta$.}
\begin{equation}
    \boxed{Z_\text{gauged}(\beta)\to \int_0^\infty \d r_h\,r_h\sinh(2\pi r_h)\,e^{\beta r_h^2}=Z_\text{dS}(\ell=-\i \beta)}\label{7.6}
\end{equation}
The boundary conditions \eqref{2.3 bc} reduce to $\sqrt{h}=\i\infty$. The $\i$ appears because we go to the real asymptotic boundary of $\d s^2_\text{AdS}$. Stepping away from DSSYK, one could consider more general boundary locations and in this setup it seems more natural to consider $\sqrt{h}=\infty$ (as in Figure 3 of \cite{Maldacena:2019cbz}). Replacing $\beta\to \i \ell$ we see that \eqref{7.6} reproduces the wavefunction of the universe in dS JT gravity, equation (2.13) in \cite{Maldacena:2019cbz}. We find this surprising because naively we would have expected a relation between the \emph{ungauged} partition function and dS JT gravity. Indeed, as explained around \eqref{fake}, it is the ungauged partition function that corresponds with ``ordinary'' black hole thermodynamics. We have
\begin{equation}
    Z_\text{ungauged}(\beta)\to \int_0^\infty \d r_h\,r_h e^{-2\pi r_h}\,e^{\beta r_h^2}\,.
\end{equation}
Indeed, the horizon area $\Phi_h=\pi-2\abs{\log q} r_h$ decreases with $r_h$, so that entropy decreases with $r_h$. This is very different from \eqref{7.6}. Nevertheless, as we have just explained, \eqref{7.6} is a known answer in dS JT gravity \cite{Maldacena:2019cbz}. This observation can maybe be used to help understand how to find a semiclassical (path integral) interpretation for the full entropy profile \eqref{3.34} of the gauged theory. It would be interesting to analyze this connection between sine dilaton gravity and dS space on the Maldacena contour further. Let us make three more comments regarding this limit.

\begin{enumerate}
    \item Just like the limits where one zooms in on the lower edge of the spectrum and near the quadratic maximum, this limit is universal in the 2d periodic dilaton gravity models discussed in section \ref{sect:genericpotential}.
    \item Sine dilaton gravity regularizes the UV divergence of dS JT gravity in the sense of equation \eqref{3.34bb} 
    \begin{equation}
        Z_\text{dS}(\ell\to 0)=\dim \mathcal{H}_\text{SYK}\,.
    \end{equation}
    The UV divergence $\ell\to 0$ in dS JT gravity made the wavefunction of the universe non-normalizable \cite{Maldacena:2019cbz}.
    \item It is interesting that in the dS JT case the Schwarzian corrections are responsible for making the density of states go to zero. It is not obvious to us how this is related with gauging $P\to P+2\pi$ and the associated null states.
\end{enumerate}
We summarize the three limits in the following picture of the periodic dilaton gravity spectrum:
\begin{equation}
    \begin{tikzpicture}[baseline={([yshift=-.5ex]current bounding box.center)}, scale=0.7]
 \pgftext{\includegraphics[scale=1]{gauge10.pdf}} at (0,0);
    \draw (3.1,-1.8) node {$\Phi_h$};
    \draw (1.8,-1.8) node {$\pi a$};
    \draw (-0,-1.8) node {$\pi a/2$};
    \draw (-1.75,-1.8) node {$0$};
    \draw (1.15,-0.1) node {$S$};
    \draw (0.1,2.7) node {\color{blue}flat space};
    \draw (0.1,2) node {\color{blue}section \ref{subsect:4.1}};
    \draw (4.4,-0.4) node {\color{red}dS space};
    \draw (4.4,-1.1) node {\color{red}section \ref{subsect:7.3dsjt}};
    \draw (-4.5,-1.1) node {\color{green}AdS space};
    \draw (-4.5,-1.8) node {\color{green}known};
  \end{tikzpicture}
\end{equation}
Periodic dilaton gravity theories are toy models for quantum gravity in {\color{green}AdS}, {\color{blue}flat space} and in {\color{red}dS space}.
\subsection{More comments on quantum cosmology}\label{sect:7.5}
We can also consider a cosmological interpretation of sine dilaton gravity at all energies. More details of this will be discussed elsewhere. Consider the sine dilaton gravity metrics \eqref{1.9} on the contour $r=\i x$\footnote{We thank Victor Gorbenko for discussions on this point.}
\begin{equation}
   \begin{tikzpicture}[baseline={([yshift=-.5ex]current bounding box.center)}, scale=0.7]
 \pgftext{\includegraphics[scale=1]{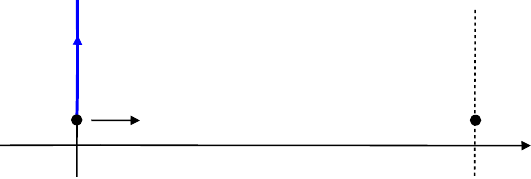}} at (0,0);
    \draw (-0.9,-0.5) node {big bang};
    \draw (4.45,-1.8) node {$r$};
    \draw (3.55,-1.8) node {$2\pi$};
    \draw (-3.2,-1.8) node {$0$};
    \draw (-3.05,1.9) node {\color{blue}$t=\infty$};
  \end{tikzpicture}
\end{equation}
Minus the effective AdS metric along these slices are Lorentzian big-bang cosmologies
\begin{equation}
    -\d s^2_\text{AdS}=-\d t^2+B^2\sinh(t)^2\d x^2\,,\quad x\sim x+1\quad \begin{tikzpicture}[baseline={([yshift=-.5ex]current bounding box.center)}, scale=0.7]
 \pgftext{\includegraphics[scale=1]{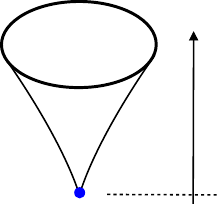}} at (0,0);
    \draw (2.2,-1.6) node {$0$};
    \draw (2.2,1.1) node {$t$};
    \draw (-1,-1.6) node {\color{blue}$B$};
  \end{tikzpicture}\label{7.12}
\end{equation}
The physical metric \eqref{1.9} itself along this contour is also an expanding real-time metric, which is more clear upon analytically continuing $\tau=\i y$ and periodically identifying in $y$. However, it is expanding in an accelerated way and becomes infinitely large after only a finite proper time. It would be interesting to study these types of contours in the context of \cite{Fumagalli:2024msi}.

In \cite{wip} we study canonical quantization of sine dilaton gravity on closed Cauchy slices.\footnote{After our analysis was completed a couple of very interesting papers appeared \cite{Alonso-Monsalve:2024oii,Held:2024rmg} which perform a similar analysis for dS JT gravity.} The dilaton zero mode $\Phi$ on this slice should be gauged $\Phi\to \Phi+2\pi$ which in turn discretizes $B=2\abs{\log q} b$, with $b$ a positive integer. The naive smooth no-boundary state $B=2\pi \i$ is projected out upon gauging. Instead, the wavefunction of the universe $\psi_\text{disk}(\ell)$ has support only on the positive integer $b$ and is normalizable, because the partition functions of periodic dilaton gravity reduces to a finite number for $\beta\to 0$ \eqref{3.34bb}.

\section*{Acknowledgments}
We thank Andreas Belaey, Leonardo Bossi, Scott Collier, Lorenz Eberhardt, Victor Gorbenko, Luca Griguolo, Oliver Janssen, Shota Komatsu, Ohad Mamroud, Lorenzo Russo, Nathan Seiberg, Domenico Seminara, Thomas Tappeiner, Herman Verlinde and Edward Witten for discussions and insights. AB was supported by ERC-COG NP-QFT No. 864583, by INFN Iniziativa Specifica GAST and by the Marvin L. Goldberger membership at IAS. AL was supported by the Heising-Simons foundation under grant no. 2023-4430, the Packard Foundation and the Department of Energy under Early Career Award DE-SC0021886. KP was supported by the US DOE DE-SC011941. TM and JP acknowledge financial support from the European
Research Council (grant BHHQG-101040024). Funded by the European Union. Views and opinions
expressed are however those of the author(s) only and do not necessarily reflect those of the European
Union or the European Research Council. Neither the European Union nor the granting authority can
be held responsible for them.

\appendix

\section{Wavefunction asymptotics}
\label{app:wf}
We discuss a simple way to extract the exact spectral density of a continuous quantum mechanics by computing the assymptotics of the gravitational wavefunctions $\psi_E(L)$ for $L\to \pm \infty$.

For $L \to +\infty$, the asymptotics are determined by the residues of the poles closest to the integration contour in the Fourier transformed wavefunction $\psi_E(p)$. This is WKB. Contributions from other poles decay exponentially in $L$, and do not contribute to the prefactor of $\delta(E_1 - E_2)$ in the inner product of wavefunctions, which comes from a volume divergence in the $L$ integral. For $L \to -\infty$, the $p$ contour is deformed into the upper half-plane where one finds a saddle-point.

To clarify these points, consider first the JT gravity  wavefunctions ($E = s^2$) with $L_\text{JT}=2\phi$ \cite{Harlow:2018tqv,Blommaert:2018oro}
\begin{equation}\label{JT}
    \psil_s(\phi) = \psir_s(\phi) = K_{2\i s}\big(2 e^{-\phi}\big) = \frac{1}{4\pi} \int_{-\infty+\i\epsilon}^{+\infty+\i\epsilon} \d k \, e^{-2\i\phi k} \, \Gamma(-\i k \pm \i s)\,.
\end{equation}
The poles of \eqref{JT} of the momentum integral are located at $k = \pm s - \i n$ with $n\geq 0$. By deforming the integration contour towards $k = -\i\infty$, one obtains the sum representation of the Bessel function. The contributions from poles with $n \neq 0$ decay as $e^{-2\phi n}$ and thus vanish in the asymptotic regime $\phi \to +\infty$. Collecting the residues from the poles at $k = \pm s$ gives the WKB approximation
\begin{equation}
    \psil_s(\phi) \to \frac{1}{2} \Gamma(-2\i s) e^{-2\i s \phi} + \frac{1}{2} \Gamma(2\i s) e^{2\i s \phi} = \frac{1}{\rho(s)^{1/2}} \frac{2}{\pi^{1/2}} \cos\big(2s(\phi - \phi_\text{min}(s))\big)\,, \quad \phi > \phi_\text{min}(s)\,.
\end{equation}
The phase of the residues determines the minimal classical size $\phi_\text{min}(s)$ of the ERB.
For the $\phi \to -\infty$ regime, we use Stirling
\begin{equation}
    \psi_s(k)=\Gamma(-\i k \pm \i s) \to 2\pi \frac{\i}{k} \exp\big(-\pi k - 2\i k \log k + 2\i k\big)\,.
\end{equation}
The $k$ integral then has a saddlepoint at $k_* = \i e^{-\phi}$, which indeed becomes large for $\phi\to-\infty$. Including the one-loop factor one recover the known assymptotics of the Bessel function
\begin{equation}
    \psil_s(\phi) = \frac{\pi^{1/2}}{4} \exp\left(\frac{\phi}{2} - 2e^{-\phi}\right)\,, \quad \phi \to -\infty\,,
\end{equation}

For ungauged sine dilaton gravity (section \ref{sect:ungauge}) we can derive the assymptotics of the left \eqref{left} and right \eqref{right} in a similar manner. Recall the locations \eqref{polesleftwf} of the left wavefunction. Introduce $s=\theta/\abs{\log q}$. In the WKB regime $\phi\rightarrow \infty$ we pick up the poles at $p=\pm s$ in the integral representations \eqref{left} and \eqref{right} for $\psil_\theta(\phi)$ and $\psir_\theta(\phi)$, resulting in\footnote{The poles $k=\pm s-2\pi m/\abs{\log q}$ naively give contributions of the form
\begin{equation}
    \delta(\theta_1=\theta_2\pm 2\pi (m_1+m_2))\,,\quad m_1+m_2>0\,.
\end{equation}
From the exact calculation of the inner product \eqref{m_element} we might also naively expect delta functions at these locations. But a more careful analysis there shows that no delta function appears due to an additional overall zero from the double sine functions. A more careful analysis of the residues here should lead to the same conclusion.}
\begin{align}
    \pi b\,\psil_{s_1}(\phi)=S_b(-\i b s_1)e^{-\i s_1(\phi-\i\pi/2)}+S_b(\i b s_1)e^{\i s_1(\phi-\i\pi/2)}+\sum_{m=1}\dots\,,\quad \phi\to+\infty\nonumber\\ \pi b\,\psir_{s_2}(\phi)=S_b(-\i b s_2)e^{-\i s_2(\phi-\i\pi/2)}+S_b(\i b s_2)e^{\i s_2(\phi-\i\pi/2)}+\sum_{m=1}\dots\,,\quad \phi\to+\infty\label{7.37ass}
\end{align}
If we compute the normalization of these states by integrating over $\phi$ (as in \eqref{spectral}) one indeed recovers the spectral density of (ungauged) sine dilaton gravity \eqref{density}. For $\phi\to -\infty$ one uses the Stirling-type approximation of the double sine function \cite{kashaev2001quantum,ip2015tensor}\footnote{For general complex $b$, there are Stokes phenomena in the assymptotics \cite{kashaev2001quantum}. However, one can check that, upon analytic continuation \eqref{bb}, the saddle \eqref{7.41kstar} remains within the region where \eqref{7.39sbass} holds, even though the saddle approaches the transition line between two asymptotic expansions.}
\begin{equation}
    S_b(x)\to \exp\bigg(-\text{sgn}(\text{Im}(x))\frac{\i \pi}{2}\bigg(x-\frac{Q}{2}\bigg)^2  \bigg).\label{7.39sbass}
\end{equation}
This results in
\begin{equation}
    \psil_s(\phi)\to \frac{1}{2\pi}\int_{-\infty}^{+\infty}\d k \exp\bigg(-\i k (\phi+\i \pi b^2/2)+\frac{1}{2}\i \pi b^2 k^2\bigg)\,.
\end{equation}
This integral has a saddle at
\begin{equation}
    \pi b^2k_*=\phi+\i \pi b^2/2\,.\label{7.41kstar}
\end{equation}
We can repeat the same analysis for $\psir_s(\phi)$. By including the integral over quadratic fluctuations around the saddle and upon analytic continuation \eqref{bb},  one finds
\begin{align}
    \psil_s(\phi)&=\left(\frac{\abs{\log q}}{2 \pi}\right)^{1/2}\exp\bigg(-\frac{1}{2 \abs{\log q}}\left(\phi-\frac{\abs{\log q}}{2}\right)^2\bigg )\,,\quad \phi\to -\infty \nonumber\\
    \psir_s(\phi)&=\i \left(\frac{\abs{\log q}}{2 \pi}\right)^{1/2}\exp\bigg(\frac{1}{2 \abs{\log q}}\left(\phi-\frac{\abs{\log q}}{2}\right)^2\bigg )\,,\quad \phi\to -\infty \label{7.42normapp}
\end{align}
We checked the asymptotics \eqref{7.42normapp} numerically in Figure \ref{fig:psil} and \ref{fig:psirpoles}. Up to a prefactor that only depends on $q$ (which we ignore throughout), this proves equation \eqref{7.42norm} in the main text.

One can use this technique also to quickly check the DSSYK density of states  \eqref{standard}, starting from from the q-Hermite representation \eqref{int_rep} for $n\to +\infty$
\begin{equation}
    H_n(\cos(\theta)|q^2) = (q^2;q^2)_n\int_{0}^{2\pi}\frac{\d p}{2\pi}\,\frac{e^{-\i pn}}{(e^{\i(p\pm \theta)};q^2)_\infty}\to \frac{1}{(e^{2\i\theta};q^2)_\infty}e^{-\i\theta n}+\frac{1}{(e^{-2\i\theta};q^2)_\infty}e^{\i\theta n}\,.
\end{equation}
\section{q-Schwarzian contours}\label{app:contours}
In this appendix we propose a description of subleading q-Schwarzian saddles in \eqref{3.42} and \eqref{Z} in terms of winding contours of the length quantum mechanics.

For the case $n=0$, the entropy was computed in \cite{Blommaert:2024ydx} using a contour integral of the symplectic term in the action with $z=e^{\i \sin(\theta)\tau}$
\begin{equation}\label{eqn:gthetaintegral}
S=\frac{1}{\abs{\log q}}\int_{1}^{e^{-4\i \theta}} \d z\, g_{\theta}(z)=-\frac{\left(\theta-\pi/2\right)^2}{\abs{\log q}}\,,\quad g_{\theta}(z) = \left( \frac{1}{2z} - \frac{1}{z-e^{-2i\theta}}\right) \log \left(e^{i\theta} \frac{z-e^{-4i\theta}}{z-e^{-2i\theta}} \right)\,.
\end{equation} 
Here the contour winds counter-clockwise along the circle at $|z| =1$ from $1$ to $e^{-4i\theta}$. We will refer to this original contour, which gives the leading thermodynamics, as $\mathcal{C}_0$. We would like to understand what a generalization of this contour looks like for saddles associated to shifted images $\theta \to \theta+\pi n$ in the formula \eqref{Z}. We will refer to such a generalized contour as $\mathcal{C}_n$.

Before doing this, we note that the integrand in \eqref{eqn:gthetaintegral} has multiple non-analyticities. It has poles at $z = 0$, $e^{-2i\theta}$ and $\infty$ as well as a branch cut running from $e^{-2i\theta}$ to $e^{-4i\theta}$. The residue around the pole at the origin is just 
\begin{align}\label{eqn:gamma0res}
\oint_{\gamma_0} \d z\, g_{\theta}(z) = \pi \theta,
\end{align}
where we take $\gamma_0$ to circle counter-clockwise around the origin. Similarly, the residue around the pole at infinity is also
\begin{align}\label{eqn:gammainfres}
\oint_{\gamma_{\infty}}\d z\,  g_{\theta}(z) = \pi \theta.
\end{align}
\begin{figure}
 \includegraphics[width=\textwidth]{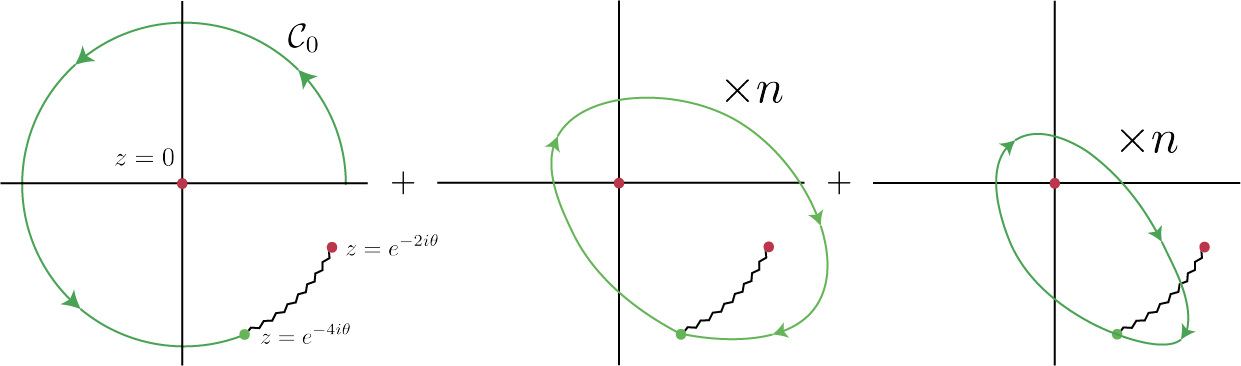}
    \caption{We illustrate the contour $\mathcal{C}_n$ which we associate to the sub-leading terms in \eqref{Z}. The first piece of $\mathcal{C}_n$ on the left is just $\mathcal{C}_0$, which is the contour for the leading saddle, first discussed in \cite{Blommaert:2024ydx} and described below \eqref{eqn:gthetaintegral}. In the middle of the figure, we illustrate the second piece of the contour, which winds $n$-times around both the pole $z=0$ and the branch cut. Finally, the contour winds $n$-times around the pole at zero, crossing the branch cut each time. When it crosses the branch cut, we have in mind that it goes to the next sheet of the logarithm in \eqref{eqn:gthetaintegral}. Due to the branch cut, this final contour is not just $n$ times the residue of the pole at the origin.}\label{fig:windingcontour}
\end{figure}

We now detail our proposal for the contour associated to the solutions with $\theta_n = \theta + \pi n$. Note that to describe a solution with the correct boundary conditions, we must preserve the condition that the contour starts at $L =0$ and ends at $L =0$. We therefore need that 
\begin{align}\label{eqn:Lzerocondition}
\int_{\mathcal{C}_n} \d\tau \, \phi'(\tau) = \int_{\mathcal{C}_n} \d z \left( \frac{1}{2z} - \frac{1}{z-e^{-2i\theta}}\right) = 0.
\end{align}
It is clear then that for every time $\mathcal{C}_n$ circles the pole at $z = e^{-2i\theta}$ then it must circle the pole at $ z= 0 $ \emph{twice}, due to the relative factor of $2$ between the two terms in \eqref{eqn:Lzerocondition}. By studying what happens to the defining contour in \eqref{eqn:gthetaintegral} as we continue $\theta \to \theta+\pi n$, one can see that the correct contour, $\mathcal{C}_n$, is just to add the original contour, $\mathcal{C}_0$, to a contour which circles the pole and branch cut a total of $n$ times in the clockwise direction. Then it circles $n$-times the pole at $z =0$. After each encircling of $z = 0$, it crosses the cut and goes to the next sheet. On each next sheet, the log just gets shifted by $-2\pi i k$. Note that this contour has the desired property that it circles the pole at $z = 0$ twice as often as the pole at $z =e^{-2i\theta}$ and so preserves the desired boundary conditions. We illustrate the contour in Fig. \ref{fig:windingcontour}.

Using the formulae \eqref{eqn:gamma0res} and \eqref{eqn:gammainfres} for the residues of the poles and the fact that on each consecutive sheet the logarithm gets shifted by $-2\pi i$, we get that the entropy for our proposed contour, $\mathcal{C}_n$, is just
\begin{align}
& \int_{\mathcal{C}_n} \d z\,g_{\theta}(z) = \left(\int_{\mathcal{C}_0} \d z\,g_{\theta}(z)\right)+ \left( - n \pi \theta\right) + \left(-n\pi \theta - \sum_{k=0}^{n-1} 2\pi^2 k \right) \nonumber \\
& =-(\theta-\pi/2)^2 -2n \pi \theta - \pi^2 n(n-1) = -(\theta + \pi (n-1/2))^2,
\end{align}
where each term in the parenthesis in the first line corresponds to each piece of $\mathcal{C}_n$ in Fig. \ref{fig:windingcontour}. Up to the requisite factor of $\abs{\log q}$, this is the correct answer for the entropy of the sub-leading terms in \eqref{Z}.

\section{More technical details}
We gather several technical details that are required to check results mentioned in the main text.
\subsection{Expressions for q-Hermite polynomials}
 The generating function of q-Hermite polynomials is
\begin{equation}
\sum_{n=0}^{+\infty} \frac{\rH_n(\cos(\theta)| q^2)}{(q^2;q^2)_n}t^{n}=\frac{1}{(te^{\pm \i\theta};q^2)_\infty}\,.
\end{equation}
Setting $t=e^{\i p}$ and inverting this expression results in the integral representation
\begin{equation}\label{int_rep}
\frac{\rH_n(\cos(\theta)|q^2)}{(q^2;q^2)_n}= \int_{0}^{2\pi} \mathrm{d}p \, \frac{e^{-\i p n}}{(e^{\i (p\pm \theta)};q^2)_\infty}\,.
\end{equation}
The orthogonality relation of q-Hermite polynomials reads, for real valued $\theta$, reads
\begin{equation}\label{standard}
    \sum_{n=0}^{+\infty}\frac{1}{(q^2; q^2)_n} \rH_n(\cos(\theta_1) | q^2)\, \rH_n(\cos(\theta_2) | q^2)
    = \abs{\sin(\theta_1)}\frac{\delta(\cos(\theta_1)-\cos(\theta_2))}{(e^{\pm 2\i\theta};q^2)_\infty}\,.
\end{equation}
\subsection{Orthogonality calculation}
In this appendix, we compute the orthogonality relation \label{app:ortho} of the ungauged left \eqref{left} and right \eqref{right} sine dilaton gravity wavefunctions, which we repeat for convenience
\begin{align}
    \psil_\theta(\phi)&=\frac{1}{2\pi}\int^{+\infty+\i \epsilon}_{-\infty+\i \epsilon}\d p\, e^{-\i p \phi}\,S_b\bigg(-\i \frac{b p}{2}\pm\i \frac{b}{2}\frac{\theta}{\abs{\log q}}\bigg)\exp\bigg(-\abs{\log q}\frac{p^2}{4}+\frac{\theta^2}{4\abs{\log q}}-\frac{\pi p}{2}\bigg)\nonumber\\
    \psir_\theta(\phi)&=\frac{1}{2\pi}\int_{\gamma}\d p\, e^{\i p \phi}\,S_b\bigg(\i \frac{b p}{2}\pm\i \frac{b}{2}\frac{\theta}{\abs{\log q}}\bigg)\exp\bigg(\abs{\log q}\frac{p^2}{4}-\frac{\theta^2}{4\abs{\log q}}+\frac{\pi p}{2}\bigg)\,.
\end{align}
We do this by computing the following limit \eqref{spectral}
\begin{equation}
\left\langle \theta_1 | \theta_2 \right\rangle = \lim_{\Delta \rightarrow 0} \bra{\theta_1}e^{-2\Delta \phi}\ket{\theta_2} = \lim_{\Delta \rightarrow 0} \int_{-\infty}^{+\infty} \mathrm{d}\phi \, e^{-2 \Delta \phi} \, \psil_{\theta_1}(\phi) \psir_{\theta_2}(\phi)\,.
\end{equation}
In the Fourier domain, the $\phi$-integral simply gives a delta-function $\delta \left(2 \i \Delta-p+k\right)$. This leaves a single momentum integral
\begin{equation}
\begin{split}
e^{\frac{1}{4 \abs{\log q}}\left(\theta_1^2-\theta_2^2\right)} (-1)^{\Delta} q^{\Delta^2} \int_{-\infty}^{+\infty} \frac{\mathrm{d}p}{2 \pi}  \ e^{-\i \Delta p \abs{\log q}} \ S_b\bigg(-\i \frac{b p}{2}\pm\i \frac{b}{2}\frac{\theta_1}{\abs{\log q}}\bigg)S_b\bigg(\i \frac{b p}{2}+b \Delta \pm\i \frac{b}{2}\frac{\theta_2}{\abs{\log q}}\bigg)\,,
\end{split}
\end{equation}
which can be evaluated exactly using the q-deformed Barnes lemma
\begin{align}
&\int_{-\infty}^{+\infty} \mathrm{d}\tau e^{\pi \tau \left(\alpha+\beta+\gamma+\delta\right)} S_{b}(\alpha+\i \tau)S_{b}(\beta+\i \tau)S_{b}(\gamma-\i \tau)S_{b}(\delta-\i \tau)\nonumber\\
&\qquad\qquad\qquad\qquad =e^{\pi \i \left(\alpha \beta-\gamma \delta\right)} \frac{S_{b}(\alpha+\gamma)S_{b}(\alpha+\delta)S_{b}(\beta+\gamma)S_{b}(\beta+\delta)}{S_{b}(\alpha+\beta+\gamma+\delta)}\,,
\end{align}
with parameters
\begin{equation}
\alpha=\i \frac{b}{2}\frac{\theta_1}{\abs{\log q}} \qquad \beta=-\i \frac{b}{2}\frac{\theta_1}{\abs{\log q}} \qquad \gamma=b \Delta +\i \frac{b}{2}\frac{\theta_2}{\abs{\log q}} \qquad \delta=b \Delta -\i \frac{b}{2}\frac{\theta_2}{\abs{\log q}}.
\end{equation}
Therefore we obtain
\begin{equation}\label{m_element}
\bra{\theta_1} e^{-2\Delta \phi} \ket{\theta_2} = \frac{2}{\pi b} (-1)^{\Delta} \frac{S_b\left(b \Delta \pm \frac{\theta_1}{2 \pi b} \pm \frac{\theta_2}{2 \pi b}\right)}{S_b\left(2b \Delta\right)}\,.
\end{equation}
We can now take the limit $\Delta\to 0$ using $S_b(\epsilon)=1/2\pi\epsilon$. The matrix element then \eqref{m_element} vanishes unless \( \theta_1 \) and \( \theta_2 \) are close together, two of the double-sine functions in the numerator should then be expanded for small arguments. This results in
\begin{align}\label{overlap}
\left\langle \theta_1 \middle| \theta_2 \right\rangle = \frac{2 S_b\left(\pm \frac{\theta_1}{\pi b}\right)}{\pi^2 b} \lim_{\Delta \to 0} \frac{b \Delta}{b^2 \Delta^2 - \frac{(\theta_1 - \theta_2)^2}{4\pi^2 b^2}}&= \frac{4 |\log q|}{\pi b^2} S_b\left(\pm \frac{\theta_1}{\pi b}\right) \lim_{\Delta \to 0} \frac{1}{2 \pi |\log q| \Delta} \frac{1}{1 + \frac{(\theta_1 - \theta_2)^2}{4 |\log q|^2 \Delta^2}} \nonumber \\
&= \frac{4 |\log q|}{\pi b^2} S_b\left(\pm \frac{\theta_1}{\pi b}\right) \delta(\theta_1 - \theta_2),
\end{align}
In the last step we have used the definition of the delta function
\[
\delta(x) = \lim_{\epsilon \to 0} \frac{1}{\pi \epsilon} \frac{1}{1 + \frac{x^2}{\epsilon^2}}\,.
\]
Finally using the property
\begin{equation}
\frac{1}{S_b\left(\pm \frac{\theta}{\pi b}\right)} = 4 \i \sin(\theta) \sinh\left(\frac{\pi \theta}{|\log q|}\right)\,,
\end{equation}
one recovers the answer \eqref{2.14} for the density of states claimed in the main text.

\subsection{Right wavefunctions near integers}
\label{app:reg}
\begin{figure}[h]
\centering
\includegraphics[width=0.25\textwidth]{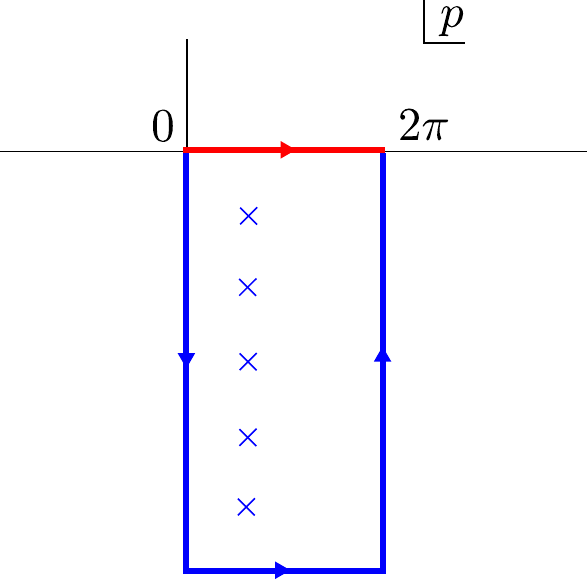}
\caption{Deformed integration contour.}
\label{fig:cdefpsiR}
\end{figure}

Consider the right eigenvectors \eqref{right1} near integers, following the same steps as in \eqref{inf}
\begin{equation}\label{refine}
    \psir_\theta(n+\i \epsilon)=\exp\bigg(-\frac{\theta^2}{2\abs{\log q}}\bigg)\int_{0}^{2\pi}\frac{\d p}{2\pi}\,\frac{e^{-\i pn}}{(e^{\i(p\pm \theta)};q^2)_\infty}\sum_{m=-\infty}^{+\infty}(q_\text{dual}^{2m-2} e^{\pi(p\pm\theta)/\abs{\log q}};q_\text{dual}^{-2})_\infty \, e^{2\pi  m \epsilon}
\end{equation}
We deform the integration contour over $p$ as in Figure \ref{fig:cdefpsiR}. The vertical segments give the same integrand up to a shift $m \to m+1$. For $\epsilon\to 0$ these two contributions cancel out. The horizontal segment near $-i\infty$ drops out due to the $e^{-ipn}$ suppression for $n>0$.

What remains is the sum over residues at non-negative integers $p = \theta-2\i k \abs{\log q}$. The terms in the sum over $m$ are independent of $k$, thus they factor out of the sum over poles, yielding a multiplicative factor
\begin{equation}
    \sum_{m=-\infty}^{+\infty}(q_\text{dual}^{2m-2};q_\text{dual}^{-2})_\infty (q_\text{dual}^{2m-2}e^{2\pi\theta/\abs{\log q}};q_\text{dual}^{-2})_\infty e^{2\pi  m \epsilon}
\end{equation}
This sum truncates to $m\leq 0$, because of the first factor. The divergence then comes from the universal contribution
\begin{equation}\label{eq:images}
    \sum_{m=-\infty}^{+\infty}(q_\text{dual}^{2m-2};q_\text{dual}^{-2})_\infty (q_\text{dual}^{2m-2}e^{2\pi\theta/\abs{\log q}};q_\text{dual}^{-2})_\infty e^{2\pi m \epsilon}\supset \sum_{m=-\infty}^0 e^{2\pi m \epsilon}=\frac{1}{1-e^{-2\pi \epsilon}}.
\end{equation}
Therefore we have obtained the more rigorous version of equation \eqref{inf}
\begin{equation}
\psir_\theta(n+\i \epsilon)=\frac{1}{1-e^{-2\pi \epsilon}}\exp\bigg(-\frac{\theta^2}{2\abs{\log q}}\bigg)\frac{\rH_n(\cos(\theta)|q^2)}{(q^2;q^2)_n}+\text{finite}\label{b.13}
\end{equation}

\subsection{Left wavefunctions near integers}\label{derivation}
The purpose of this appendix is to show analytically that the left wavefunctions \eqref{left1}
\begin{equation}
\psil_\theta(n) = \frac{1}{2 \pi}\int_{-\infty+\i \epsilon}^{+\infty+\i \epsilon} \d p \, e^{-\i p n} \, \frac{(q_\text{dual}^{-2} e^{\pi(p\pm\theta)/\abs{\log q}}; q_\text{dual}^{-2})_\infty}{(e^{\i (p \pm  \theta)}; q^2)_\infty} \exp\left(-\frac{p^2}{2 |\log q|}\right)\,,
\end{equation}
reduce on integer to equation \eqref{left2}
\begin{equation}
\psil_\theta(n) = \frac{(q^{-2}_\text{dual}; q^{-2}_\text{dual})_\infty}{(q^2; q^2)_\infty} \exp\left(-\frac{\theta^2}{2 |\log q|}\right) \rH_n\left(\cos(\theta)| q^2\right)\,,\quad n\in \mathbb{N}\,.\label{b.15}
\end{equation}
The proof goes in two steps. First, we find a different solution to the left recursion relation in \eqref{7.2} and show that is is related to the right eigenvector \eqref{right1} as
\begin{equation}\label{compare}
\bar{\psil}_\theta(n)=\frac{(e^{2\pi \i n};q^{-2}_\text{dual})_\infty}{(q^{2n+2};q^2)_\infty} \exp \left(\frac{\theta^2}{2 \abs{\log q}}\right)\psir_\theta(n)\,.
\end{equation}
Taking the limit to integer values this becomes\footnote{The prefactor in \eqref{compare} vanishes for non-negative integers. Near integers $n+\i \epsilon$ one indeed finds
\begin{equation}\label{factors}
    \frac{(e^{-2\pi \epsilon};q^{-2}_\text{dual})_\infty}{(q^{2n+2};q^2)_\infty}=(q^2;q^2)_n \frac{(q^{-2}_\text{dual};q^{-2}_\text{dual})_\infty}{(q^2;q^2)_\infty} (1-e^{-2\pi\epsilon})\,.
\end{equation}
The first and last term cancel with similar terms in the right wavefunction \eqref{b.13}.}
\begin{equation}
    \bar{\psil}_{\theta}(n)=\frac{(q^{-2}_\text{dual};q^{-2}_\text{dual})_\infty}{(q^2;q^2)_\infty}\rH_n(\cos(\theta)|q^2)\,.\label{7.71}
\end{equation}
The second step is to exploit the fact that $\psil_{\theta}(n)$ and $\bar{\psil}_{\theta}(n)$ satisfy the same recursion relation to show that 
\begin{equation}
    \psil_{\theta}(n)=\exp\left(-\frac{\theta^2}{2\abs{\log q}}\right)\bar{\psil}_{\theta}(n)\,.
\end{equation}
In combination with \eqref{7.71} this proves \eqref{b.15}. In the remainder we prove these two steps.
\begin{enumerate}
    \item In \cite{Lenells:2021zxo} non-polynomial generalizations of the polynomials of the q-Askey scheme were constructed. The non-polynomial generalization of the q-Hermite polynomial is\footnote{The contour $\Gamma$ stays above $x=\pm \sigma_s$ and $\Im(x)<\Im(\eta)/2-Q/4$ as $\Re(x)\rightarrow +\infty$. We wrote the form presented in \cite{Lenells:2021zxo} using Faddeev's quantum dilogarithm using \eqref{master}.}:
    \begin{equation}\label{Q}
        \mathcal{Q}(b,\sigma_s,\eta)=\int_{\Gamma} \mathrm{d}x \, \frac{e^{\frac{\pi \i}{6}}e^{\frac23 \pi \i Q^2} e^{-\eta \pi Q}}{\phi_b(\eta)}e^{\pi Q x} \frac{1}{\phi_b(x\pm \sigma_s)}\, e^{2 \pi i x \eta}\,,\quad Q=b+b^{-1}\,.
    \end{equation}
In \cite{Lenells:2021zxo} (Proposition 7.7) it is proven that  $\mathcal{Q}(b,\sigma_s,\eta)$ satisfies the following difference equation
\begin{equation}\label{diff2}
\left(e^{\i b \partial_{\eta}}+\left(1+e^{2\pi b (\eta-\i b/2)}\right)e^{-\i b \partial_{\eta}}\right)\mathcal{Q}(b,\sigma_s,\eta)=2 \cosh(2 \pi b \sigma_s) \mathcal{Q}(b,\sigma_s,\eta).
\end{equation}
This is identical to the left-recursion relation \eqref{7.2} upon substituting
\begin{equation}
    \eta=-\frac{\phi}{b \pi}+\frac{\i}{2b}\,,\quad\sigma_s=\frac{\i \theta}{2 \pi b}\,.
\end{equation}
Therefore \eqref{Q} is a solution to the left-recursion relation \eqref{7.2}
\begin{equation}
\bar{\psil}_\theta(\phi)=\mathcal{Q}(b,\frac{\i \theta}{2 \pi b},-\frac{\phi}{b \pi}+\frac{\i}{2b})\,.
\end{equation}
Using \eqref{7.33} to define $n$ and equation \eqref{master}, we can manipulate \eqref{Q} into
\begin{equation}
\bar{\psil}_\theta(n)=\frac{e^{\frac{\pi \i}{6}(1+Q^2)} }{\abs{\log q}}\frac{(e^{2\pi \i n};q^{-2}_\text{dual})_\infty}{(q^{2n+2};q^2)_\infty}\int_{\gamma} \mathrm{d}x \  e^{-\i n x }  \frac{(q_\text{dual}^{-2} e^{\pi(x\pm\theta)/\abs{\log q}}; q_\text{dual}^{-2})_\infty}{(e^{\i (x \pm  \theta)}; q^2)_\infty}\label{9.9math}
\end{equation}
The contour $\gamma$ is the same one as in \eqref{right}.\footnote{Equation \eqref{Q} was defined in \cite{Lenells:2021zxo} for real $b$. We can analytically continue \( b \), as described in \eqref{bb}. We can smoothly deformed the contour \( \Gamma \) in \eqref{Q} into the contour \( \gamma \) as long as no poles are crossed during the process and the convergence at \( \Re(x) \to +\infty \) is maintained, both of which are indeed the case.}
Comparing \eqref{9.9math} with \eqref{right1} we complete the first step of proving \eqref{compare}.
\item Both $\bar{\psil}_{\theta}(\phi)$ and $\psil_\theta(\phi)$ satisfy two difference equations, as they both have $b\to 1/b$ symmetry \cite{Lenells:2021zxo}. Therefore their $\phi$ dependence is uniquely determined up to a $\phi$-independent prefactor
\begin{equation}
    \frac{\psil_{\theta}(\phi)}{\bar{\psil}_{\theta}(\phi)}=f(\theta)\,,\label{7.72}
\end{equation}
This holds for any $\phi$, so we can compute $f(\theta)$ for instance at $\phi\to-\infty$
\begin{equation}\label{f}
    f(\theta) = \lim_{\phi \to -\infty} \frac{\psil_{\theta}(\phi)}{\bar{\psil}_{\theta}(\phi)} = \lim_{\phi \to -\infty} e^{\frac{\i\pi}{6}(1+Q^2)} \exp\left(-\frac{\theta^2}{2\abs{\log q}}\right) \phi_b \left(-\frac{\phi}{\pi b} + \frac{i}{2b}\right) \frac{\psil_{\theta}(\phi)}{\psir_\theta(\phi)},
\end{equation}
where we substituted the expression from \eqref{compare} and used \eqref{master}. The asymptotics of \( \psil_\theta(\phi) \) and \( \psir_\theta(\phi) \) for \( \phi \rightarrow -\infty \) have already been computed in \eqref{7.42norm}. Additionally, the asymptotic behavior of \( \phi_b(x) \) is known \cite{ip2015tensor}
\begin{equation}
    \phi_b \left(-\frac{\phi}{\pi b} + \frac{i}{2b}\right) \to  \exp \left(-\frac{\i\pi}{12}(Q^2-2) + \frac{\pi \i}{4 b^2} - \phi - \frac{\i \phi^2}{\pi b^2}\right), \qquad \phi \rightarrow -\infty.
\end{equation}

Substituting the above expansion and \eqref{7.42norm} into \eqref{f} we obtain up to a constant prefactor (that we ignore)
\begin{equation}\label{f2}
    f(\theta) = \exp\left(-\frac{\theta^2}{2\abs{\log q}}\right).
\end{equation}

A consistency check for this approach, as well as for the asymptotic calculations, is that the same result is obtained in the regime \( \phi \rightarrow +\infty \), using formula \eqref{7.37ass}. This completes the proof.
\end{enumerate}
\subsection{No discretized wavefunctions with negative renormalized lengths}\label{app:B5nonegativelengths}
In this appendix we point out that there are no reasonable wavefunctions in gauged sine dilaton gravity with support at $n<0$. Namely, a normalizable wavepacket with support on $n<0$ can be decomposed into complex energy modes, which are unphysical.

Consider the left-and right recursion relations \eqref{7.2} in terms of the $n$ and $p$ variables of section \ref{sect:gauged}
\begin{align}
\label{eq:rec2}
    2\cos(\theta)\, \psil_\theta(n)&=\psil_\theta(n+1)+(1-q^{2n})\psil_\theta(n-1)\nonumber\\
    2\cos(\theta)\, \psir_\theta(n)&=(1-q^{2n+2})\psir_\theta(n+1)+\psir_\theta(n-1)\,.
\end{align}
We considered in section \ref{sect:gauged} solutions which vanish at negative integers
\begin{equation}
    \psir_\theta(n)=\frac{\rH_n(\cos\theta|q^2)}{(q^2;q^2)_n}\,.\label{b.26}
\end{equation}
But there are also solutions to the recursion relations which vanish instead at all non-negative integers
\begin{equation}
    \psir_\theta(n)=\rH_{-n-1}( \cos(\theta)|q^{-2})\,,\quad \psil_\theta(n)=\frac{\rH_{-n-1}(\cos\theta|q^{-2})}{(q^{-2};q^{-2})_{-n-1}}\,.\label{b.27}
\end{equation}
General solutions to \eqref{eq:rec2} are uniquely specified by their values at two points, say $n=-1$ and $n=0$. Therefore they are a linear combination of \eqref{b.26} and \eqref{b.27}. We will now argue that the second set of solutions \eqref{b.27} should be disregarded.

The point is that the solutions \eqref{b.27} should be viewed as linear combinations of complex energy solutions. To appreciate this we consider the solutions \eqref{b.27} with
\begin{equation}
    \theta=\frac{\pi}{2}-\i 2 k \abs{\log q}\,,\quad E=\i \sinh(2 k \abs{\log q})\,,\quad k\in \mathbb{Z}\,.\label{b.28}
\end{equation}
Hermite polynomials of the type \eqref{b.27} were studied by
\cite{askey89} and \cite{ismail94}. The normalization for generic $\theta$ is equation (2.8) in \cite{ismail94}
\begin{equation}\label{realov}
    \langle \theta_1|\theta_2\rangle=  \sum^{\infty}_{n=-\infty} \langle \theta_1|n\rangle\langle n|\theta_2\rangle = \frac{(q^2e^{\pm\i\theta_1\pm \i\theta_2};q^2)_\infty}{(q^2;q^2)_\infty}\,.
\end{equation}
From this we see that the discrete set \eqref{b.28} is orthogonal and normalizable. They are also complete.\footnote{Completeness is often stated as an integral along the entire energy axis \cite{askey89}, however the non-uniqueness of the solution to the Stieltjes measure problem allows for many different choices of measure, including one located at our discrete energies \eqref{b.28}.} Indeed, according to equations (6.27)-(6.31) in \cite{ismail94} (we checked also numerically)
\begin{equation}
    \sum_{k=-\infty}^{+\infty}\frac{\langle n_1 |k\rangle\langle k|n_2\rangle}{\langle k |k\rangle} = \delta_{n_1n_2}\,.
\end{equation}
Any normalizable wavepacket with support on negative integers can thus be decomposed into complex energy modes.\footnote{This is different from quasinormal modes which are also a discrete set of imaginary frequency modes found in systems with continuous real energy spectrum. These quasinormal modes are not normalizable and arise for a Hermitian Hamiltonian, to be contrasted with the case considered above, where at negative length the Hamiltonian is $\i$ times a Hermitian Hamiltonian.} This includes the formal solutions \eqref{b.27} for real $\theta$. Naively one might be tempted to associate a real energy $-\cos(\theta)$ to \eqref{b.27}, but this would be wrong. This can also be appreciated from the overlap \eqref{realov}. A basic fact of linear algebra is that eigenstates with real left-and right eigenvalues are orthogonal \cite{yao2018edge}, therefore \eqref{b.27} are not real energy eigenstates.\footnote{What happens is that in deriving the relation between the left-and right recursion relation \eqref{eq:rec2} one does a summation version of integration by parts (as usually when one intends to find out how an operator acts on the bra). The boundary term in this case does not vanish, because the Hamiltonian blows up when $n\to -\infty$ due to the $q^{2n}$ term in the recursion relation, unless if integrated (summed) against wavefunctions that decay even faster towards $n=-\infty$. The wavefunctions in \eqref{b.27} do not, therefore they are not actually Hamiltonian eigenvectors.} 

The conclusion is that all reasonable (not complex energy) Hamiltonian eigenvectors in the gauged theory are the usual q-Hermite \eqref{b.26} with support on positive integers. This is related with the fact that the Lorentzian phase space \eqref{weyl} is limited to $n\geq 0$. The fact that the Hermitian version of the sine dilaton Hamiltonian \eqref{6.18} is purely complex for $n<0$ explains the complete set \eqref{b.28} of complex energy modes.
\subsection{Flat space wavefunctions}\label{app:flatwave}
We consider the flat space limit of section \ref{sect:flathermite} on the right- and left wavefunctions of sine dilaton gravity. First, we consider $\psir_\theta(n)$ defined in \eqref{right1} at integer $n$. We will rescale it by $|\log q|^\frac{n}{2}$, while simultaneously rescaling $\psil_\theta(n)$ by the opposite so that the overlaps do not change. As explained around \eqref{inf}, evaluating $\psir_\theta(n)$ at integer $n$, gives an $\infty$ that cancels with the volume of the gauge group times the contour integral of the denominator. In this case, we are then interested in the limit 
\begin{equation}\label{rightlim}
    \lim_{q\to 1} \,|\log q|^\frac{n}{2}\int_{-\pi}^\pi \frac{\d p}{2\pi}\, \frac{e^{-\i p n}}{(e^{\i(p\pm \theta)};q^2)_\infty}  = \lim_{q\to 1} \,\frac{1}{2\pi \i} \oint \d z\, \frac{z^{-n-1}}{(ze^{\pm \i \theta}\sqrt{|\log q|};q^2)_\infty}\, ,
\end{equation}
with $e^{\i p} = z |\log q|^{\frac12}$, and considering $\theta=\frac{\pi}{2}-\sqrt{|\log q|}x$ while keeping $x$ fixed. Using
\begin{equation}
   \lim_{q\to 1}\, (ze^{\pm \i \theta}\sqrt{|\log q|};q^2)_\infty =  e^{z x-\frac{z^2}{4}}\,,
\end{equation}
\eqref{rightlim} becomes the generating function for the Hermite polynomials
\begin{equation}
       \frac{1}{2\pi \i} \oint \d z z^{-n-1}e^{-z x + \frac{z^2}{4}} = \frac{\rH_n(x)}{2^n n!}\,,
\end{equation}
which are thus our relevant gauged right-eigenfunctions in section \ref{subsect:4.3}.

For the left-eigenvectors $\psil_\theta(n)$ we aim to show \eqref{eq:psilqto1}, which we reproduce here for convenience:
\begin{equation}
    \lim_{q\to 1} \, \frac{(q^2;q^2)_\infty}{|\log q|^{\frac{n}{2}}} \,e^{\theta^2/2|\log q|}\,\psil_\theta(n) = 2^\frac{n}{2}e^{x^2/2}\mathsf{D}_n(\sqrt{2}x)\,.\label{b.35}
\end{equation}
We start by picking up the poles of the momentum wave function $\psil(p)$ at $p = \pm \theta +2\i m \log q$ (others are suppressed as $q\to 1$). Rewriting the residues using
\begin{equation}
    \frac{1}{(q^{-2m},q^2)_m} = \frac{(-1)^m q^{m(m+1)}}{(q^2;q^2)_m} = \frac{(-1)^m q^{m(m+1)}(q^{2m+2};q^2)_\infty}{(q^2;q^2)_\infty}\, ,
\end{equation}
the LHS of \eqref{b.35} becomes
\begin{equation}
    \lim_{q\to 1} \sum^\infty_{m=0} \frac{q^{2m(n+1)}(q^{2m+2};q^2)_\infty (q^{2m+2}e^{-2\i \theta};q^2)_\infty (1-e^{-2\i \theta}) e^{-\i n \theta}}{|\log q|^{\frac{n}{2}}(e^{\pm 2\i \theta};q^2)_\infty (q^2;q^2)_\infty}+ \text{c.c.}\label{b.37}
\end{equation}
Useful limits are
\begin{align}
  \log (e^{\pm 2 i \theta};q^2) &\to \frac{\pi^2}{12\abs{\log q}}-x^2+\log 2\,\\
     \log (q^2;q^2)_\infty &\to -\frac{\pi^2}{12\abs{\log q}} - \frac12\log(\frac{\abs{\log q}}{\pi})\,.
\end{align}
For $\abs{\log q}\to 0$ the poles become a continuum, therefore the sum over $m$ in \eqref{b.37} is to be replaced by an integral. An intermediate step in the limit gives
\begin{equation}
   \lim_{q\to 1} e^{x^2} \int_0^\infty \d \tau\, \tau^n e^{-\i n \theta}(1-e^{-2\i \theta})(\tau \sqrt{\abs{\log q}};q^2)_\infty(\tau \sqrt{\abs{\log q}}\,e^{-2\i \theta};q^2)_\infty + \text{c.c.}
\end{equation}
Taking the limit gives an integral representation of the parabolic cylinder functions, as was to be shown
\begin{equation}
     \frac{e^{x^2}}{\sqrt{\pi}}\int^\infty_0 \d \tau\,  \tau^n e^{-\tau^2/4} \cos(\frac{n \pi}{2} - x\tau) = 2^\frac{n}{2} e^{\frac{x^2}{2}} \,\mathsf{D}_n(\sqrt{2}x)\,.
\end{equation}
Here we used 9.241.1 in \cite{gradshteyn2014table}

\bibliographystyle{utphys}
\bibliography{Refs}

\end{document}